\documentclass[prx,aps,10pt,groupedaddress,twocolumn,superscriptaddress,longbibliography]{revtex4-2}

\usepackage{float}
\usepackage{makecell}

\usepackage{amsmath,amssymb,amsthm} 

\usepackage{mathtools}

\usepackage{graphicx}
\usepackage[colorlinks=true, allcolors=blue]{hyperref}
\usepackage{hyperref}
\usepackage{mathrsfs}
\usepackage{amsfonts}
\usepackage{caption} 
\usepackage{subcaption}
\usepackage{algorithm}
\usepackage{algorithmicx}
\usepackage{algpseudocode}
\usepackage{subcaption}
\usepackage{cancel}
\usepackage{url} 
\usepackage{hyperref}
\usepackage{braket}
\usepackage{bbm}

\usepackage{bm} 
\usepackage{adjustbox}

\begin{document}


\title{Metric response of relative entropy: a universal indicator of quantum criticality}


\author{Pritam Sarkar}
\email[]{prikarsartam@gmail.com}
\affiliation{School of Physical Sciences, Indian Association for the Cultivation of Science, Jadavpur, Kolkata 700032, India}
\author{Diptiman Sen}
\email[]{diptiman@iisc.ac.in}
\affiliation{Center for High Energy Physics, Indian Institute of Science, Bengaluru 560012, India}
\author{Arnab Sen}
\email[]{tpars@iacs.res.in}
\affiliation{School of Physical Sciences, Indian Association for the Cultivation of Science, Jadavpur, Kolkata 700032, India}


\date{\today}

\begin{abstract}
The information-geometric origin of fidelity susceptibility and its utility as a universal probe of quantum criticality in many-body settings have been widely discussed. Here we explore the metric response of quantum relative entropy (QRE), by tracing out all but $n$ adjacent sites from the ground state of spin chains of finite length $N$, as a parameter of the corresponding Hamiltonian is varied. The diagonal component of this metric defines a susceptibility of the QRE that diverges at quantum critical points (QCPs) in the thermodynamic limit, even for small $n$. We study two spin-$1/2$ models as examples, namely, the integrable transverse field Ising model (TFIM) and a non-integrable Ising chain with three-spin interactions. We demonstrate distinct scaling behaviors for the peak of the QRE susceptibility as a function of $N$ using $n \leq 3$: namely, a square logarithmic divergence in TFIM and a power-law divergence in the non-integrable chain. 
We further show that this susceptibility diverges even at finite $N$ if the subsystem size, $n$, exceeds a certain value when the Hamiltonian is tuned to its classical limits due to the rank of the corresponding reduced density matrices (RDMs) being finite; unlike the divergence associated with the QCPs which require $N \rightarrow \infty$. 
\end{abstract}


\maketitle

\section{Introduction}

The highest sensitivity of qualitative properties near some points in the parameter space is the signature of a critical phenomenon, classical or quantum. Various theoretical methods of determining the nature of criticality in quantum many-body systems have previously been attempted using ideas like order parameters \cite{Heyl2017SpeedLimits}, renormalization group (RG) theory \cite{Tsai2001DMRG}, fidelity susceptibility \cite{damski, polkovnikov}, concurrence, and some entanglement measures \cite{Nielsen2005Entanglement,ZhukovEvolutionOfSingleSiteEntanglement, LarssonSingleSiteEntanglement} to name a few.

The finite-size scaling (FSS) of fidelity susceptibility or, more generally, the geometric tensor \cite{damski, Kolodrubetz_2013, Zanardi2007Geometry, Zanardi2006Fidelity} has previously been shown to reflect the universality of quantum critical phenomena in spin chains and free-fermion systems. The fidelity of the ground state in such situations naturally defines a Riemannian metric in the parameter space \cite{Kolodrubetz_2013, Zanardi2007Geometry,Zanardi2006Fidelity}. For systems with one parameter, its turning points and value at the critical point is known to be given by power laws in system size for integrable theories like the transverse field Ising model (TFIM) \cite{damski} and free-fermions \cite{Zanardi2006Fidelity} where the ground state can be analytically obtained.

More generally, consider a Hamiltonian of the form 
\begin{equation}
\hat{H} = \sum_{i \in \mathcal{L}} \hat{H}_{0i}+h \sum_{i \in \mathcal{L}} \hat{H}_{1i},
\label{eq:Hform}
\end{equation}
where periodic boundary conditions are assumed, $N$ denotes the number of lattice sites, and $\mathcal{L}$ indicates the sites of the lattice. Here, $\hat{H}_{0i}$ and $\hat{H}_{1i}$ are non-commuting local operators such that tuning the parameter $h$ to $h_c$ drives the system to a quantum critical point (QCP) in the thermodynamic limit. Consider the fidelity defined by 
\begin{equation}
 F(h,dh) = |\langle \psi_0(h+dh)|\psi_0(h)\rangle|,
 \label{eq:fid}
\end{equation}
where $|\psi_0(h)\rangle$ refers to the ground state of $\hat{H}$ for a given system size $N$. The fidelity susceptibility is then defined as
\begin{equation}
 F(h,dh \rightarrow 0) = 1-\frac{1}{2} \chi_F(h)dh^2+ \cdots.
\label{eq:fidsus} \end{equation}
At a QCP, $\chi_F \sim N^{2/\nu}$ as $N \gg 1$ where $\nu$ refers to the correlation length exponent. 


Here we propose a {\it different} model-agnostic measure to determine the metric response of 
the QRE of a quantum many-body system, inspired by the geometric interpretation of fidelity susceptibility and ground state manifold \cite{polkovnikov, Kolodrubetz_2013, Zanardi2006Fidelity}. We demonstrate how its FSS directly indicates the thermodynamic criticality through the convergence of its turning points and the divergence of its global maximum. "Universal" bears a two-fold meaning here, first as a method that can be used across different models and second as a quantity containing the information of a certain universality class. The robustness of this response function lies in its origin in information geometry \cite{nielsen, EunJinKimGeodNonEqComplex,Kim2011InfoGeometry} as it can be formally considered a quantum generalization of the Fisher-Rao information metric. 

A concrete notion of \textit{how two entanglement spectra differ} from each other can be established by a geometric distance between the spectra at their respective parameter values. Given two density matrices $\hat{\rho}$ and $\hat{\sigma}$, a measure of their distinguishability is given by the QRE defined as 
\begin{equation}
 S( \hat{\rho}\ || \ \hat{\sigma}) = \text{tr}[ \ \hat{\rho} \cdot ( \ln \hat{\rho} - \ln \hat{\sigma} )].
 \label{eq:def_of_metric_response_QRE}
\end{equation}
Given the ground state $|\psi_0(\vec{\lambda})\rangle$ for a generalization of Eq.~\ref{eq:Hform} such that $\hat{H}=\sum_{i \in \mathcal{L}} (\hat{H}_{0i}+\lambda_1 \hat{H}_{1i}+\lambda_2 \hat{H}_{2i}+\cdots)$, for a lattice with $N$ sites, a RDM for $\hat{\rho}_{\lambda}$ can be obtained by integrating out 
all but $n$ sites. Given such RDMs of any subsystem with a size not greater than the Schmidt rank of the pure state $|\psi_0(\vec{\lambda)}\rangle$ (such that $\hat{\rho}^{-1}$ is well-defined in Eq.~\ref{eq:def_of_susQRE}), we may define the QRE between RDMs at two infinitesimally separated parameter values in the following manner:
\begin{align}
&S( \hat{\rho} _{\lambda} \ || \ \hat{\rho} _{\lambda+\delta \lambda} ) = \Sigma_{ij} (\vec{\lambda}) \ d \lambda^i \ d \lambda^j \ + \ \mathcal{O}(\delta \lambda^3), \ \text{where} \ \nonumber \\
&\Sigma_{ij} (\vec{\lambda}) = \frac{1}{2} \ \text{tr}[ \ \hat{\rho} \cdot \partial_i \ln \hat{\rho} \cdot \partial_j \ln \hat{\rho} \ ] =\frac{1}{2} \ \text{tr}[ \ \hat{\rho}^{-1} \cdot \partial_i \hat{\rho} \cdot \partial_j \hat{\rho} \ ] \nonumber \\
& \implies \frac{\partial^2}{\partial x_i \partial x_j} S_{}(\rho_{\vec{\lambda}+\vec{x}} \ || \ \rho_{\vec{\lambda}}) \Big \vert_{\vec{x}=0}= \ \Sigma_{ij}(\vec{\lambda}). \label{eq:def_of_susQRE}
\end{align}
with $\partial_i \equiv \partial/\partial \lambda_i$ and $\vec{\lambda}=(\lambda_1,\lambda_2,\cdots)$. Since the QRE between subsystems at different parameter values is non-negative and attains its global minima when their arguments match, $S( \hat{\rho}_{\vec{\lambda}} \ || \ \hat{\rho}_{\vec{\lambda }+ \vec{\delta \lambda}}) \geq 0, \ \forall \ |\vec{\delta \lambda}| \to 0 \ \& \ \vec{\lambda} \in \mathbb{R}^k$, the linear contribution with 
respect to $\vec{\delta \lambda}$ vanishes in 
Eq.~\eqref{eq:def_of_susQRE}. This naturally endows the parameter space with an intrinsic distance $ds$ defined by the first fundamental form \cite{doCarmoDiffGeo}. $\Sigma_{ij}(\vec{\lambda})$ satisfies all the conditions for locally endowing a Riemannian metric on the parameter space $\vec{\lambda}$ and can be viewed as a metric response of QRE. The role of 
Kullback-Leibler divergence in Fisher information, as studied in non-equilibrium classical physics \cite{Kim2011InfoGeometry, EunJinKimGeodNonEqComplex} and quantum geometric tensor \cite{polkovnikov, Zanardi2007Geometry} serve the motivation to seek for an exact susceptibility of any entanglement spectrum in a general quantum system; using a non-perturbative response function rooted in information theory. 

 In this work, we will focus on two one-dimensional (1D) spin $S=1/2$ lattice models, both of which have a Hamiltonian that can be expressed as $\hat{H}$ given in Eq.~\ref{eq:Hform}, where tuning $h=h_c$ drives the QCP. Importantly, to get the RDMs to define $\Sigma$ (second line of Eq.~\ref{eq:def_of_susQRE}), we choose the subsystem to be small and contain consecutive spins $n$ with $n \leq 3$. For such models with a single parameter, one needs to consider the other parameters as constants and take the diagonal $\Sigma_{hh}$ as the response of QRE 
 with respect to the parameter $h$. This is similar to the relation between fidelity susceptibility and geometric tensor \cite{Zanardi2006Fidelity, Zanardi2007Geometry, polkovnikov}; hence we dub $\Sigma_{hh}$ as the susceptibility of QRE. To the best of our knowledge, no study has been made before on the intrinsic geometry of thermodynamic QRE and corresponding FSS of QCPs, from the perspective of an information theoretic measure of how sensitive the relative entropy is in the subsystems. We will show that $\Sigma_{hh}$ develops maxima as a function of $h$ for fixed $N$ where the turning points approach $h_c$ and the peak heights diverge as $N \rightarrow \infty$ for any $n$, in particular when it is small. The nature of the divergence of $\Sigma_{hh}$ as a function of $N$ in the vicinity of $h_c$ is controlled by the critical exponents of the QCP and not $n$. On the other hand, approaching the two classical limits of $h \rightarrow 0$ and $h \rightarrow \infty$, $\Sigma_{hh}$ diverges as a function of $1/h$ and $h$ respectively, but with a different power law, even for finite $N$ for subsystems of size $n$ beyond a fixed value. This latter divergence is related to the rank deficiency of singular RDMs near classical limits that leads to a non-universal divergence even for finite systems due to the presence of $\hat{\rho}^{-1}$ in the definition of $\Sigma_{ij}(\vec{\lambda})$ (Eq.~\ref{eq:def_of_susQRE}), a pathology that is absent at a QCP.


The rest of the article is organized as follows: Section \ref{sec:info_geo_origin} establishes the general connection of this response to the uncertainty of the entanglement Hamiltonian gradients, and the one-parameter generalization of relative entropy, namely, the Petz-R\'enyi relative entropy. Section \ref{sec:Scaling_susQRE} elaborates the scaling of the susceptibility of QRE at the critical points using both RG theory (Section ~\ref{subsec:RG}) and conformal field theory (CFT) (Section ~\ref{subsec:CFT}). While the results of Sec.~\ref{subsec:RG} are valid for general $d$-dimensional QCPs, Sec.~\ref{subsec:CFT} is specific to 1D QCPs. Section \ref{sec:quantum_spin_chains} introduces the quantum spin chain models under investigation, presenting the integrable TFIM and the non-integrable three-spin Ising chain with their symmetries and solutions including an efficient exact diagonalization (ED) routine for evaluating the response in non-integrable models.
Section \ref{sec:suscept_sec} presents the concrete route to constructing the RDMs using the non-vanishing local correlations present in the subsystem and calculating the response of QRE for the aforementioned models. Section \ref{sec:sig_crit__spin_chains} presents the results on FSS and critical behavior, demonstrating universal scaling laws and their connection to critical exponents. Section \ref{sec:sec_local_perturbations} analyzes the nature of the response function on approaching the classical limits ($h \rightarrow 0$ and $h \rightarrow \infty$) of the two Ising chains and examines the behavior of nearly-singular RDMs, elucidating the role of degeneracies in symmetry-broken ground states. Section \ref{sec:discussions} discusses the broader implications of our findings, including connections to strong sub-additivity of quantum entropies and how this response encodes the geometry of the relative-entropy landscape before Section \ref{sec:conclusions} concludes with a summary of contributions and future directions. 

Several additional technical details are given in the appendices. Appendices ~\ref{subsec:jordanwignerfromLSM}, ~\ref{subsec:DiagQuadHam}, and ~\ref{subsec:twopointcorrelationsinTFIM} elaborate the exact solution of the general theory of free fermions in one
dimension in association with TFIM using Jordan-Wigner transformation, diagonalization of quadratic Hamiltonians and discusses the method of calculating the two-point correlation functions in such systems. Appendix ~\ref{subsec:singular_contributions_section} presents the thermodynamic singularities of correlation functions and their relation to the response of relative entropy in TFIM. Appendix ~\ref{subsec:intrinsic_geom_TFIM} elucidates the intrinsic geometry of relative entropy in TFIM, in the thermodynamic limit. Appendix ~\ref{subsec:bitmask_algo} contains an efficient routine for ED of local Ising chains using maximal basis reduction with bitmasking, and lastly, Appendix ~\ref{subsec:derivativeops} presents the behaviors of the derivatives of local observables in the three-spin Ising model.

\section{Relation of $\Sigma_{ij} (\vec{\lambda})$ to other measures}
\label{sec:info_geo_origin}

The diagonal elements $\Sigma_{ii}(\vec{\lambda})$ (Eq.~\ref{eq:def_of_susQRE}) turn out to be the uncertainty of the gradient of entanglement Hamiltonian $\hat{H}_E := - \ln \hat{\rho}$ with respect to $\lambda_i$, because whenever $\hat{\rho}_{ \vec{\lambda}}$ is non-singular $\big\langle \partial \hat{H}_E / \partial \lambda_i \big\rangle = \partial \text{tr}\hat{\rho}/ \partial \lambda_i = 0$ so that
\begin{align}
\Delta \frac{\partial \hat{H}_E}{\partial \lambda_i} = \Big\langle \Big( \frac{\partial \hat{H}_E}{\partial \lambda_i} \Big)^2\Big\rangle = \frac{1}{2} \Sigma_{ii}(\vec{\lambda}).
\label{eq:uncertainty_interpretation}
\end{align}

This means that the response of QRE encodes fluctuations of the variation of entanglement Hamiltonian with respect to
the parameters, which is why it is the exact susceptibility of entanglement spectra contained within a fixed subsystem. On the other hand, RDMs are an ensemble of configurations of the subsystem, $A$, when information from the rest of the system, $\bar{A}$, is integrated out. Due to being positive semi-definite and symmetric, the RDM $\hat{\rho}_A$ of a subsystem $A$ is always diagonalizable. We denote $\hat{\rho}_A \ket{\xi_{A, i }} = \xi_{A, i } \ \ket{\xi_{A, i }}$, where $i = 1,2, \cdots, \dim(\mathcal{H}_A)]$, as the eigensystem of $\hat{\rho}_A$, i.e., $\{ \ket{\xi_{A, i }} \}$ then spans $\mathcal{H}_A$. So being the ensemble of configurations $\ket{\xi_{A, i }} \in \mathcal{H}_A$, the density matrix naturally induces an interpretation of "surprise" to the entanglement Hamiltonian $H_A^E$ of that subsystem because $\hat{H}^E_A = - \ln \hat{\rho}_A$ (\textit{cf.} \cite{EunJinKimGeodNonEqComplex} and Sec. 3.1 of Ref. \cite{strimmer2023statistics}). Note that, $\hat{\rho}_A$ and $\hat{H}^E_A$ are simultaneously diagonalizable \footnote{$[\hat{\rho}_A, \hat{H}^E_A] = [e^{-\hat{H}^E_A}, \hat{H}^E_A]=[\sum_{j=0}^{\infty}\frac{(-\hat{H}^E_A)^n}{n!}, \hat{H}^E_A] = 0$}, therefore in the eigenbasis of $\hat{\rho}_A$, eigenvalues of $\hat{H}^E_A$ are $\braket{\xi_{A, i} | \hat{H}^E_A |\xi_{A, \ j}}=- \ln \braket{\xi_{A, i} | \hat{\rho}_A |\xi_{A, \ j}} = - \delta_{ij} \ln \xi_{A, i}$ ; establishing that when $\xi_{A, i}$ is the probability of $\ket{\xi_{A, i}} \in \mathcal{H}_A$ to occur in the subsystem, its eigenvalue $\hat{H}^E_A \ \ket{\xi_{A, i}} = (-\ln \xi_{A, i}) \ \ket{\xi_{A, i}}$ of the entanglement Hamiltonian evaluates how surprising is its occurrence in the ensemble; while $\Sigma_{jj}$ is the fluctuation of how the surprise varies with
respect to $\lambda_j \in \mathbb{R}$ for all $j$.

Furthermore, the Petz-R\'enyi relative entropy generalizes the QRE as a one parameter family of entropies interpolating numerous operational scenarios due to capturing all the moments of a density matrix \cite{Seshadreesan_2018}. It is defined as
\begin{align}
 & S_\alpha(\hat{\rho}||\hat{\sigma}) = \frac{1}{\alpha-1} \ln \text{tr}\left[\hat{\rho}^\alpha \hat{\sigma}^{1-\alpha}\right], \ \alpha > 0,\, \alpha \neq 1, \\
 & \text{and} \ \ \lim_{\alpha \to 1} S_\alpha(\hat{\rho}||\hat{\sigma}) = S(\hat{\rho}||\hat{\sigma}). 
\end{align}

Now consider $S_{\alpha}(\rho_{\lambda+\delta \lambda} | \rho_{\lambda})$. It is non-negative and has its global minima $=0$ when $\delta \lambda = 0$.
Now using generalized binomial theorem $(x+y)^r = x^r + r x^{r-1} \ y+ (r/2)(r-1) x^{r-2} \ y^2 + \mathcal{O}(y^3)$ for any $r > 0$ and $\ln (1+y) = y+\mathcal{O}(y^2)$ when $y \to 0$, we find for all $\alpha, \vec{\lambda}$ that
\begin{align}
\lim_{|\delta \vec{\lambda}| \to 0}\frac{ S_{\alpha}(\rho_{\vec{\lambda}+\delta \vec{\lambda}} | \rho_{\vec{\lambda}})}{\delta \lambda_i \ \delta \lambda_j} = \frac{\partial^2 S_{\alpha}(\rho_{\vec{\lambda}+\vec{x}} || \rho_{\vec{\lambda}})}{\partial x_i \ \partial x_j} \Big \vert_{\vec{x}=0}= \alpha \Sigma_{ij}(\vec{\lambda}). \label{eq:RenyiRelativeToOurResponse}
\end{align}

Recent progress in direct experimental measurement of R\'enyi entropy of quantum systems \cite{Islam_2015, AbaninDemlerPRL2012} and estimation of quantum relative entropies in quantum computer \cite{lu2025estimatingquantumrelativeentropies}
paves a plausible route to measuring the metric response of QRE in physical setup.

\section{Scaling of $\Sigma_{hh}$ at a QCP}
\label{sec:Scaling_susQRE}

Let us now discuss the scaling of $\Sigma_{hh}$ near the QCPs of models where the Hamiltonian is of the form given in Eq.~\ref{eq:Hform}. We assume that the lattice has $N$ sites and periodic boundary conditions in all directions with a linear extent of $L$ (such that $N=L^d$) so that the unique ground state $|\psi_0 (h)\rangle$ (since $N$ is finite) has translational symmetry in the lattice. Since our subsystem $A$ consists of $m$ consecutive spins , we require $A$ to be smaller than or at most equal to its complement $\bar{A}$, so that the RDM may be invertible in what follows. 

\subsection{Scaling of $\Sigma_{hh}$ from a renormalization group perspective}
\label{subsec:RG}

Since RDMs are composed of all non-vanishing correlations within a connected subsystem, $\partial_i \hat{\rho}$ is composed of derivatives of all non-trivial local expectations $\partial_h \braket{\mathcal{\hat{O}}_{l}}$ due to the following identity for a subsystem (denoted by $l$) of $m-$local sites of quantum spin-$1/2$ degrees \cite{Horodecki_2009}.
\begin{align}
 \hat{\rho}_{m} = \frac{1}{2^m} \sum_{\{{\alpha}_i\} }^{[0,x,y,z]} \text{tr}[ \ \hat{\rho}_{m}\bigotimes_i\hat{\sigma}_{i}^{\alpha_i} \ ] \ \bigotimes_i\hat{\sigma}_{i}^{\alpha_i},
 \label{eq:operator_product_expansion_general}
\end{align}
where $\sum_l\braket{\mathcal{\hat{O}}_{l}}=N\braket{\mathcal{\hat{O}}_{l}}$ from translational symmetry. Importantly, any lattice operator $\sum_i \mathcal{\hat{O}}_i$ can be expanded in terms of the coarse-grained "scaling fields" at the QCP from the RG formalism. The set of allowed scaling fields that appear in the expansion is determined on the basis of global symmetries. For a generic lattice operator, a singularity develops in $\partial_h \langle \sum_i \mathcal{\hat{O}}_i \rangle$ at the QCP and is controlled by the scaling dimension of the most relevant scaling field that appears in its expansion. Since only operators with $\langle \mathcal{\hat{O}}_l \rangle \neq 0$ contribute to RDMs at finite $h$, the set of allowed scaling fields that have a finite overlap with such lattice operators have the same global symmetries as the ground state $|\psi_0(h)\rangle$ at finite $N$. On the other hand, lattice operators with different global symmetries compared to the ground state have $\langle \mathcal{\hat{O}}_l \rangle=0$ for any finite $h$. 

Typically, $\langle \hat{H}_{1i} \rangle \neq 0$ at any $h$. Furthermore, it clearly has an overlap with at least one relevant scaling field since $h$ needs to be tuned to $h_c$ to reach the QCP. The behavior of $\partial_h \langle \hat{H}_{1i} \rangle$ at the QCP for $N \gg 1$ can be understood from the following argument. The thermodynamic scaling of the second derivative of the ground state energy and its universality have been investigated in a general setting in Ref. \cite{Albuquerque__2010}. For a Hamiltonian $H = \sum_{i \in \mathcal{L}} H_{0i} + h \sum_{i\in \mathcal{L}} H_{1i}$ with periodic boundary condition, where $H_1 = \sum_{i \in \mathcal{L}}H_{1i}$ is driving a system across some QCP at $h_c$; the second derivative of the ground state energy $\chi_E = - \partial^2E_0/\partial h^2 = L^d \ \partial \braket{H_{1i}}/\partial h$ ($\forall \ i$)
has a universal scaling given by
\begin{align}
\chi_E \sim L^{2/\nu -z} \implies \frac{\partial \braket{H_{1i}}}{\partial h} \sim L^{ \ 2/\nu -(d+z)},
\end{align}
where $\nu$ and $z$ are the critical exponent of the correlation function and the dynamical critical exponent, respectively, associated with the $d$-dimensional QCP. 

From the above arguments about scaling fields and symmetry, we expect $\partial_h \langle \hat{O}\rangle \sim L^{ \ 2/\nu -(d+z)}$ for {\it generic} operators that satisfy $\langle \hat{O} \rangle \neq 0$ for finite $h$. We will demonstrate this property for both spin chains that we will study as examples in the following sections. Since $\langle \hat{O} \rangle$ themselves are $\mathcal{O}(1)$ at the critical point in the thermodynamic limit, the thermodynamic scaling of the response of QRE in Eq.~\ref{eq:def_of_susQRE} is due to the square of $\partial_h \langle \hat{O} \rangle$, all of which scale identically as long as $\hat{O}$ is not finetuned so that it does not overlap with the most relevant scaling field. We therefore conclude that
\begin{align}
 & \Sigma_m(h_c, N\to \infty) \sim L^{ \ 2 [2/\nu -(d+z) ] } \ 
 \label{eq:thermo_scaling_of_metric_response_gen}
 \\
 & \implies \Sigma_m(h \to h_c, N = \infty) \sim |h -h_c|^{- 2 [2-\nu(d+z)]}. \nonumber
\end{align}
Clearly from Eq.~\ref{eq:thermo_scaling_of_metric_response_gen}; for a $d$-dimensional QCP where $2/\nu > (d+z)$, $\Sigma_{m}(h_c, N \to \infty) \to \infty$. Moreover the monotonicity of relative entropy \cite{SSA_QuantumEntropy_Lieb} induces monotonicity of its response, giving 
\begin{align}
 \Sigma_{m' \geq m}( h, N) \geq \Sigma_{m}( h, N), \ \ \forall ~m, N, h,
\end{align}
as long as $m$ denotes the size of the smaller subsystem $A$ used to calculate the corresponding RDM. Lastly, if $H_{1i}$ corresponds to a $q$-body operator driving the system across a QCP as $h$ is tuned across $h_c$, then the singular part of all $m \geq q$- body operators overlap with the most relevant operator at the critical point.

The two examples of QCPs that we consider later in quantum spin chains $(d=1)$ correspond to an Ising universality class and a four-state Potts universality class, for which $z=\nu=1$ and $z=1, \ \nu=2/3$, respectively, while $N=L$ in both cases. This gives $\Sigma_m(h_c, N\to \infty) \sim N^0$ for the former case and $\Sigma_m(h_c, N\to \infty) \sim N^2$ for the latter case from Eq.~\ref{eq:thermo_scaling_of_metric_response_gen}. In both these examples, $H_{1i}$ corresponds to a $1$-body operator and all $1 \leq m \leq \lfloor N/2 \rfloor$ RDMs carry information about the corresponding 1D critical points (where $\lfloor\rfloor$ is the floor function to consider even and odd $N$ on equal footing).

\subsection{Scaling of $\Sigma_{hh}$ from a conformal field theory perspective in one
dimension}
\label{subsec:CFT}

For models with one parameter $h$, the Hamiltonian is always $\hat{H} = \hat{A}+h\hat{B}$ for some $\hat{A}$ and $ \hat{B}$, where both $\hat{A}$ and $\hat{B}$ are sums of local operators and are themselves extensive. Assuming a non-degenerate ground state $\ket{\psi_0(h)}$ with energy $E_0(h)$ and representation $\mathcal{O}_{ij}(h) = \braket{\psi_i(h)|\mathcal{\hat{O}}|\psi_j(h)}$ for a finite system, using first-order perturbation theory, one obtains
\begin{align}
 & \ket{\psi_0(h + dh)} = \ket{\psi_0(h)} + \sum_{n \ne 0} \frac{B_{n0}(h) \ \ket{\psi_n(h)}}{E_0(h) - E_n(h)}\ dh \nonumber \\
 & \implies \partial_h \ket{\psi_0(h)} = \sum_{n \ne 0} \frac{B_{n0}(h)}{E_0(h) - E_n(h)} \ket{\psi_n(h)}. \label{eq:psi0_derivative}
\end{align}

Any operator $\mathcal{\hat{O}}$ acting on the Hilbert space of $m$-local sites $\mathcal{H}_m$ with non-vanishing expectation in the ground state contributes in the local description of the subsystem through $\hat{\rho}_m(h)$, but for any such operator, one has
\begin{align}
 \partial_h \braket{\mathcal{\hat{O}}} &= (\partial_h \bra{\psi_0(h)}) \mathcal{\mathcal{\hat{O}}} \ket{\psi_0(h)} + \bra{\psi_0(h)} \mathcal{\mathcal{\hat{O}}} (\partial_h \ket{\psi_0(h)}) \nonumber \\
 &= \sum_{n \ne 0} \frac{B_{0n}(h)\mathcal{O}_{n0}(h)+B_{n0}(h)\mathcal{O}_{0n}(h)}{E_0(h) - E_n(h)}. \label{eq:derivative_of_local_operator_expectation}
\end{align}

At $h_c$, assuming a continuous quantum
phase transition (QPT), there is an accumulation of low-energy eigenstates near the nondegenerate ground state for any large but finite $N$. and the lowest energy gap $\Delta_c(N) = \min_n (E_n(h_c)-E_0(h_c)) \sim N^{-z}$ where $z$ is the dynamical critical exponent of the QCP. Although our analysis to arrive at Eq.~\ref{eq:derivative_of_local_operator_expectation} is general so far, we now restrict ourselves to one dimension
where $z=1$ QCPs are described by appropriate CFTs in the low-energy limit. We analyze the dominant contribution to Eq.~\ref{eq:derivative_of_local_operator_expectation} from this picture. 

For $z=1$ QCPs in 1D systems, low-lying eigenstates (including the ground state) with $E_n(h_c)-E_0(h_c) \sim 1/N$ for $N \gg 1$ and PBC are described by a CFT through the state-operator correspondence~\cite{Cardy_1996, BELAVIN1984_CFT, Cardy_Nightingale_1986_CFT, Vidal_extraction_CFT_data_2017, Vidal_CFT_&_OPEs_in_spin_systems_2020,Fradkin_2013_field_thery_in_condensed_matter}. More specifically, 
\begin{eqnarray}
 E_{\alpha}(h_c)- E_0(h_c)&=&\frac{2\pi v_{\mathrm{ex}}}{N} \left( \Delta_{\alpha}^{\mathrm{CFT}}\right) +O(N^{-x}), \nonumber \\ 
 k_{\alpha} &=& \frac{2\pi}{N} s_\alpha,
 \label{eq:CFT1}
\end{eqnarray}
where $E=v_{\mathrm{ex}}|k|$ for $|k| \rightarrow 0$ at the QCP (where $k$ is measured with respect to the momentum of $|\psi_0(h_c)\rangle$), $k_\alpha$ refers to the momentum of the state $|E_\alpha \rangle$, $x>1$ is nonuniversal while $\Delta_{\alpha}^{\mathrm{CFT}}$, $s_\alpha=0, \pm 1, \pm 2, \cdots$ are properties of the specific CFT with 
\begin{eqnarray}
 \Delta_{\alpha}^{\mathrm{CFT}} &=& (h_p+\bar{h}_p)+(N_p+\bar{N}_p), \nonumber \\
 s_{\alpha} &=& (h_p-\bar{h}_p)+(N_p-\bar{N}_p),
 \label{eq:CFT2}
\end{eqnarray}
where $(h_p,\bar{h}_p)$ refer to the conformal dimensions of a "primary field", with $\Delta_p = h_p+\bar{h}_p$ being its scaling dimension and $s_p=h_p-\bar{h}_p$ being its conformal spin, while $\Delta_{\alpha}^{\mathrm{CFT}}$ and $s_\alpha$ with $N_p, \bar{N}_p \in \mathbb{Z}^{+}$ refer to "descendants" derived from the primary field indicated by $p$. 

QCPs of lattice models are often described by CFTs with a finite number of primary fields. The eigenstates described by Eq.~\ref{eq:CFT1} can then be viewed as a tower of states created by the action of the primary fields and their descendants on the ground state. One of these primary fields is denoted by $\mathbb{I}$ (identity) with $\Delta_p=s_p=0$. The action of $\mathbb{I}$ on the ground state leaves it invariant, while the tower of states is created by its descendants. All other primary fields have $\Delta_p>0$ and may have either $s_p=0$ or $s_p \neq 0$, with each of such primary fields acting on the ground state to give a different nondegenerate excitation that acts as a base on which a tower of states arises due to the action of its descendants. The lightest such primary excitation has the smallest nonzero $\Delta_p$. 

Returning to Eq.~\ref{eq:derivative_of_local_operator_expectation}, the inverse of $E_\alpha(h_c)-E_0(h_c)$ acts as a weight of contributions from the excited states, with the least-excited states contributing the most. Thus, the nondegenerate states produced by the (non-identity) primary fields give the dominant contributions, and the descendants give contributions that are strictly less relevant. Since the lattice operator $\hat{B}$ in Eq.~\ref{eq:derivative_of_local_operator_expectation} is translationally invariant and produces excitations with the same $k$ and other global symmetries as the ground state, we need to only consider (non-identity) primary fields with conformal spin $s_p=0$ and the same global symmetry as the ground state. This generically picks out the energy density $\epsilon$ to be the lightest such primary field. 

For example, for the simplest case of an Ising CFT with central charge $c=1/2$, there are only two non-identity primaries, $\epsilon$ and $\sigma$ with $\Delta_\epsilon=1$ and $\Delta_\sigma = 1/8$ with both being scalars ($s_\epsilon=s_\sigma=0$). Although $\Delta_\sigma < \Delta_\epsilon$, $\sigma$ is odd, while $\epsilon$ and $|\psi_0(h_c)\rangle$ are even under an appropriate $Z_2$ symmetry. For more complicated CFTs, there may be multiple primary fields satisfying such symmetry constraints (this is analogous to the case of multiple scaling fields with the same global symmetries in the previous subsection), and we continue to denote the lightest such primary field by $\epsilon$. For any CFT, the matrix element of a primary field $\hat{\phi}_p$ between the ground state and the lowest non-degenerate excitation that the primary field creates is also universal and follows
\begin{equation}
 \langle E_p|\hat{\phi}_p|E_0 \rangle = \left(\frac{2\pi}{N}\right)^{\Delta_p}.
 \label{eq:CFT3}
\end{equation}
Given that $\hat{B}$ is extensive and $\hat{O}$ is local in Eq.~\ref{eq:derivative_of_local_operator_expectation} and both have a finite overlap with $\epsilon$ (the lightest non-identity primary field with $\Delta_\epsilon >0$ and the same global symmetries as the ground state) assuming a generic case, we obtain the following
\begin{eqnarray}
 \langle E_p|\hat{B}|E_0 \rangle &\sim& N a_b \left(\frac{2\pi}{N}\right)^{\Delta_\epsilon},\nonumber \\
 \langle E_p|\hat{O}|E_0 \rangle &\sim& a_o \left(\frac{2\pi}{N}\right)^{\Delta_\epsilon},
 \label{eq:CFT4}
\end{eqnarray}
as the leading behavior as $N \gg 1$, where $a_b, a_o$ denote the overlaps of the corresponding local lattice operators $\hat{H}_{1i}$ and $\hat{O}_i$ with the primary field $\epsilon$. Using Eq.~\ref{eq:CFT1} and Eq.~\ref{eq:CFT4} in Eq.~\ref{eq:derivative_of_local_operator_expectation}, we get 
\begin{equation}
 \partial_h \langle \hat{O} \rangle \sim \frac{a_b a_o}{2\pi v_{\mathrm{ex}}\Delta_\epsilon} \left( \frac{2\pi}{N}\right)^{2\Delta_\epsilon -2},
 \label{eq:derfromCFT}
\end{equation}
which implies that 
\begin{equation}
 \Sigma_{ii}\sim \frac{1}{4\pi^2 v^2_{\mathrm{ex}}\Delta^2_\epsilon} \left( \frac{2\pi}{N}\right)^{4(\Delta_\epsilon -1)}.
 \label{eq:responsefromCFT}
\end{equation}
For the Ising CFT, $\Delta_\epsilon=1$ while for the 
four-state Potts CFT, $\Delta_\epsilon=1/2$ which gives $\Sigma_{ii} \sim N^0$ in the former case and $\Sigma_{ii} \sim N^2$ in the latter case, exactly in agreement with what we obtained by applying 
Eq.~\eqref{eq:thermo_scaling_of_metric_response_gen} to these two QCPs in the previous subsection.

\section{Quantum spin chains}
\label{sec:quantum_spin_chains}

We will now concentrate on two specific interacting spin $S=1/2$ chains, namely, the transverse field Ising model (TFIM) and another Ising model with three-spin interaction terms. While the former model is integrable, which allows analytic calculations for $N \gg 1$, the latter model is not and we will resort to ED on small chains to calculate the response $\Sigma_{hh}$. In the former (latter) case, we calculate $\Sigma_{hh}$ using $n=1,2$ ($n=1,2,3$) subsystems to investigate the behavior of this quantity for small subsystems.

\subsection{An Integrable system: TFIM }

Consider the Hamiltonian of $N$ spin-$1/2$ degrees representing the quantum Ising chain in a transverse magnetic field. This exactly solvable model is given by the Hamiltonian
\begin{equation}
 H = - \sum_{j=1}^{N} \Big[ J \ \sigma_j^x \sigma_{j+1}^x + h \ \sigma_j^z \Big].
 \label{eq:TFIM_ham}
\end{equation}
This can be mapped to a fermionic system described by a quadratic Hamiltonian by performing a Jordan-Wigner transformation connecting the algebra of spin-$1/2$'s and spinless fermions \cite{polkovnikov, Lieb1961SolubleModels, Sachdev2011QPT}. This section makes use of Appendices \ref{subsec:jordanwignerfromLSM}, \ref{subsec:DiagQuadHam} and \ref{subsec:twopointcorrelationsinTFIM} in presenting the results of the ground state for $J=1$ in the TFIM. For a general Hamiltonian that is
quadratic in fermionic operators, we have 
\begin{align}
 H = \sum_{i,j} ~[c_i^{\dagger} A_{ij} c_j ~+~ \frac{1}{2}(c_i^{\dagger} \ B_{ij} \ c_j^{\dagger} + \text{H.c.})],
\end{align}
with $A_{ij}$ and $B_{ij}$ containing the complete spatial interactions as in 
Eq.~\ref{eq:A_and_B_TFIM} and Eq.~\ref{eq:PhiPsiSS}. Now a transformation to the momentum space yields the following \cite{Lieb1961SolubleModels, polkovnikov, Zanardi2006Fidelity,Zanardi2007Geometry, quantumisingforbeginners},
\begin{align}
&H = - \sum_k \Psi_k^{\dagger} \begin{pmatrix} (h- \cos k) & - \sin k \ \\ 
- \sin k \ & -(h- \cos k) 
\end{pmatrix} \Psi_k \\
& ~~~ = \sum_k \epsilon_{\mathbf{k}} \ ( \ \gamma_k^\dagger \gamma_{k} + \gamma_{-k}^\dagger\gamma_{-k} \ - 1 \ ), \ \ \ 
\text{with} \ 
\label{XYChainWithTransverseField} 
\nonumber \\
&\Psi_k =\begin{pmatrix} c_k \\ c^\dagger_{-k} \end{pmatrix}= \begin{pmatrix} \ \cos\frac{\theta_k}{2} & i \ \sin\frac{\theta_k}{2} \\ i \ \sin \frac{\theta_k}{2} & \cos\frac{\theta_k}{2} \end{pmatrix} \begin{pmatrix} \ \gamma_k \\ \gamma^\dagger_{-k} \end{pmatrix}, 
\nonumber \\
& \theta_k = \tan^{-1} \left(\frac{ \sin k}{h - \cos k} \right), \nonumber \\
& \epsilon_{\mathbf{k}} = \pm \sqrt{(h- \cos k)^2 + \sin^2 k}.
\end{align}

In this fermionic language the ground state is given as a spinless BCS ground state \cite{polkovnikov, Sachdev2011QPT} 
\begin{align}
&| \ 0 \ \rangle = \bigotimes_{k} ( \cos\frac{\theta_k}{2} +\sin\frac{\theta_k}{2} \ c^{\dagger}_{k} c^{\dagger}_{-k}\ ) | 0_k \rangle, \label{BCSgroundState} \\
&\text{so that } ~\langle 0 | 0 \rangle = 1, 
~~{\rm and}~~ \gamma_k | 0 \rangle = 0. \nonumber
\end{align}

We note that the original Hamiltonian of a periodic spin chain transforms into the Hamiltonian of a free-fermionic system differently for different parity sectors of the fermionic spectrum as parity emerges as a constant of motion of this Hamiltonian; therefore, the quantization of the Brillouin zone is different for even/odd parity of ground state and even/odd number of sites. These different cases can be compactly summarized in $\mathcal{K}^{p=0,1}_{N = \text{even/odd}}$ as in Table \ref{tab:allowed_momentum_quanta_ALL}. This does not manifest anyway at the thermodynamic limit, since the discrete sum becomes an integral over $k \in [-\pi, \pi]$, but any deviation from the allowed momentum quanta would result in an erroneous calculation of all the correlations of finite systems (\textit{cf.} Appendix \ref{subsec:jordanwignerfromLSM}). 

We are interested in the exact correlations in this spin chain, particularly for one and two sites. The elegant calculation of long-range correlation in the antiferromagnetic $XY$ chain in Ref. \cite{Lieb1961SolubleModels} is the most suitable for analytically finding the corresponding density matrices (\textit{cf.} Appendix \ref{subsec:twopointcorrelationsinTFIM}). In this language, the correlations of spin-$1/2$ degrees can be exactly cast into expectation of fermionic operators, which are often easier to calculate. Due to the Wick theorem in Eq.~\eqref{WickTheorem}, the two-point spin correlations described in 
Eq.~\eqref{eq:derivedTwoPointCorrelations} 
can be analytically expressed in terms of the solved $\{\mathbf{\phi_{k}}, \mathbf{\psi_{k}}\}$'s for the model, which are the essential building blocks of all the 
two-point correlations in this model due to 
Eq.~\eqref{basicCorrelations}. Symmetry ensures that for a single site subsystem, only $m_z = \braket{\sigma^z}$ survives as nonzero correlation function of TFIM. Due to the same symmetry again, the only non-vanishing nearest neighbor correlations are $\hat{\rho}^{\alpha}_{l, \ l+1} = \langle \hat{\sigma}_{i}^{\alpha} \hat{\sigma}_{i+1}^{\alpha} \rangle$, $\alpha = x, y, z$, for the analysis of $\Sigma_2(N,h)$ for nearest neighbors.
\begin{eqnarray}
& & G_0(h):= m_z (h, N)= \frac{1}{N}\sum_{k \in \mathcal{K}} \frac{1}{\epsilon_{\mathbf{k}}}(h- \cos k), \nonumber \\
& & G_{+1}(h) := -\frac{1}{N}\sum_{k \in \mathcal{K}} \frac{1}{\epsilon_{\mathbf{k}}}[ (h -\cos k )\cos k + \sin^2 k ], \nonumber \\
& & G_{-1}(h) := - \frac{1}{N}\sum_{k \in \mathcal{K}} \frac{1}{\epsilon_{\mathbf{k}}}[ \ (h -\cos k )\cos k - \sin^2 k ]. \nonumber \\
&& \label{eq:AllTheGs} \end{eqnarray}

The above equations are all we need to construct the exact RDMs of any adjacent sites in a finite chain with an arbitrary $N$. Furthermore, it is easy to make use of 
Eq.~\eqref{eq:derivedTwoPointCorrelations} to calculate all such correlation functions analytically. The correct discretizations of the momentum space $\mathcal{K}$ should be considered to maintain the exactness of this analysis (\textit{cf.} Table \ref{tab:allowed_momentum_quanta_ALL}) with identification $G_{0} = \braket{\sigma^z}$, $\langle \hat{\sigma}_{i}^{x} \ \hat{\sigma}_{i+1}^{x} \rangle = G_{-1}$, $\langle \hat{\sigma}_{i}^{y} \ \hat{\sigma}_{i+1}^{y} \rangle = G_{+1}$ and $\langle \hat{\sigma}_{i}^{z} \ \hat{\sigma}_{i+1}^{z} \rangle = G_{0}^2 - G_{-1}G_{+1}$ \cite{Lieb1961SolubleModels}.

\subsection{Non-integrable systems: 
Three-spin Ising chain and beyond}
\label{subsec:3SI_and_its_self_duality} 

Most of the systems of real interest are usually not integrable, and therefore their universal features are far from being analytically tractable. For such systems, the RDMs can be computed using various known methods, but for an exact analysis, we consider ED. This analysis demonstrates the potential of $\Sigma_{hh}$ for any non-integrable system and therefore further facilitates the investigation of universal features and critical phenomenon of such theories with a measure rooted in information theory. The Ising model with three-body interactions in a transverse magnetic field is 
non-integrable but it has a \textit{self-dual} critical point~\cite{sen3spinising, L_Turban_1982, Pfeuty_1982__PhaseTrans_w_MultipleInteractions,Maritan_1st_2nd_PhaseTrans_w_MultipleInteractions}. Therefore, even though the spectrum of $H_{3}$ is not solvable, the location of its QCP is exactly known. It is described by the Hamiltonian of spins-$1/2$ degrees in one dimension as 
\begin{align}
 H_3 = - \sum_{j=1}^{N} [ J \ \sigma_j^{x} \sigma_{j+1}^{x} \sigma_{j+2}^{x} + h \ \sigma_j^{z} \ ],
 \label{eq:3SpinIsing}
\end{align}
with $\sigma_j^{\alpha}, ~\alpha = x,y,z$
denoting the standard Pauli matrices at the site $j$, and we consider PBC such that $\sigma_{N+i}^{\alpha} = \sigma_{i}^{\alpha}$ for all 
$\alpha$. We fix $J=1$ in what follows.
Similar to the TFIM, the three-spin Ising model also exhibits duality on an infinitely large system \cite{sen3spinising,L_Turban_1982}. We show this by starting from the original lattice with sites labeled by an integer $j$ that goes from $-\infty$ to $+\infty$ while the sites of the dual lattice also lie at $j$. This is in contrast to the TFIM where the sites of the dual lattice lie at $j + 1/2$ so that the sites transform to edges in every dimension. The transformation of the Pauli matrices going from the original lattice $\sigma^a_j$ to the dual lattice $\tilde{\sigma}^a_j$ is given by
\begin{align}
&\tilde{\sigma}^z_{j+1} = \sigma^x_j \sigma^x_{j+1} \sigma^x_{j+2} ~~{\rm and}~~ \tilde{\sigma}^x_{j-1} \tilde{\sigma}^x_j \tilde{\sigma}^x_{j+1} = \sigma^z_j, \ \nonumber \\
&\implies \ \ \tilde{H}_3 = - \sum_{j=-\infty}^{\infty} [\tilde{\sigma}^z_{j+1} + h \ \tilde{\sigma}^x_{j-1}\tilde{\sigma}^x_j\tilde{\sigma}^x_{j+1}].
\end{align}
Here $\tilde{H}_3$ is the Hamiltonian in the dual lattice. Thus, going from $H_3$ to $\tilde{H}_3$, the transverse field $h$ gets mapped to $1/h$. So due to the self-duality $(h \leftrightarrow 1/h)$, if $H_3$ (and equivalently $\tilde{H}_3$) has a phase transition it must be at $|h| = 1$. The self-dual critical points $h_c=\pm1$ lie in the universality class of the 
four-state Potts model, different from the Ising universality class for the critical phenomenon in 
the TFIM.

The ground state of $H_3$ for a periodic chain is computed by an efficient ED routine by diagonalizing the smallest possible relevant block of the Hamiltonian using Alg. (\ref{alg:maximal_basis_reduction_with_bitmasks}) from Appendix \ref{subsec:bitmask_algo}. When the size is a multiple of $3$, $H_3$ has the $(D_1, D_2, D_3)$ symmetries, each of which admits $(\pm 1)$ eigenvalues, where 
$D_1, D_2, D_3$ are defined by 
\begin{align} &D_1 = \bigotimes_{j=1}^{N} \sigma^z_{A_j} \sigma^z_{B_j}, D_2 = \bigotimes_{j=1}^{N} \sigma^z_{B_j} \sigma^z_{C_j}, D_3 = \bigotimes_{j=1}^{N} \sigma^z_{C_j} \sigma^z_{A_{j+1}}. \nonumber
\end{align}
The ground state is known to be in the $(+1,+1,+1)$ sector \cite{sen3spinising} and also has lattice momentum $k=0$. We want the basis that block-diagonalizes $H_3$ in the $\{(+1, +1, +1), (-1, -1, +1), (-1, +1, -1), (+1, -1, -1)\}$ sectors of $(D_1, D_2, D_3)$ operators, and zero momentum sector using $(+1)$ eigenvalue of the translation operator. To maximize efficiency for exact diagonalization with finite resources, bitmasking is used to represent the action of the symmetry operators, and it generalizes to other Ising chains for different extent of locality, which is the range of interactions between bounded contiguous lattice sites. Alg. (\ref{alg:maximal_basis_reduction_with_bitmasks}) presents a way to construct the maximally reduced basis for any finite chain in order to be able to exact diagonalize largest possible Hamiltonian, given a fixed machine.

Local correlation functions are constrained by the symmetry $\mathcal{S}_{\text{3SI}}= \{ (-\sigma^x \to \sigma^x)_{{A_i, B_i}}, (-\sigma^y \to \sigma^y)_{{A_i, B_i}}, (\sigma^z \to \sigma^z)_{{A_i, B_i}},(\sigma^{z,y,z} \to \sigma^{x,y,z})_{C_i}: 
~~{\rm for} i = 1,2,\cdots,N/3]\}$. In addition to that, the translational symmetry of a periodic three-spin Ising chain ensures $\braket{\sigma^x_i} = \braket{\sigma^y_i} = 0 = \braket{\sigma^x_i \sigma^x_{i+1}} = \braket{\sigma^y_i\sigma^y_{i+1}} \neq \braket{\sigma^z_i} \ \& \ \braket{\sigma^z_i\sigma^z_{i+1}}$ for all $h \neq 0 \ \& \ N = 3L \in \mathbb{Z}$. Due to $\partial_h \hat{\rho}$, derivatives of the non-vanishing correlations are responsible for thermodynamic singularity of the response of QRE at the critical point due to 
Eq.~\eqref{eq:def_of_metric_response_QRE}. We have further observed that the self-duality of the TFIM and three-spin Ising chain extends to Ising chains with any higher degree of locality. Using the maximally reduced basis set, one is able to run the ED routine for larger systems in any fixed machine and find the same trend of criticality in any local Ising chain. We will elaborate on this in a future investigation. For this context, it should be noted that representing the symmetries as bitmasks reduces the $\mathcal{O}(N^2)$ operations to enact a symmetry on $N$ sites to $\mathcal{O}(1)$, 
for all $N$.

\section{Response of quantum relative entropy} \label{sec:suscept_sec}

Density matrices of any subsystem of a quantum system is composed of all the non-zero correlations across that subsystem. First, we calculate all the necessary correlation functions to construct density matrices of connected subsystems in an integrable theory (TFIM) and then compute the density matrices by an efficient ED routine for a non-integrable theory, the three spin Ising chain - both admitting a continuous QCP. 
Furthermore, Eq.~\eqref{eq:def_of_metric_response_QRE} naturally defines a Riemannian metric on the parameter space such that the absence of any geodesic of finite length between parameters corresponding to two distinct quantum phases characterizes the QCP. Appendix \ref{subsec:intrinsic_geom_TFIM} elaborates the simplest of such cases for a thermodynamic chain of TFIM in one dimension.

\subsection{Integrable case}

\begin{figure}[ht]
\centering
\includegraphics[width=\linewidth]{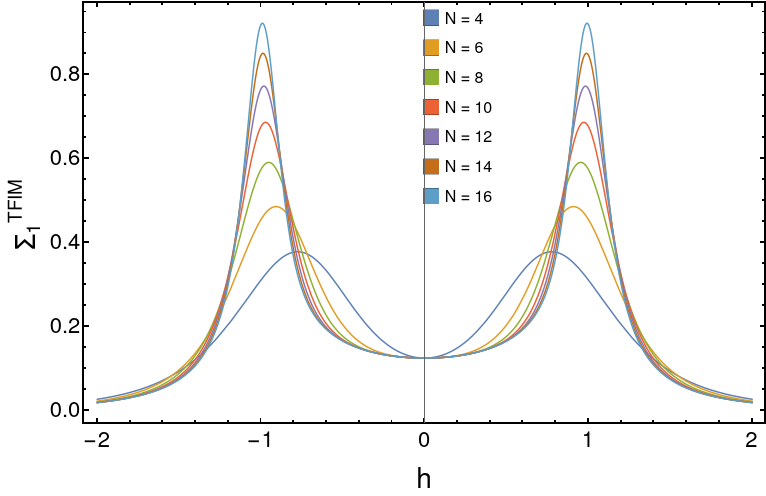}
\caption{Response of QRE for a single site, $\Sigma_1^\text{TFIM}(h,N)$, in the TFIM for some small values of $N$.}
\label{fig:tfim_response_one_site}
\end{figure}

\begin{figure}[h]
\centering
\includegraphics[width=\linewidth]{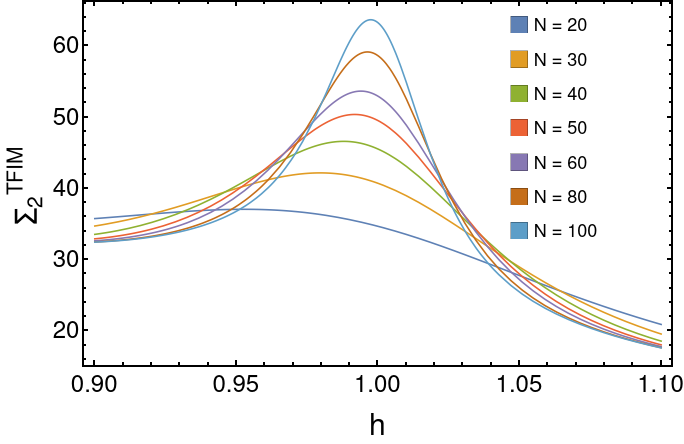}
\caption{Response of QRE for two adjacent sites, $\Sigma_2^\text{TFIM}(h,N)$, in the TFIM for some small values of $N$.}
\label{fig:tfim_response_two_sites}
\end{figure}

The free-fermionic solution of the TFIM \cite{polkovnikov, Jordan1928Wigner,Lieb1961SolubleModels} allows one to analytically calculate all the correlation functions necessary to construct the density matrix for any subsystem. We use that to calculate the response of QRE for the single-site and nearest-neighbors ($n=1$ and $n=2$) as in Figs. \ref{fig:tfim_response_one_site} and \ref{fig:tfim_response_two_sites}. Being the smallest local description of the entanglement spectra in a many body quantum system, the single site RDM is the ensemble of possible configurations for it to be a subsystem of a given state. So proceeding with the calculation for the ground state, for a single spin $1/2$ degree in a translationally invariant theory, we have
\begin{align}
& \hat{\rho} = \frac{e^{ - \beta \hat{H}}}{\text{tr}[e^{ - \beta \hat{H}}]} \to \hat{\rho}^{(1)} = \frac{1}{2} \sum_{\alpha}^{ [ 0,x, y, z] } \text{tr}[\hat{\rho} \ \hat{\sigma}^{\alpha}] \ \hat{\sigma}^{\alpha}. 
\end{align}
In our context, $\beta \to \infty$ since we will examine the ground state without any thermal fluctuation. The formalism in this section holds equally for finite and thermodynamically large systems, as it requires no more change than replacing finite sums with integrals in the expressions of correlation functions. The Hamiltonian possesses the symmetry $ \mathcal{S}_{\text{TFIM}} :=\{ \hat{\sigma^{x}} \to -\hat{\sigma^{x}}, \ \hat{\sigma^{y}} \to - \hat{\sigma^{y}}, \ \hat{\sigma^{z}} \to \hat{\sigma^{z}} \}$, which is maintained by the ground state of every finite system, except for the one with infinitely many spins which spontaneously breaks the underlying $\mathbb{Z}_2$ symmetry for $|h|<1$. This makes the expectation of $\hat{\sigma}^{x, y}$ vanish in any finite chain. The response of QRE for a single site in TFIM for some sizes is given in Fig. \ref{fig:tfim_response_one_site} corresponding to the following reduced density matrix
\begin{align}
& \hat{\rho}^{(1)}_{\text{TFIM}}(h,N) = \frac{1}{2} ( \mathbf{1}_{2} + m_z(h,N) \ \hat{\sigma}^{z} \ ) , \ ~\text{where} \label{eq:singleSiteReduceddensityMatrix} \\
& m_{z}(h,N) = \langle \ \sigma^{z} \ \rangle_{\text{TFIM}} \equiv G_0. \nonumber
\end{align}
The ground state $| 0 \rangle$ is given by the spinless BCS ground state of free fermions as in Eq.~\eqref{BCSgroundState}. Therefore for a single-site the only non-vanishing one-site correlation function $G_0$, and its derivative are behind the finite-size scaling and thermodynamic singularity of this response function.
\begin{equation}
 \Sigma_{1}^\text{TFIM}(h,N) = \frac{1}{2} \frac{(\partial_{h}m_z(h,N))^2}{1- m^2_z(h,N)}.\label{eq:TFImOneSiteSuscept}
\end{equation}

Similarly, the nearest-neighbor RDM for any spin chain is given 
by~\cite{Nielsen2005Entanglement} 
\begin{align}
\hat{\rho}_{\text{NN}}^{(2)}:= \hat{\rho}_{i, i+1} = \frac{1}{4} \sum_{\alpha, \beta }^{[0,x,y,z]} \text{tr}[ \ \hat{\sigma}_{i}^{\alpha} \ \hat{\sigma}_{i+1}^{\beta} \ ] \ \hat{\sigma}_{i}^{\alpha} \otimes \hat{\sigma}_{i+1}^{\beta},
\end{align}
with $\{ \hat{\sigma}_{i}^{\alpha} \ | \ \alpha \in [0, x, y, z] \}$ spanning the space of operators acting on the Hilbert space of a single site. For our case the non-vanishing correlation functions contributing in this expression are $\langle \sigma^z_i\rangle, \ \langle \sigma^x_i \sigma^x_{i+1}\rangle, \ \langle \sigma^y_i \sigma^y_{i+1}\rangle, \ \langle \sigma^z_i \sigma^z_{i+1}\rangle$ because rest of them vanish due to the symmetry $\mathcal{S}_\text{TFIM}$. Now directly from 
Eq.~\eqref{eq:AllTheGs} we can read off the 
nearest-neighbor reduced density matrix (\textit{cf.} Appendix \ref{subsec:twopointcorrelationsinTFIM})
\begin{align}
\hat{\rho}_{\text{TFIM}}^{(2)}(h) & = \frac{1}{4}[ \ \braket{\sigma^z_i} (\sigma^z \otimes \mathbf{1} + \mathbf{1} \otimes \sigma^z) \nonumber \\
&~~~~ + \braket{\sigma^x_i\sigma^x_{i+1}} \ \sigma^x \otimes \sigma^x 
+ \braket{\sigma^y_i\sigma^y_{i+1}} \ \sigma^y \otimes \sigma^y \nonumber \\
&~~~~ + \braket{\sigma^z_i\sigma^z_{i+1}} \ \sigma^z \otimes \sigma^z \ ].
\label{eq:NNReducedMatrixOfTFIM}
\end{align}
This expression is exact for any finite system, and due to the exact solvability, we can calculate these for very large systems by performing a sum of $\mathcal{O}(N)$ terms. It is also crucial to use the correct quantization of the Brillouin zone as in Table \ref{tab:allowed_momentum_quanta_ALL} for obtaining the correct results for finite systems, although these are equivalent in the thermodynamic limit. For periodic chains in the case of TFIM, the ground state of a chain of even number of sites is in the even parity sector for all $h$, while that parity flips as $h$ changes sign for all chains of odd sizes. So we use $\mathcal{K}_{\text{even}}^0(N)$. These allow us to determine the response function using the nearest-neighbor RDMs as in 
Fig.~\ref{fig:tfim_response_two_sites}.


Although the analysis of the previous section indicates that the derivatives of all the non-vanishing correlation functions $\braket{\sigma^z_i}, \braket{\sigma^x_i\sigma^x_{i+1}}, \braket{\sigma^y_i\sigma^y_{i+1}}$ and $\braket{\sigma^z_i\sigma^z_{i+1}}$ within two local sites scale as $N^0$ and thus do not have a power-law divergence in $N$, the exact results indicate a divergence with $N$ in finite systems as in Fig. \ref{fig:tfim_response_one_site} and (\ref{fig:tfim_response_two_sites}). An exact analysis later reveals a slower logarithmic divergence of the derivatives of the correlations (for
example, see Eq.~\eqref{eq:correlation_and_their_derivaties_in_TFIM}) due to the massless free-fermionic spectrum at the critical point resulting from $\epsilon_{k} \sim k$.

\subsection{Non-integrable case}

For the non-integrable case, we can \textit{exact diagonalize} the Hamiltonian in its computational basis, solve for the eigenvector corresponding to the smallest eigenvalue, create the projector of it to obtain the full density matrix of the system and then partially trace out all the sites except $m$ consecutive spins to compute the appropriate RDMs at any $h$ (denoted by ${}^3\hat{\rho}_{m}$). We do this for $m = 1,2,3$
to compute the response of QRE using 
Eq.~\eqref{eq:def_of_metric_response_QRE} by plugging in the ${}^3\hat{\rho}_{m}$'s for each $m$ as a function of the parameter $h \in \mathbb{R}$. Since this model has a single parameter, it makes it an easier platform to test our hypothesis that the proposed measure is universally applicable. So we have
\begin{equation}
 {}^3\Sigma_m(N,h) := \frac{1}{2} \ \text{tr}[ \ {}^3\hat{\rho}_{m} (\partial_h \ln ({}^3\hat{\rho}_{m}))^2 \ ]. 
 \label{eq:susceptfor3SpinIsing}
\end{equation}
The steps are implemented using an open-source Python package QuSpin for computations involving ED and quantum dynamics \cite{Quspin_2017}, \footnote{Because QuSpin allows to (1) define the computational basis as the spectrum of $ \bigotimes_{i}\sigma^z_i$, (2) define the Hamiltonian in this basis easily for any local operator, and (3) take trace of reduced density matrices using \textit{in-built methods} in the library, for simple and straightforward implementation of this idea. See their \href{https://quspin.github.io/QuSpin/}{documentation} for more details. But in principle one can write the whole routine in any programming language} using the algorithm discussed in Appendix~\ref{subsec:bitmask_algo}. In Fig. ~\ref{fig:three_spin_Ising_three_RDMs}, we present the ${}^3\Sigma_m(N,h)$ for $m = 1,2,3$ over a range of
$h$ around $h_c=1$. Using the same prescription, one can exactly analyze the FSS of this measure for any non-integrable quantum system. 

We evaluated all relevant correlations and their derivatives numerically using the QuSpin library~\cite{Quspin_2017}. Since the Hamiltonian in Eq.~\eqref{eq:3SpinIsing} is real-Hermitian in the eigenbasis of $\bigotimes_{i} \hat{\sigma}^z_i$, any eigenstate can be chosen to be in $\mathbb{R}^{2^N}$ because if $v$ is a eigenvector of $H_3$ with eigenvalue $\lambda$ then due to $\lambda v^\dagger=(\lambda v)^\dagger= (H_3 v)^\dagger= H_3 v^\dagger$, one always has $(v+v^\dagger) \in \mathbb{R}^{2^N}$ as an eigenstate. 
Therefore, in this basis, the expectation of any purely imaginary operator $\mathcal{O}_{ij}$ vanishes due to $\langle \psi | \hat{\mathcal{O}} | \psi \rangle = \langle \psi | \hat{\mathcal{O}} | \psi \rangle^* = -\langle \psi | \hat{\mathcal{O}} | \psi \rangle \Rightarrow \langle \psi | \hat{\mathcal{O}} | \psi \rangle = 0$. This ensures that expectation of any Hermitian operator including an odd number of $\hat{\sigma}^y_i$ vanishes. Furthermore, the symmetry $\mathcal{S}_{\text{3SI}}$ ensures that all the non-vanishing correlations within three consecutive sites are: $G_0 = \braket{\sigma^z_i}, G_{zz} = \braket{\sigma^z_i \sigma^z_{i+1}}, G_{z0z}=\braket{\sigma^z_i \sigma^z_{i+2}}, G_{zzz}=\braket{\sigma^z_i \sigma^z_{i+1}\sigma^z_{i+2}}, G_{xxx}=\braket{\sigma^x_i \sigma^x_{i+1}\sigma^x_{i+2}}, G_{yyx}=\braket{\sigma^y_i \sigma^y_{i+1}\sigma^x_{i+2}}=G_{xyy}$, and $G_{yxy}=\braket{\sigma^y_i \sigma^x_{i+1}\sigma^y_{i+2}}$. These correlations
and their derivatives are all that is behind the response of QRE within at most three sites of three-spin Ising chain. Using Eq.~\eqref{eq:operator_product_expansion_general}, the RDMs of $[1,2,3]$-local sites for the three-spin Ising chain are found to be
\begin{widetext}
 \begin{align}
& {}^3\hat{\rho}_1 = \frac{1}{2} \Big[ \mathbf{1}_2 + G_0 \ \hat{\sigma}^z \Big], \ {}^3\hat{\rho}_2 = \frac{1}{4} \Big[ \mathbf{1}_4 + G_0 ( \hat{\sigma}^z \otimes \mathbf{1} + \mathbf{1} \otimes \hat{\sigma}^z) + G_{zz} \hat{\sigma}^z\otimes \hat{\sigma}^z \Big],\label{eq:3SI__rdm___formulae} \\
& {}^3\hat{\rho}_3 = \frac{1}{8} \Big[ G_0 \ ( \hat{\sigma}^z \otimes \mathbf{1}\otimes \mathbf{1} + \mathbf{1} \otimes \hat{\sigma}^z\otimes \mathbf{1} + \mathbf{1}\otimes \mathbf{1}\otimes \hat{\sigma}^z) \ + G_{zz} \ (\hat{\sigma}^z\otimes \hat{\sigma}^z\otimes \mathbf{1} + \mathbf{1} \otimes\hat{\sigma}^z\otimes \hat{\sigma}^z) \ + \ G_{z0z} \ \hat{\sigma}^z\otimes\mathbf{1} \otimes \hat{\sigma}^z \ + \nonumber \\
& (G_{yyx}+G_{xyy}) \ \hat{\sigma}^y \otimes\hat{\sigma}^y\otimes \hat{\sigma}^x \ + \ G_{yxy} \ \hat{\sigma}^y \otimes\hat{\sigma}^x\otimes \hat{\sigma}^y + G_{xxx}\ \hat{\sigma}^x \otimes\hat{\sigma}^x\otimes \hat{\sigma}^x + G_{zzz} \ \hat{\sigma}^z \otimes\hat{\sigma}^z\otimes \hat{\sigma}^z + \mathbf{1}_8 \Big].
\nonumber
 \end{align}
\end{widetext}

For each of these subsystems, the corresponding response functions are unique analytic functions of all the correlations contained within them. 
The analytical expressions for $m=1,2$ are given below, while we omit the long expression for ${}^3\Sigma_3$ for brevity.

\begin{widetext}
\begin{align}
& {}^3\Sigma_1 = \frac{1}{2}\frac{(\partial_hG_0)^2}{1-G_0^2}, \ \ \ {}^3\Sigma_{2} = \frac{
4G_{0}\ \partial_{h}G_{0}\ \partial_{h}G_{zz}\ ( 1 - G_{zz} )
+ \left( \partial_{h}G_{zz} \right)^{2} \left( 2\,G_{0}^{2} -1 - G_{zz} \right)
+ 2 \left( \partial_{h}G_{0} \right)^{2} \left( -1 + G_{zz}^{2} \right)
}{
2\left( -1 + G_{zz} \right)\left( -4\,G_{0}^{2} + 
\left( 1 + G_{zz} \right)^{2} \right)} \ .
\end{align}
\end{widetext}

\section{Signature of criticality in quantum spin chains}
\label{sec:sig_crit__spin_chains}

\begin{figure*}[htbp]
\centering
\makebox[\textwidth]{ 
 \begin{minipage}[t]{0.325\textwidth}
 \centering
 \includegraphics[width=\textwidth, keepaspectratio]{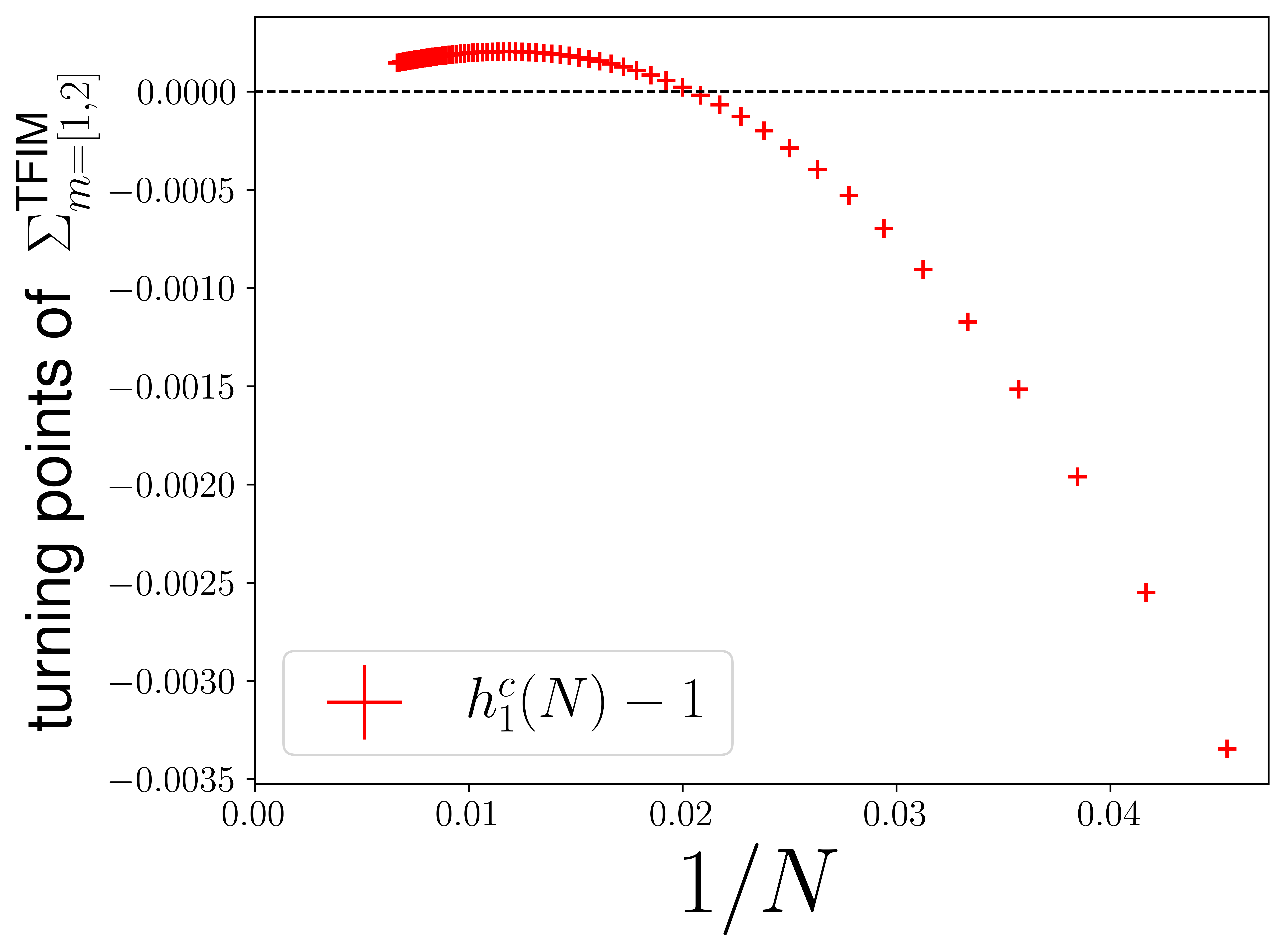}
 \subcaption{}
 \label{fig:nonMonotonicPeakLocationIn_TFIM}
 \end{minipage}
 \hfill
 \begin{minipage}[t]{0.33\textwidth}
 \centering
 \includegraphics[width=\textwidth, keepaspectratio]{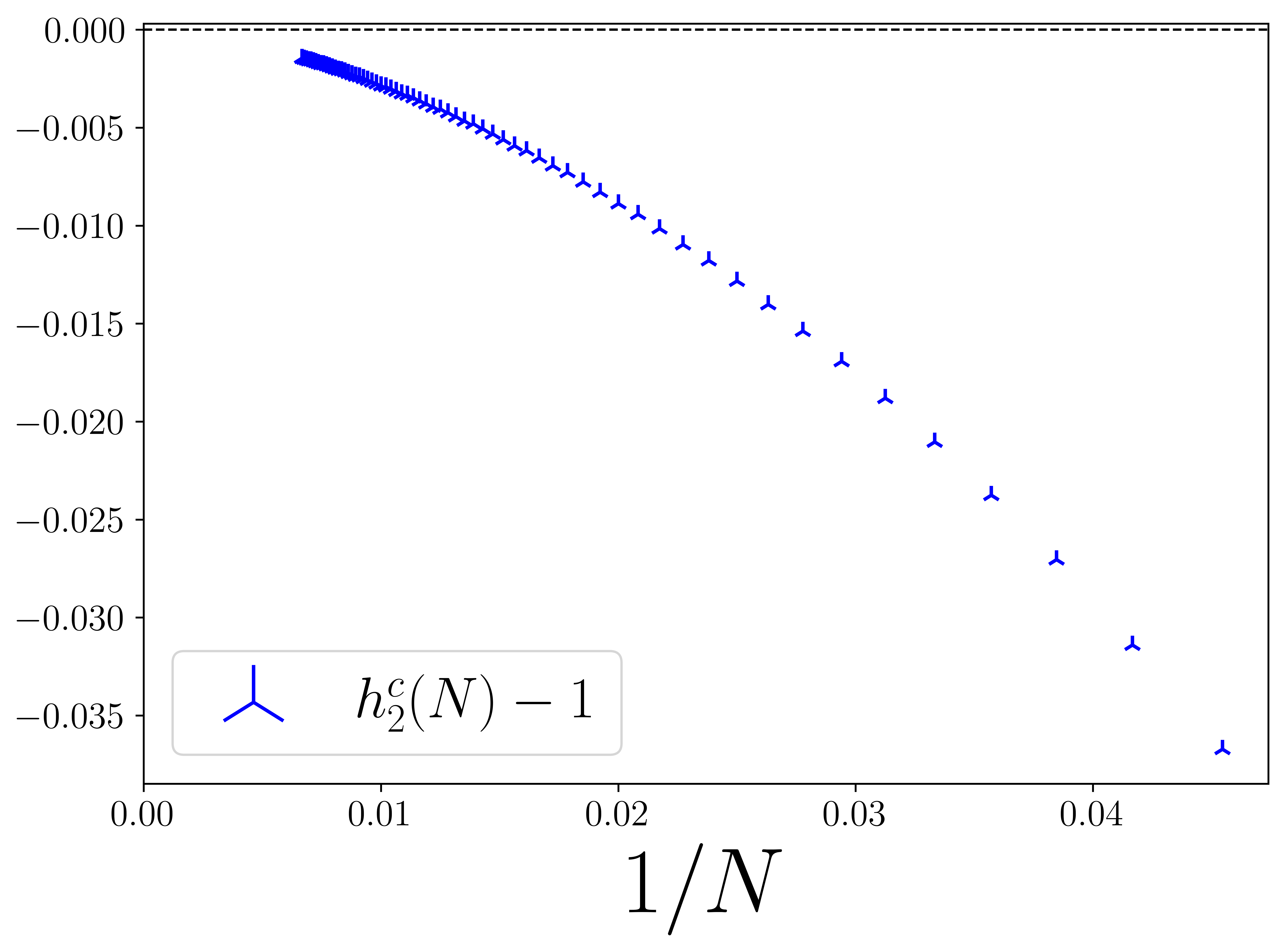}
 \subcaption{}
\label{fig:MonotonicPeakLocationIn_TFIM_two_sites}
 \end{minipage} 
 \hfill
 \begin{minipage}[t]{0.335\textwidth}
 \centering
 \includegraphics[width=\textwidth, keepaspectratio]{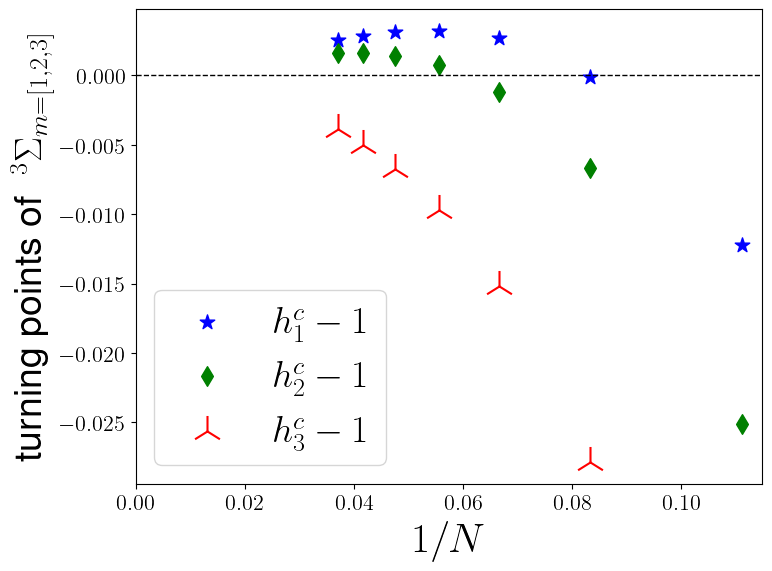}
 \subcaption{}
 \label{fig:3SpinIsing__all_peak_locations}
 \end{minipage} 
 }
 \hfill
 \makebox[\textwidth]{ 
 \begin{minipage}[t]{0.48\textwidth}
 \centering
 \includegraphics[width=\textwidth, keepaspectratio]{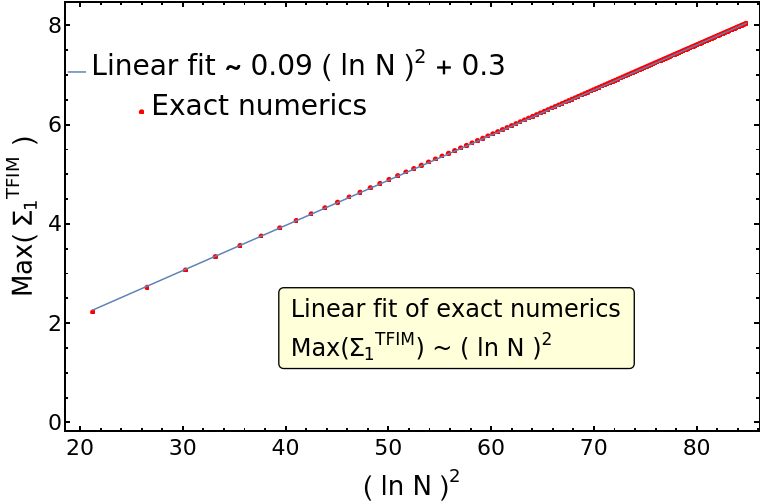}
 \subcaption[]{}
 \label{fig:tfim_peak_heights}
 \end{minipage}
 \hfill
 \begin{minipage}[t]{0.49\textwidth}
 \centering
 \includegraphics[width=\textwidth, keepaspectratio]{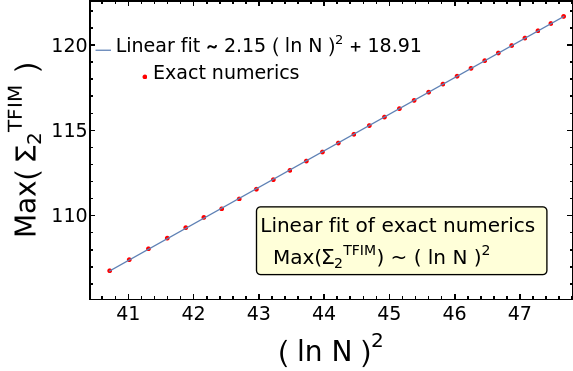}
 \subcaption[]{}
 \label{fig:tfim_peak_heights2}
 \end{minipage}}
\caption{Turning points of the response of QRE for (a) a single site, (b) neighboring sites in the TFIM, (c) local 
regions with 1, 2, 3 sites in the three-spin Ising chain, (d) scaling of the global maxima of the response of QRE $\max[\Sigma_1^{\text{TFIM}}(N)]$ with $N$, and (e) scaling of the global maxima of the response of QRE $\max[\Sigma_2^{\text{TFIM}}(N)]$ with $N$.}
\end{figure*}

Given $m \leq \lfloor N/2\rfloor $ and a bipartition $(m,N-m)$ of a 1D quantum system of length $N$, consider $\text{Sch}(\ket{\psi^{gs}_{h}},m)=$ Schmidt rank of that bipartition of the ground state at the point $h$ in the parameter space. Then a RDM $\hat{\rho}_m(h)$ is full-rank and thus invertible if and only if $\dim[\hat{\rho}_m(h)] = \text{Sch}(\ket{\psi^{gs}_{h}},m)$. Since diverging quantum fluctuations at a QCP ensure that $\text{Sch}(\ket{\psi^{gs}_{h_c}},m) = \dim[\hat{\rho}_{m}(h_c)] $, RDMs of every subsystem is invertible at a QCP. Furthermore, for a system $H = H_0 + h \mathcal{V}$, with $\mathcal{V}$ being a $q$-body operator driving the system across a QCP for some value of $h_c$; singular part of all $(m\geq q)$-body operators overlap with the most relevant operator for the critical point. Since TFIM and three-spin Ising models are driven by a $(q=1)$-body operator, RDMs for all $1 \leq m \leq \lfloor N/2 \rfloor$ subsystems can be used to access the thermodynamic QCPs using the data at finite $N$ and performing appropriate FSS analysis. 


With $t=|h-h_c|$ and $\zeta = 2 [2-\nu(d+z)]$ 
Eq.~\ref{eq:thermo_scaling_of_metric_response_gen} gives us $\Sigma_m \sim t^{-\zeta}$, which implies 
\begin{eqnarray}
\Sigma_m = L^{\zeta/\nu} \ \psi_m(L^{1/\nu} t)
\label{eq:universalscaling}
\end{eqnarray}
for a universal function $\psi_m(y)$ \cite{cardy1988FiniteSizeScaling} from a standard scaling ansatz. If $\psi_m(y)$ has a peak at $y_0$, then a plot of $\Sigma_m$ vs $h$ will have a peak at $h = h_c + y_0 \ L^{-1/\nu}$ and a maximum value of $L^{\zeta/\nu} \psi_m(0)$. However, the shift exponent can differ from $1/\nu$ in certain cases as reported in Ref. \cite{cardy1988FiniteSizeScaling}. For example, this happens for the turning points of fidelity susceptibility in TFIM where $\nu=1$ but the shift exponent of its turning point is $2$ instead of $1/\nu=1$, as calculated in Ref. \cite{damski}. We will see a similar feature for the turning points of $\Sigma_{hh}$ for $n=1,2$ for the TFIM below (e.g., see Eq.~\ref{eq:turningpointTFIM}). Nevertheless, Figs. ~\ref{fig:Scaling_Collapse__3SpinIsingRDM__1} and ~\ref{fig:Scaling_Collapse__3SpinIsingRDM__2} demonstrate that the shift exponent derived from the universal scaling ansatz (Eq.~\ref{eq:universalscaling}) agrees with the turning points of the response of relative entropy in the three-spin Ising model.

\subsection{Scaling for the TFIM}

In the following, we report the FSS analysis of the turning points and the maxima of the response of QRE for the TFIM. For the analysis, we will use both the single-site reduced density matrix $\hat{\rho}^{(1)}$ 
(Eq.~\eqref{eq:singleSiteReduceddensityMatrix}) and the nearest-neighbor reduced density matrix $\hat{\rho}^{(2)}_{NN}$ 
(Eq.~\eqref{eq:NNReducedMatrixOfTFIM}). The allowed momentum modes are $\mathcal{K}^0_\text{even}(N)$ as we stick to even sizes (\textit{cf.} Table \ref{tab:allowed_momentum_quanta_ALL}).

\subsubsection{Numerical results}


There is no phase transition for any finite system, as the susceptibility remains finite. For an infinite system, it diverges at the thermodynamic critical points $h=\pm 1$. Starting from smallest sizes $\Sigma_{s}^{\text{TFIM}}(N,h)$ converges to the thermodynamically singular behavior as $h \to 1$ with its turning points approaching to the QCP, maxima diverging as $\ln^2 N$ and monotonic decrease in the full-width at half-maxima (FWHM) (Fig \ref{fig:tfim_response_one_site}). For a single site, the two turning points near $h=\pm 1$ cross the thermodynamic critical point for $N=48$ to $N= 50$, from the ferromagnetic side to the paramagnetic one, and eventually approaches to the QCP asymptotically as $N \to \infty$ (Fig. \ref{fig:nonMonotonicPeakLocationIn_TFIM}). But for a pair of neighboring sites the 
finite-size turning points approach the QCP strictly from the ferromagnetic side as in Fig. \ref{fig:MonotonicPeakLocationIn_TFIM_two_sites}. Since the analysis is identical with
respect to the sign of $h$, we stick to the FSS near $h=1$. Then we have
\begin{equation}
\max_h[\Sigma_{1}^\text{TFIM}] \sim \ln^2 N \sim \max_h[\Sigma_{2}^\text{TFIM}]. \label{eq:exact_numerics_TFIM_metric_response} \end{equation}

As is evident from Figs.~ \ref{fig:tfim_response_one_site} and \ref{fig:tfim_response_two_sites}, the maximum response scales as $\ln N$ with an exponent $2$ (Fig.~\ref{fig:tfim_peak_heights}). This is due to the singular contributions of the derivatives of correlation functions involved as a reduced version of fidelity susceptibility has been studied to diverge $\sim \ln^2 N$ in the Ref. \cite{reduced_fidelity_susceptibility} (\textit{cf.} Appendix \ref{subsec:singular_contributions_section}). Similarly we can harness Eq.~\eqref{eq:NNReducedMatrixOfTFIM} to calculate the nearest-neighbor reduced density matrix $\hat{\rho}_{NN}$ for any given size $N$ and we find the same trend as observed in Fig. \ref{fig:tfim_peak_heights2}. 

\subsubsection{Analytical results for the thermodynamic singularity}

The derivatives of the correlation functions cause the response of QRE to become singular in the thermodynamic limit of $N = \infty$ where the system changes its phase at $h=1$. However, for any finite $N$, the response maximizes at $h^c_d(N)$. Due to the asymptotic convergence $h_d(N) \to 1$, the value of the response at $h=1$ captures its square-logarithmic divergence. For chains of even sizes, the quantization of the Brillouin zone $\mathcal{K}^0_{\text{even}}$ from Table \ref{tab:allowed_momentum_quanta_ALL} transforms $\sum_{k} \to N/(2\pi) \int^{\pi}_{-\pi}dk$ where all correlation functions remain finite at $h=1$ but not any of their derivatives remain finite because they all contain poles at $k=0$. Furthermore, due to the symmetry of the correlations $G_0, G_{xx}, G_{yy}$ and $G_{zz}$ around $k=0$, the summation runs twice over $\mathcal{K}^{0,+}_{\text{even}}(N) \in [\pi/N, \pi]$, so in the thermodynamic limit, integrals with singularity at $k=0$ are ill-defined for $\mathcal{K}^{0,+}_{\text{even}}(N =\infty)$; therefore, it is necessary to evaluate $\int_{\pi/N}^\pi dk$ and then allow $N \to \infty$ so that $|k_{\text{min}}| = \pi/N$ corresponds to the source of singularity. Closely investigating the Laurent series of the summands of the correlations at $h=1$, one finds a pole of order $1$ as $|k| \to 0$ therefore the first derivatives contribute as $ (-\ln |1/k_{\text{min}}|) \sim \ln N$ (\textit{cf.} Appendix \ref{subsec:singular_contributions_section}) - leading to a divergence in the thermodynamic limit. TFIM features $n$-th order pole at $k=0$ for any $n$-th derivative of a correlation function when $n$-is odd, but admits an $(n-1)$-th order pole when $n$-is even, so all of them diverge. For a single site with $h\ge0$, only the non-vanishing correlation $\braket{\sigma^z}$ takes the form of 
Eq.~\eqref{eq:thermo_mz_def_EliptInt} with $J=1$. This leads to $d m_z/dh \vert_{h \to 1} \sim \ln |h-1|/\pi$ and $m_z(1, \infty) = 2/\pi$. Therefore, using 
Eq.~\eqref{eq:TFImOneSiteSuscept} one can obtain the singular contribution of the response of QRE as

\begin{align}
\Sigma_{1}^\text{TFIM}(h \to 1,\infty)_{\text{sing}} = \frac{\ln^2 |h-1|}{ 2(\pi^2-4)}. 
\end{align}

\begin{figure*}[htbp]
 \centering
 \makebox[\textwidth]{ 
 \begin{minipage}[t]{0.33\textwidth}
 \centering
 \includegraphics[width=\textwidth, keepaspectratio]{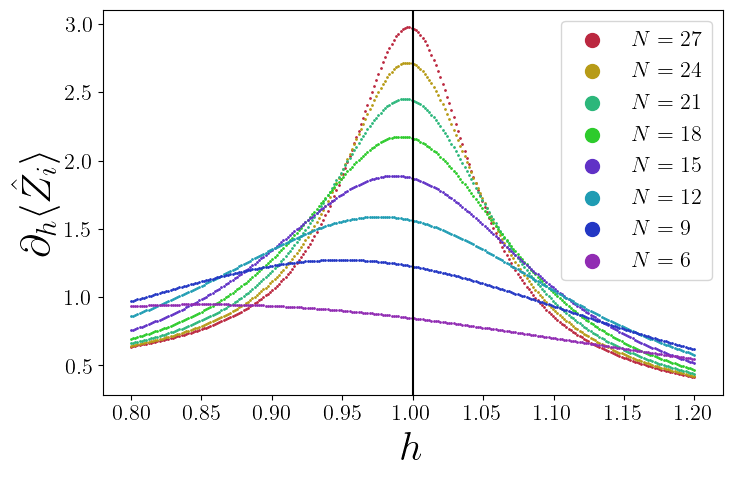}
 \subcaption{}
 \label{fig:dh_Z_3SI}
 \end{minipage}
 \hfill
 \begin{minipage}[t]{0.33\textwidth}
 \centering
 \includegraphics[width=\textwidth, keepaspectratio]{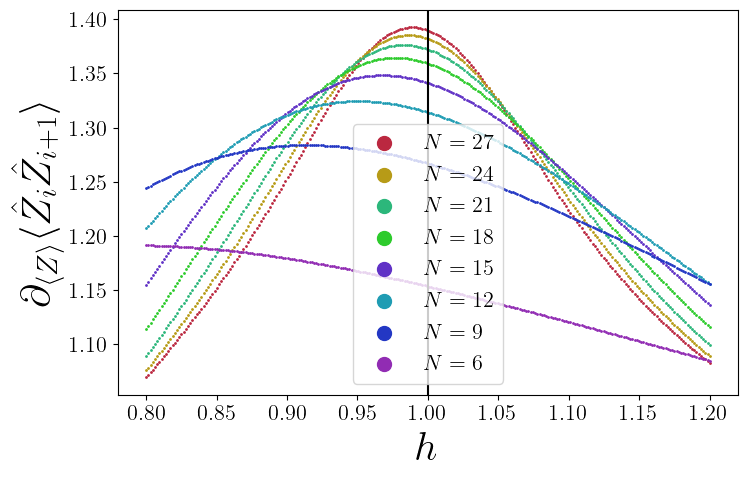}
 \subcaption{}
 \label{fig:dh_ZZ_3SI_overlap}
 \end{minipage}
 \hfill
 \begin{minipage}[t]{0.33\textwidth}
 \centering
 \includegraphics[width=\textwidth, keepaspectratio]{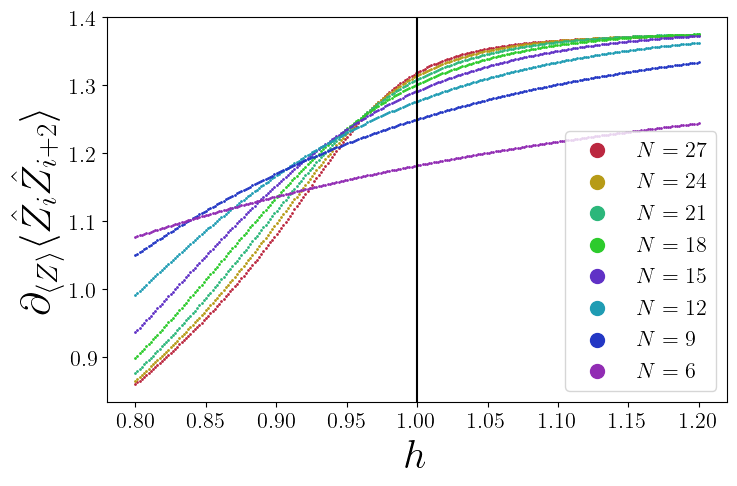}
 \subcaption{}
 \label{fig:dh_Z0Z_3SI_overlap}
 \end{minipage}
 \hfill
 
 }
 \makebox[\textwidth]{ 
 \begin{minipage}[t]{0.33\textwidth}
 \centering
 \includegraphics[width=\textwidth, keepaspectratio]{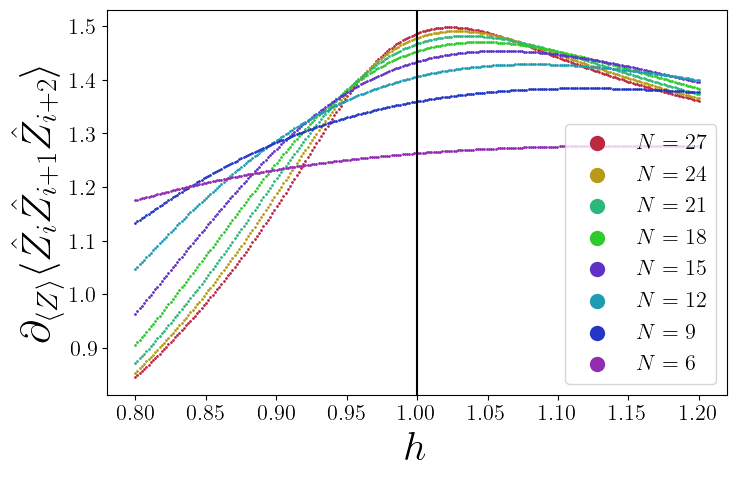}
 \subcaption{}
 \label{fig:dh_ZZZ_3SI_overlap}
 \end{minipage}
 \hfill
 \begin{minipage}[t]{0.33\textwidth}
 \centering
 \includegraphics[width=\textwidth, keepaspectratio]{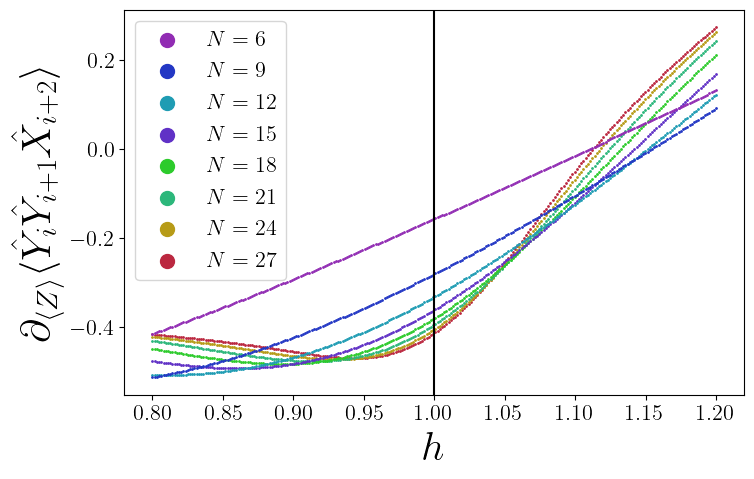}
 \subcaption{}
 \label{fig:dh_YYX_3SI_overlap}
 \end{minipage} 
 \hfill
 \begin{minipage}[t]{0.33\textwidth}
 \centering
 \includegraphics[width=\textwidth, keepaspectratio]{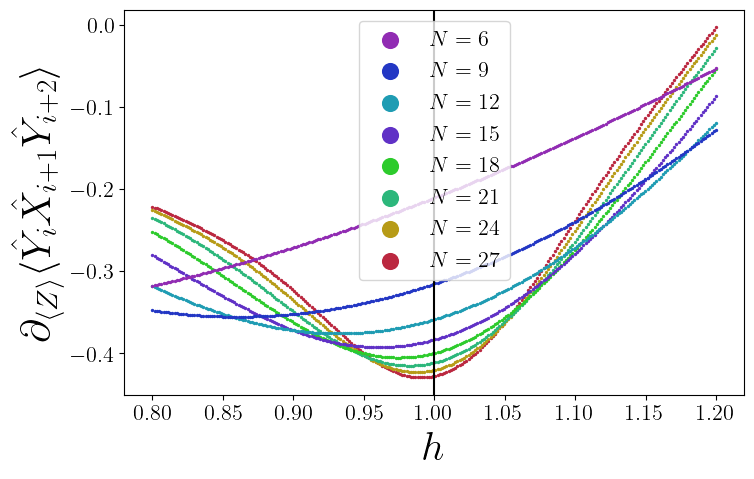}
 \subcaption{}
 \label{fig:dh_YXY_3SI_overlap}
 \end{minipage} 
 }
 \makebox[\textwidth]{ 
 \begin{minipage}[t]{0.33\textwidth}
 \centering
 \includegraphics[width=\textwidth, keepaspectratio]{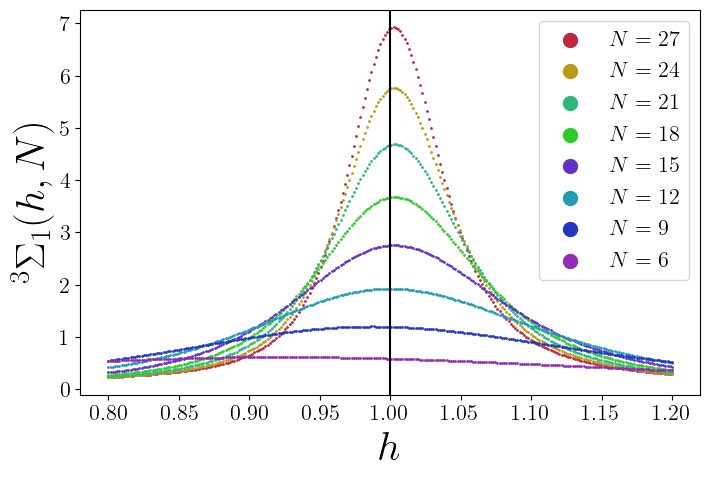}
 \subcaption{}
 \label{fig:response_3SI_rdm_1}
 \end{minipage}
 
 \begin{minipage}[t]{0.33\textwidth}
 \centering
 \includegraphics[width=\textwidth, keepaspectratio]{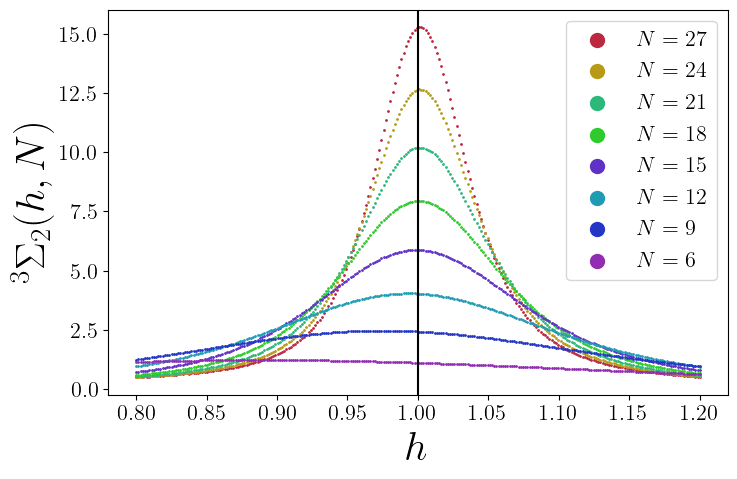}
 \subcaption{}
 \label{fig:response_3SI_rdm_2}
 \end{minipage} 
 \hfill
 \begin{minipage}[t]{0.33\textwidth}
 \centering
 \includegraphics[width=\textwidth, keepaspectratio]{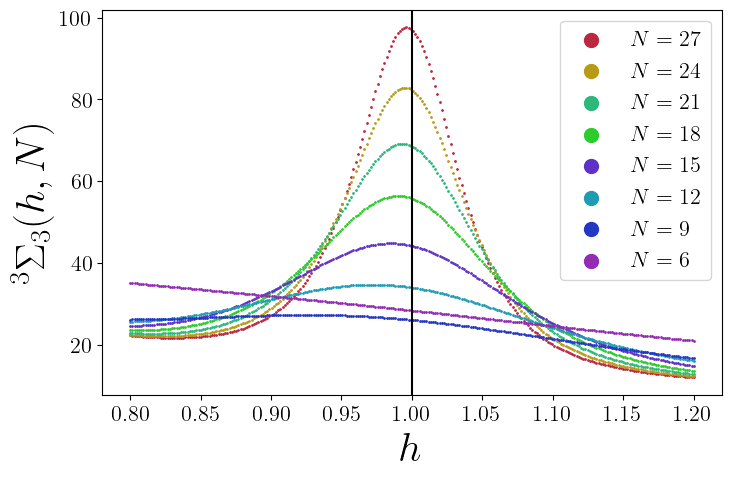}
 \subcaption{}
 \label{fig:response_3SI_rdm_3}
 \end{minipage} 
 }
\caption{(a-f) Derivatives of some local expectation values (which contribute to every local reduced density matrix) with respect to the expectation value of the most relevant operator $Z_i$, and (g-i) the corresponding responses of QRE in the three-spin Ising chain with subsystem sizes $m =1,2,3$. Here $X_i, ~Y_i, ~Z_i$ denote the Pauli operators $\sigma^x_i, ~\sigma^y_i~ \sigma^z_i$ respectively.}
\label{fig:three_spin_Ising_three_RDMs}
\end{figure*}

For finite but large $N$, the singular contributions can be estimated by the making use of the divergence from the derivative of correlations and the thermodynamic values of the correlations at $h=1$.
\begin{align}
& G_0\Big\vert^{h=1}_{N = \infty} = \frac{2}{\pi}, \ \ G_{xx}\Big\vert^{h=1}_{N = \infty} = \frac{2}{\pi},\label{eq:correlation_and_their_derivaties_in_TFIM}\\
& G_{yy}\Big\vert^{h=1}_{N = \infty} = -\frac{2}{3\pi}, \ \ G_{zz}\Big\vert^{h=1}_{N = \infty} = \frac{16}{3\pi^2}, \nonumber \\
& \frac{d \ G_0}{dh}\Big\vert^{h=1}_{N \to \infty, \ \text{sing}} = \frac{\ln N}{\pi} = \frac{d \ G_{xx}}{dh}\Big\vert^{h=1}_{N \to \infty, \ \text{sing}}, \nonumber \\
& \frac{d \ G_{yy}}{dh}\Big\vert^{h=1}_{N \to \infty, \ \text{sing}} = \frac{\ln N}{\pi} , \ \frac{d \ G_{zz}}{dh}\Big\vert^{h=1}_{N \to \infty, \ \text{sing}} = \frac{16 \ln N}{3 \pi^2}. \nonumber
\end{align}
Therefore, the divergence of this response function is solely due to the logarithmic divergence of the first derivative of relevant correlation functions. 

\begin{eqnarray}
\Sigma_{1}^\text{TFIM}(h=1,N \to \infty)_{\text{sing}} &=& A_1 \ (\ln N)^2, \nonumber \\
\Sigma_{2}^\text{TFIM}(h=1,N \to \infty)_{\text{sing}} &=& A_2 \ (\ln N)^2 ,
\end{eqnarray}
with $A_1 \approx 0.085$ and $A_2 \approx 1.819$. Away from the critical point, both $|h-1|$ and $N < \infty$ manifest non-trivially.

In the divergence of the response of QRE in a single site, the leading role is played by the singular contributions of $(\partial_h G_0)^2$ in 
Eq.~\eqref{eq:singularity_of_two_site_response} (\textit{cf.} Appendix \ref{subsec:singular_contributions_section}). It diverges as $\ln^2N$ and turning points approach the QCP at $h_c=\pm1$ from the ferromagnetic side as
\begin{align}
 h_m(N) = 1 - \frac{\pi^2 \ln N}{N^2} + \mathcal{O}(\frac{1}{N^4}).
 \label{eq:floating_exponent_of_peak_value}
\end{align}

However, even though the denominator $\propto 1/(1-G_0^2)$ remains regular in the thermodynamic limit, its presence in the response of a single site introduces another layer of complexity, as now the turning points approach the QCP from the paramagnetic side of $|h|>1$ because the denominator has the maximum change within the full-width at half-maxima of $(\partial_h G_0)^2$. This causes the maxima of the resultant $\Sigma_1^{\text{TFIM}}$ shift towards $|h|>1$, thus resulting in the observed FSS in the 
Eq.~\eqref{eq:exact_numerics_TFIM_metric_response}. Linear fit in log-log scale of the shift of turning points from $h=1$ resulted in monotonically decreasing exponent for a larger window of sizes, lower-bounded by $(-2)$; reassures the floating nature of the exponent as in 
Eq.~\eqref{eq:floating_exponent_of_peak_value}. Note that for a single site the turning point $h^c_1(N)$ is the solution of $\partial_h \Sigma_1^{\text{TFIM}}=0$ or
Eq.~\eqref{eq:single_site_response_turning_point_equation} such that $\partial_h^2 \Sigma_1^{\text{TFIM}} (h^c_1(N), N)<0$ for all $N$. Therefore by making use of the singular contributions near $h \to 1$ for systems with $N \to \infty$ we find (\textit{cf.} Eqs.~\eqref{eq:singularity_of_dh_G0}, \eqref{eq:singularity_of_ddh_G0}, and \eqref{eq:floating_exponent_of_single_site_response}) that
\begin{equation}
h^c_1(N) = 1+ \frac{4\pi^2 \ln^2 N}{3(\pi^2-4)N^2} - \frac{\pi^2 \ln N}{N^2} + \mathcal{O}(\frac{1}{N^3}).
\label{eq:turningpointTFIM}
\end{equation}

Similarly, the analysis of floating exponents for a pair of neighboring sites is straightforward but is immensely involved due to the presence of a $4 \times 4$ reduced density matrix resulting in a combination of four 
non-vanishing correlation functions and their derivatives in the corresponding response of QRE; 
we leave this for future investigations.

\subsection{Scaling in three-spin Ising chain} \label{subsec:scaling_in_3SI_criticality}

Here we present the finite-size behaviors of the turning points and the maxima of the response of QRE in a non-integrable theory, namely, the three-spin Ising model \cite{sen3spinising} described by the Hamiltonian Eq.~\eqref{eq:3SpinIsing}. As is evident from Figs.~\ref{fig:response_3SI_rdm_1}, \ref{fig:response_3SI_rdm_2} and \ref{fig:response_3SI_rdm_3}, we can calculate the response and make a polynomial fit very close to the positive critical point $h_c=1$ which we know to contain exactly one extrema. For our analysis, we consider $m = 1,2,3$ to be the number of local sites for constructing the RDMs. Unlike integrable models, determining the RDMs of higher rank is computationally simpler due to fewer number of trace operations over the remaining sites; but for determining the response of QRE ${}^3\Sigma_{m}(N,h)$ the final tracing ensures that working with RDM of any $m$ is computationally equally costly. Our numerics indicate that the turning points for finite sizes cross the thermodynamic critical point from below without touching it and then turns back to approach $h_c=1$ as in Fig.~\ref{fig:3SpinIsing__all_peak_locations}. This is reminiscent of what we observed in TFIM for small systems when the turning points cross $h_c=1$ for $h_c(N=48)$ and $h_c(N=50)$, as shown in Fig.\ref{fig:nonMonotonicPeakLocationIn_TFIM}. 

The underlying symmetry $\mathcal{S}_{\text{3SI}}$ ensures that only $\braket{\sigma^z_i} \neq 0 \neq \braket{\sigma^z_i \sigma^z_{i+1}}$ for a subsystem with $n=2$, therefore $\partial_h \braket{\sigma^z_i}$ and $\partial_h \braket{\sigma^z_i\sigma^z_{i+1}}$ cause the responses for a single and two sites to diverge. The ED analysis for sizes $N=6,9,12,15, 18, 21, 24,27$ shows monotonic growth in the derivative of the correlations, resulting in the divergence of the response of the QRE in local subsystems. The source of the divergence in the smallest few local sites are captured as, one-site: Fig. [\ref{fig:dh_Z_3SI}] $\to$ [\ref{fig:response_3SI_rdm_1}], two-site: Figs. [\ref{fig:dh_Z_3SI}, \ref{fig:dh_ZZ_3SI}] $\to$ [\ref{fig:response_3SI_rdm_2}] and three-site: Figs. [\ref{fig:dh_Z_3SI},\ref{fig:dh_ZZ_3SI},\ref{fig:dh_Z0Z_3SI},\ref{fig:dh_ZZZ_3SI},\ref{fig:dh_YYX_3SI},\ref{fig:dh_YXY_3SI}] $\to$ [\ref{fig:response_3SI_rdm_3}]. It has been studied before that local entanglement measures diverge due to containing the susceptibility of the most relevant operator driving the critical phenomenon in the sense of RG \cite{Venuti_PRA_div_loc_ent_measures__}, which for Hamiltonian in Eq.~\eqref{eq:3SpinIsing} is $\hat{\sigma}^z$ and the corresponding susceptibility is $\partial_h \braket{\sigma^z_i}$. Translation symmetry makes the site-index $i$ immaterial here. Although the disconnected two-point correlation function $\braket{\sigma^z_i\sigma^z_{i+1}} = \braket{\sigma^z_i\sigma^z_{i+1}}_c+\braket{\sigma^z_i}^2$ contributes in the response of two-site while its derivative causes the response to diverge; the derivative of the connected two-point correlation function, i.e., $\partial_h \braket{\sigma^z_i\sigma^z_{i+1}}_c$ too, diverges monotonically near the QCP. 


We note further that for a Hamiltonian $\hat{H} = \hat{A} + h\hat{B}$ the ground state energy $E_0 = \braket{\hat{A}}+h\braket{\hat{B}}$ where the expectation $\langle \rangle$ is taken in ground state. This, along with $\partial_h E_0 = \braket{\hat{B}}$ due to nondegenerate first-order perturbation theory, implies $\partial_h \braket{\hat{A}} = -h \partial_h \braket{\hat{B}}$. For the Hamiltonian of three-spin Ising chain in Eq.~\eqref{eq:3SpinIsing} one has $\hat{A} = - \sum_i \hat{\sigma}^x_i\hat{\sigma}^x_{i+1}\hat{\sigma}^x_{i+2}, \& \ \hat{B} = - \sum_i \hat{\sigma}^z_i$. Thus due to translation symmetry $\partial_h\braket{ \hat{\sigma}^x_i\hat{\sigma}^x_{i+1}\hat{\sigma}^x_{i+2}}(h,N)=-h \ \partial_h\braket{ \hat{\sigma}^z_i}(h,N)$ for all
$N$. 

For a larger subsystem with three consecutive sites, $G_{yyx}, G_{xyy}, G_{yxy}, G_{xxx}$ do not have any disconnected components and still their derivatives diverge (\textit{cf.} Figs. \ref{fig:dh_YYX_3SI} and \ref{fig:dh_YXY_3SI}). 
Eq.~\eqref{eq:derivative_of_local_operator_expectation} promptly generalizes a finding of Ref. \cite{Venuti_PRA_div_loc_ent_measures__} where the authors pointed out that every local operator can be expressed as a linear combination of the scaling operators, therefore, all of which will contain singular contributions, resulting in the divergence of their derivatives. 

In fact, we can numerically compute a measure of the "overlap" of a lattice operator of the form $\hat{O}_i$ with $\hat{Z} = \sigma_i^z$ by calculating $\partial_h\langle \hat{O} \rangle/\partial_h \langle \hat{Z} \rangle$ which is equal to $\partial_{\braket{Z}}\braket{ \hat{O}}$ since $\langle \hat{Z} \rangle$ is a monotonic function of $h$ for a fixed $N$. Monitoring the behavior of this overlap in the vicinity of $h_c$ as $N$ grows larger is illuminating. If $\partial_{\braket{Z}}\braket{ \hat{O}} \rightarrow a$ as $N \rightarrow \infty$, where $a \sim O(1)$, then $a$ is the overlap between the local operators $\hat{O}$ and $\hat{Z}$, while if $a=0$, then the scaling fields that contribute to $\hat{O}$ are strictly less relevant compared to the most relevant scaling field that contributes to $\hat{Z}$. The behavior of different $\partial_{\braket{Z}}\braket{ \hat{O}}$ for the nonzero $\langle O \rangle$ in a $n=3$ subsystem is shown in the panels (b) to (f) of Fig. \ref{fig:three_spin_Ising_three_RDMs} from which it is evident that the overlap with $\hat{Z}$ is nonzero in all these cases. Since $\partial_{\braket{Z}}\braket{ \hat{\sigma}^x_i\hat{\sigma}^x_{i+1}\hat{\sigma}^x_{i+2}}=-h$ from the previous discussion, this operator has been omitted from the panel in Fig. \ref{fig:three_spin_Ising_three_RDMs}. 

\begin{figure}[h]
 \centering
 \begin{minipage}[t]{\linewidth}
 \centering
 \includegraphics[width=\textwidth, keepaspectratio]{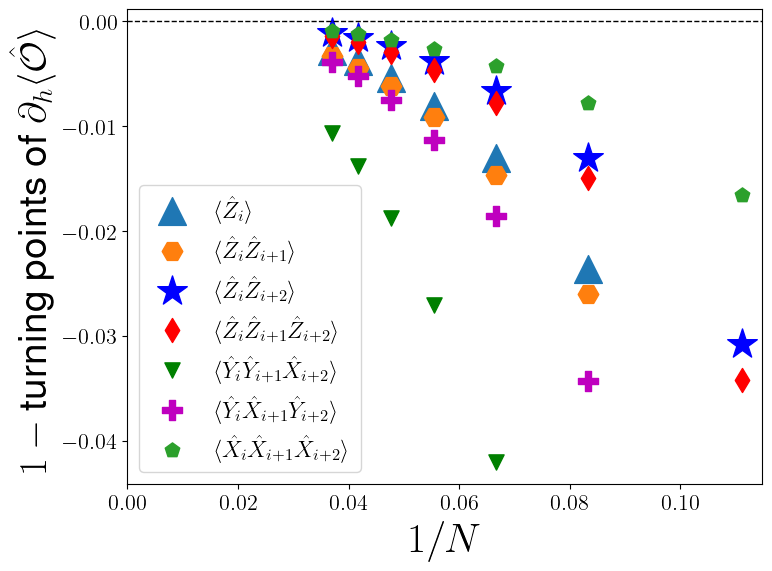}
 \subcaption{}
\label{fig:one_minus_turning_points__of_all_correlations}
 \end{minipage}
 \hfill
 \begin{minipage}[t]{\linewidth}
 \centering
 \includegraphics[width=\textwidth, keepaspectratio]{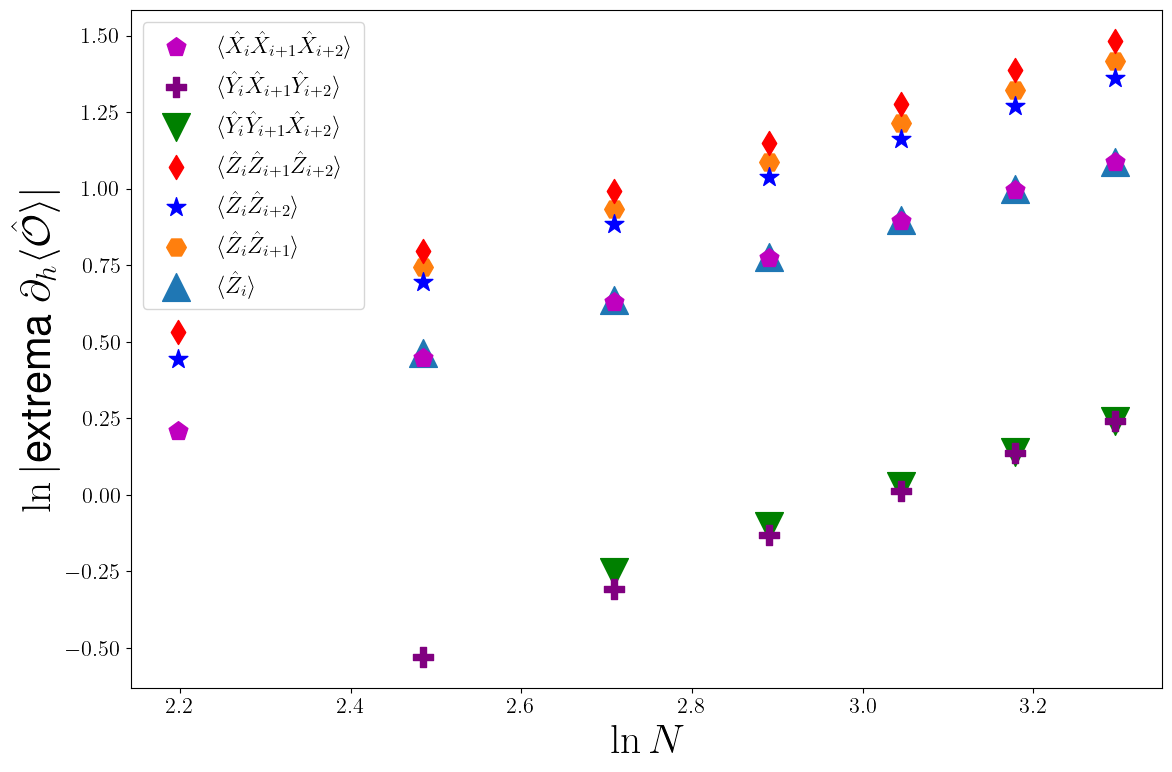}
 \subcaption{}
\label{fig:log_log_scaled_extrema_values__of_all_correlations}
 \end{minipage} 
\caption{(a) Turning points and (b) extrema of the derivatives of correlation functions which contribute to every local reduced density matrix of sizes $m =1,2,3$ in the three-spin Ising model, as shown in 
Figs.~\ref{fig:three_spin_Ising_three_RDMs}
and \ref{fig:dh_local_correlations_apndx}. Here $X_i,Y_i$ and $Z_i$ denote the Pauli operators $\sigma^x_i, \sigma^y_i$ and $\sigma^z_i$ respectively.}
\label{fig:turning_points_and_extrema_scaling_of_dervatives_of_correlations_in_3SI}
\end{figure}

\begin{figure*}[htbp]
 \centering
 \makebox[\textwidth]{ 
 \begin{minipage}[t]{0.33\textwidth}
 \centering
 \includegraphics[width=\textwidth, keepaspectratio]{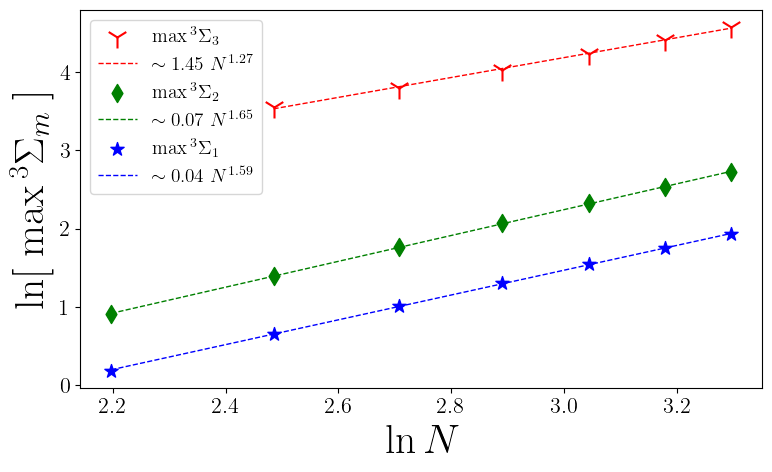}
 \subcaption{}
 \label{fig:3SpinIsing__all_peak_values}
 \end{minipage}
 \hfill
 \begin{minipage}[t]{0.33\textwidth}
 \centering
 \includegraphics[width=\textwidth, keepaspectratio]{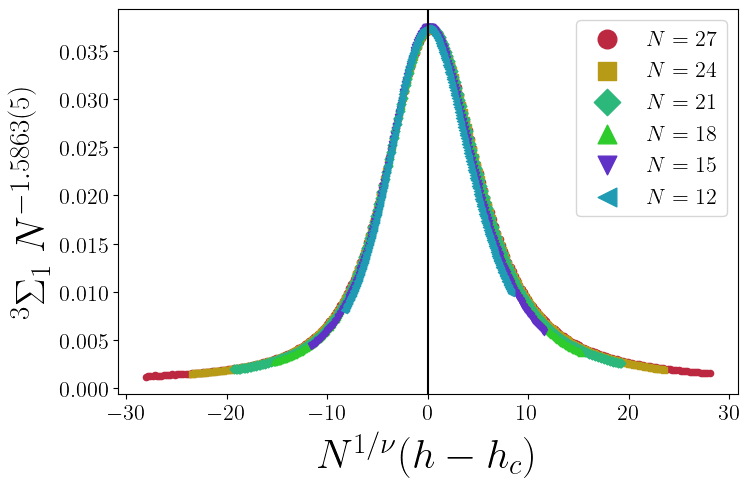}
 \subcaption{}
 \label{fig:Scaling_Collapse__3SpinIsingRDM__1}
 \end{minipage} 
 \hfill
 \begin{minipage}[t]{0.33\textwidth}
 \centering
 \includegraphics[width=\textwidth, keepaspectratio]{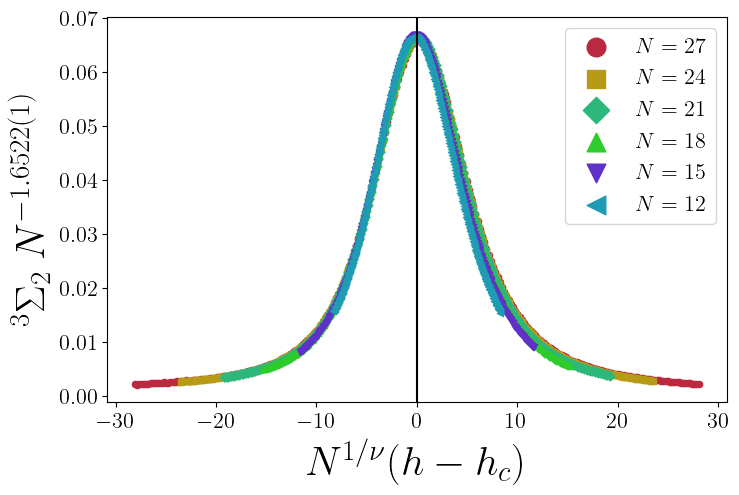}
 \subcaption{}
 \label{fig:Scaling_Collapse__3SpinIsingRDM__2}
 \end{minipage} 
 }
\caption{(a) Global maxima of the response of QRE $\max [{}^3\Sigma_m]$ for $m =1,2,3$ for
the three-spin Ising model. (b) Scaling collapse for the response of QRE for (b) $m=1$
and (c) $m=2$, using $\nu = 2/3$ and $h_c=1$ in the three-spin Ising model.}
\end{figure*}


However, the derivatives of all correlation functions do not contribute to the thermodynamic singularity. At the critical point of the three-spin Ising model, the most relevant primary operator that preserves symmetry in the 
four-state Potts model has conformal dimensions $(h, \bar{h}) = (1/4, 1/4)$ so that $\Delta_\epsilon\vert_{\text{3SI}} =h + \bar{h} = 1/2$. Therefore, the response of QRE scales as $\Sigma_h \sim N^2$ at the critical point due to 
Eq.~\eqref{eq:responsefromCFT}. The field-theoretic descendants have higher scaling dimensions as $\Delta_{\partial^p A} = p+\Delta_{A}$ for the $p$-th spatial derivative, making them increasingly more irrelevant. As representative examples, consider $\mathcal{O}_1 = (\sigma^y_{i} \sigma^y_{i+1} \sigma^x_{i+2} - \sigma^y_{i} \sigma^x_{i+1} \sigma^y_{i+2})=$ $\sigma^x_{i} \sigma^x_{i+1} \sigma^x_{i+2}(\sigma^z_{i} \sigma^z_{i+1}-\sigma^z_{i} \sigma^z_{i+2}) \ \& \ \mathcal{O}_2 = (\sigma^z_{i} \sigma^z_{i+2}-\sigma^z_{i} \sigma^z_{i+1})$ in the three-spin Ising model where the presence of one spatial derivative tells $\Delta_{\mathcal{O}_1} = \Delta_{Z} +1 = \Delta_{\mathcal{O}_2} = 3/2$ which is relevant but strictly less relevant than either of their parts (\textit{cf.} Figs. [\ref{fig:dh_YYX_3SI}, \ref{fig:dh_YXY_3SI}] $\to$ [\ref{fig:dh_O1_3SI}] and Figs. [\ref{fig:dh_ZZ_3SI},\ref{fig:dh_Z0Z_3SI}] $\to$ [\ref{fig:dh_O2_3SI}] in Appendix~\ref{subsec:derivativeops}). Again, according to this perspective, another spatial derivative, say $\mathcal{O}_3 = \sigma^z_{i}\sigma^z_{i+3} - 2\sigma^z_{i}\sigma^z_{i+2}-\sigma^z_{i}\sigma^z_{i+1}$ would render it irrelevant as $\Delta_{\mathcal{O}_3} = \Delta_{Z}+2=5/2>2$. The noticeably weaker divergence of $\partial_h \braket{\mathcal{O}_3}$ compared with $\partial_h \braket{\mathcal{O}_2}$ in the $-$ve $y$ direction reflects that even for smallest systems (\textit{cf.} Figs. \ref{fig:dh_O2_3SI} and \ref{fig:dh_O3_3SI} in Appendix~\ref{subsec:derivativeops}). It also demonstrates the limitation of 
Eq.~\eqref{eq:derivative_of_local_operator_expectation} where $\mathcal{O}_3$ would render a $\mathcal{O}(1)$ number after the sum over all the excited states so that concluding about divergence of any local operator is sensitive to the detail of the theory and the corresponding operator as that amounts to identifying the relevance of it in RG sense because there is nothing that guarantees an absence of sign fluctuation of the numerator in 
Eq.~\eqref{eq:derivative_of_local_operator_expectation}.

Operators involving spatial derivatives in a lattice theory can scale differently in the thermodynamic limit than are anticipated from a field theoretic point of view because only near zero 
momentum ($k=0$) will a spatial derivative take the form $\partial \mathcal{O} \equiv (\mathcal{O}_{i+1} - \mathcal{O}_{i})$ such that it scales consistently under rescaling of the coordinates. As an example, consider in a lattice theory with PBC, the operator $\mathcal{O} = \sum_{j}O_j (O_{j+1}-O_{j}) = \sum_k (e^{ik}-1)O_k O_{-k}$. Here $x \to bx \implies k \to b^{-1}k$ reproduces the correct field theoretic scaling dimension for modes near $k=0$, but for modes away from $k\neq 0$, $(e^{ik}-1) \to (e^{ik/b}-1)$ which is 
of order 1 instead of being close to zero.
Hence, the scaling of an operator defined in a lattice theory will vary depending on the most dominant momentum mode which leads to a thermodynamic singularity. For example, the three-spin Ising model is known to have gapless modes at $k=0$ and at $k=2\pi/3$ and $k=4\pi/3$ at $h_c$ in the thermodynamic limit.

The turning points of the derivative of all the correlation functions wrt $h$ approach the critical point $h_c=1$ from the ferromagnetic side of $|h|<1$ as in Fig. \ref{fig:one_minus_turning_points__of_all_correlations}, while the absolute values of their extrema become a straight line on $\log-\log$ scale with a strictly positive slope as in Fig.~\ref{fig:log_log_scaled_extrema_values__of_all_correlations}. This demonstrates the role of singularity of the expectations of scaling operators in all the non-trivial correlation functions in any continuous QPT, anticipated from Eq.~\eqref{eq:derivative_of_local_operator_expectation}. This is responsible for the divergence of the maxima of the response function which displays a power-law in system size $N$, unlike the power-law in $\ln N$ as in the case of TFIM (Figs. (\ref{fig:tfim_peak_heights}) and (\ref{fig:tfim_peak_heights2})). This strongly indicates that the underlying universality class manifests itself in the nature of this divergence. Using the critical exponent of the correlation length $\nu=2/3$ for the three-spin Ising chain \cite{sen3spinising}, the scaling collapse (using Eq.~\ref{eq:universalscaling}) is presented in Figs. [\ref{fig:response_3SI_rdm_1}, \ref{fig:response_3SI_rdm_2}] $\to$ [\ref{fig:Scaling_Collapse__3SpinIsingRDM__1},\ref{fig:Scaling_Collapse__3SpinIsingRDM__2}] for $12 \leq N \leq 27$ available from our ED studies. For $m=1,2$, the turning points should follow $(h-h_c) \propto N^{-1/\nu}$ as the scaling collapses (Eq.~\eqref{eq:universalscaling}) of the figures [~\ref{fig:Scaling_Collapse__3SpinIsingRDM__1} and ~\ref{fig:Scaling_Collapse__3SpinIsingRDM__2}] also indicate. The fact that we need to use different values of $\zeta/\nu$ in Fig.~\ref{fig:Scaling_Collapse__3SpinIsingRDM__1} and in Fig.~\ref{fig:Scaling_Collapse__3SpinIsingRDM__2} indicates that larger system sizes are needed to see the correct asymptotic scaling form. The response obtained from RDMs of the $m=3$ and $4$ nearest-neighbor sites also shows similar scaling, and turning points approach faster for higher $m$, but the inversion of a nearly singular density matrix as $h \rightarrow 0$ leaves a different qualitative behavior in the ferromagnetic region of $|h|<1$ (\textit{cf.} Fig. \ref{fig:qualitative_comparison_of__different_sites_responses__N_15__3SI}). This behavior will be discussed further in the next section.



An efficient ED routine can be written with the Algorithm \ref{alg:maximal_basis_reduction_with_bitmasks} in Appendix \ref{subsec:bitmask_algo} to determine the scaling of the response of QRE in the subsystems of Ising chains of any higher locality. Using this, we also observed the critical features in Ising chains of up to $10$-local interactions. Even though for $2-$local and $3-$local interactions the quantum phase transition at $|h|=1$ is known to be second order, and first order for range of interaction larger than $4$, it is unknown for the case of $4-$ local interactions~\cite{sen3spinising}. This measure has the potential to shed light on the nature of the phase transition in such systems; we leave this for future investigations.

\section{Response of relative entropy near the classical limits} \label{sec:sec_local_perturbations}

\begin{figure*}[htbp]
 \centering
 \begin{minipage}{0.29\textwidth}
 \centering
 \includegraphics[width=\linewidth, keepaspectratio]{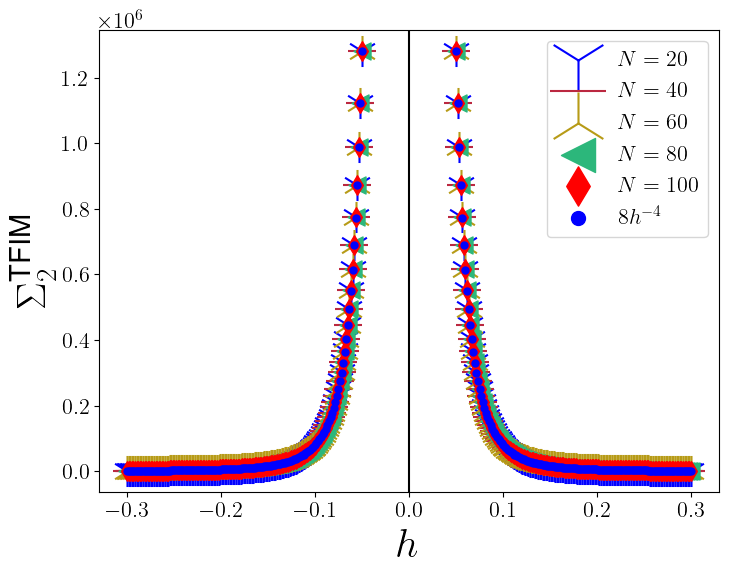}
 \subcaption[]{}
 \label{fig:TFIM_two_site_near_0__divergence}
 \end{minipage}
 \begin{minipage}{0.36\textwidth}
 \centering
 \includegraphics[width=\linewidth, keepaspectratio]{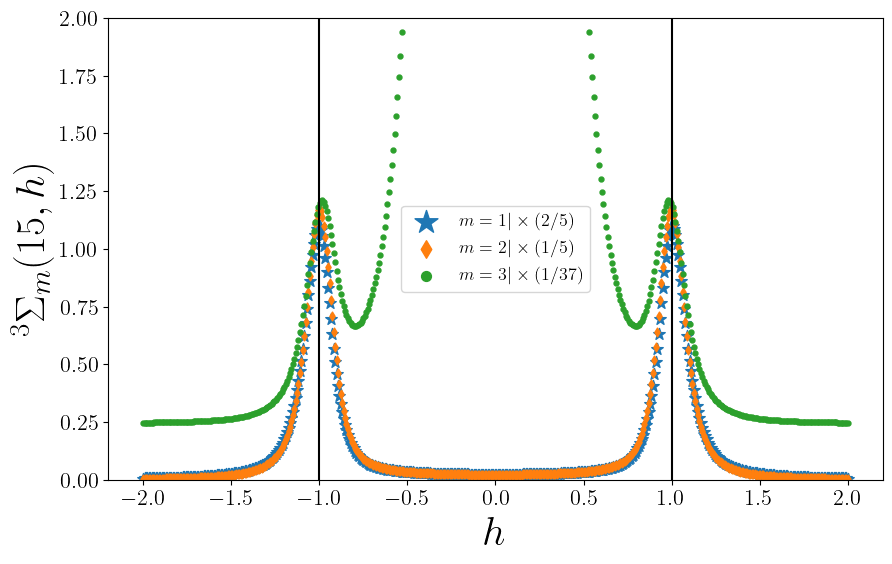}
 \subcaption[]{}
\label{fig:qualitative_comparison_of__different_sites_responses__N_15__3SI}
 \end{minipage}
 \begin{minipage}{0.29\textwidth}
 \centering
 \includegraphics[width=\linewidth, keepaspectratio]{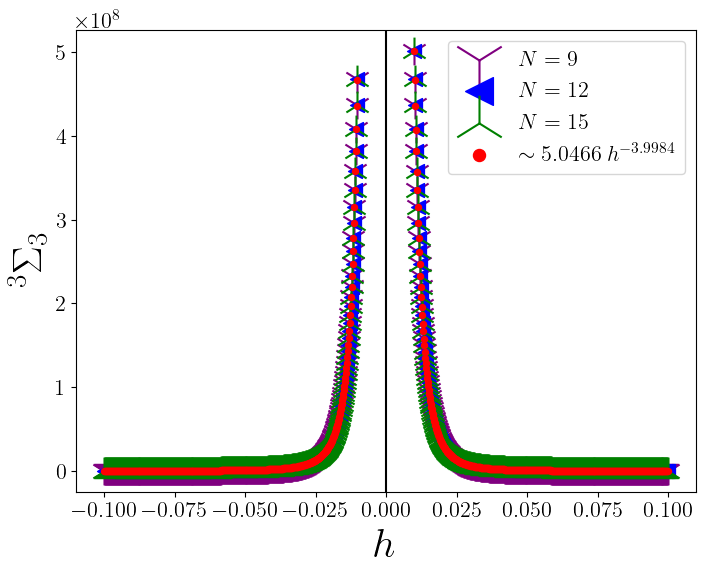}
 \subcaption[]{}
\label{fig:three_spin_three_site_near_0__divergence}
 \end{minipage}
 \hfill
 \begin{minipage}{0.32\textwidth}
 \includegraphics[width=\linewidth]{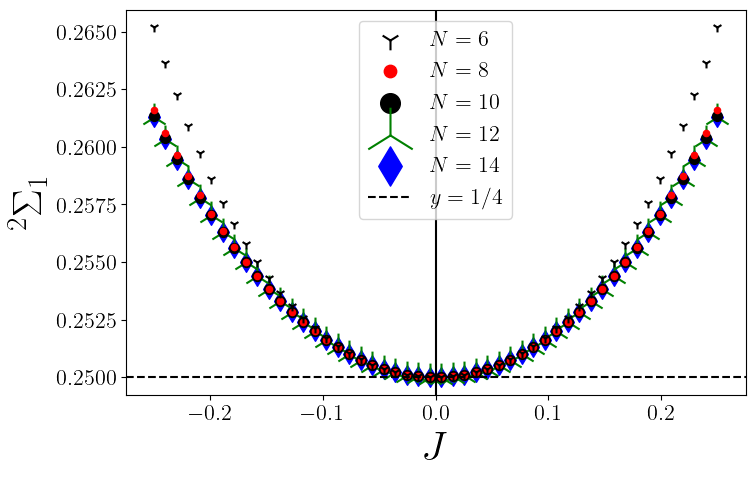}
 \subcaption[]{}
 \label{fig:near_J_0__2_Sigma_1__}
 \end{minipage}
 \begin{minipage}{0.32\textwidth}
 \includegraphics[width=\linewidth]{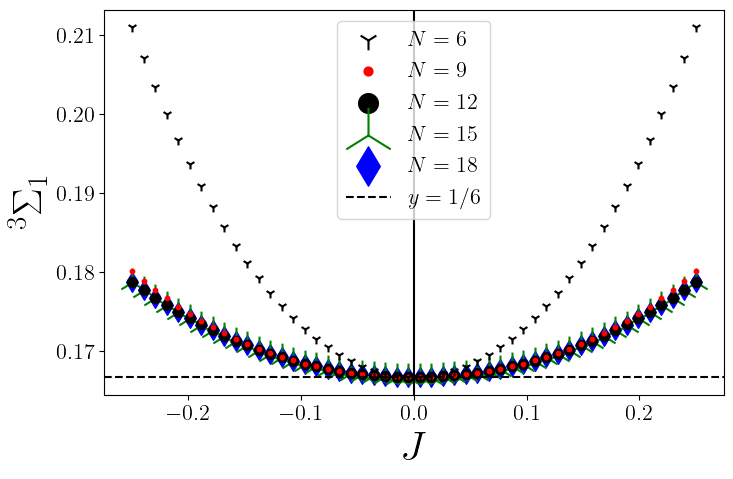}
 \subcaption[]{}
 \label{fig:near_J_0__3_Sigma_1__}
 \end{minipage}
 \begin{minipage}{0.32\textwidth}
 \includegraphics[width=\linewidth]{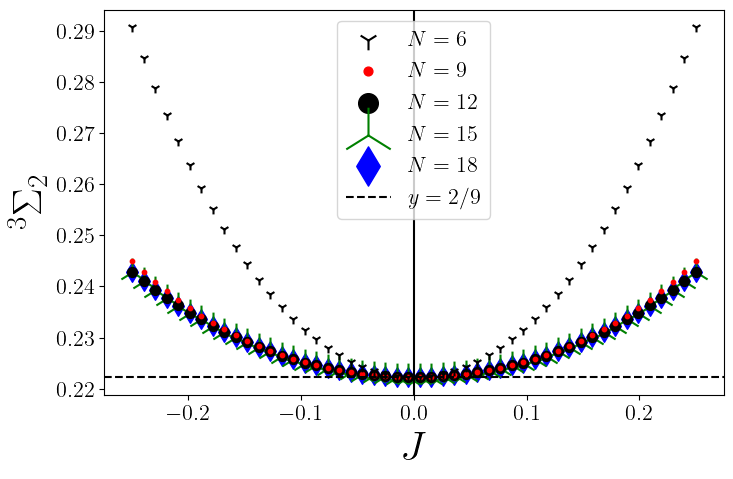}
 \subcaption[]{}
 \label{fig:near_J_0__3_Sigma_2__}
 \end{minipage}
 \hfill
 \begin{minipage}{0.49\textwidth}
 \includegraphics[width=\linewidth]{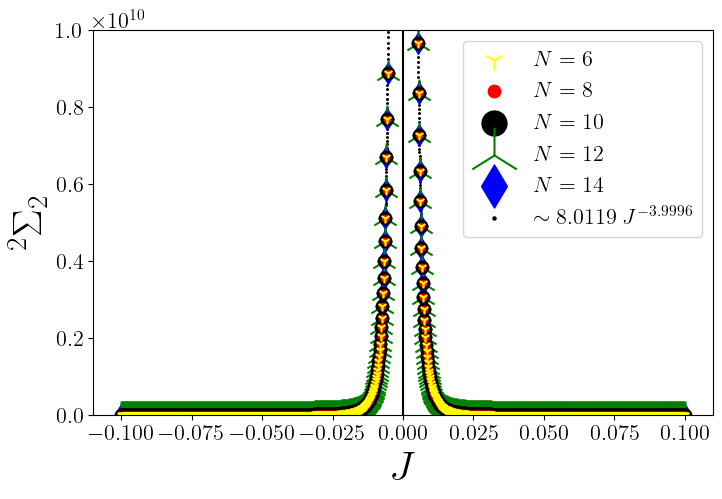}
 \subcaption[]{}
 \label{fig:near_J_0__2_Sigma_2__}
 \end{minipage}
 \begin{minipage}{0.49\textwidth}
 \includegraphics[width=\linewidth]{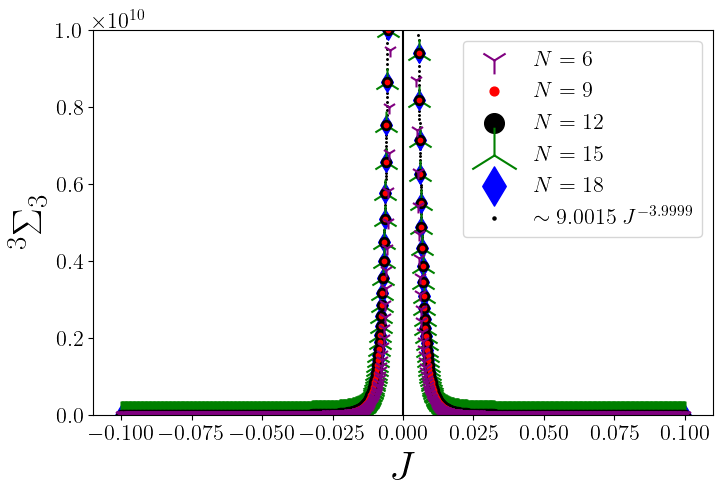}
 \subcaption[]{}
 \label{fig:near_J_0__3_Sigma_3__}
 \end{minipage}
 \caption{Response of QRE $^n\Sigma_m$, where $n=$ locality and $m=$ subsystem size, under perturbations in the case of deformations around a classical limit for finite-sized systems: (a) TFIM for $h \to 0$, (b) and (c) three-spin Ising model for $h \to 0$, 
 (d) and (g) for $H'_2$ when $J \to 0$, and (e), (f) and (h) for $H'_3$ when $J \to 0$. Plots (g) and (h) show a good fit to 
 $1/J^4$ for $J \to 0$ in agreement with 
 Eq.~\eqref{2233}.}
\end{figure*}

Given a bipartition $(m,N-m)$, $\dim[\hat{\rho}_m(\lambda')]>\text{Sch}(\ket{\psi^{gs}_{\lambda'}},m)$ imply that $\hat{\rho}_m(\lambda')$ is singular because at least one eigenvalue of $\hat{\rho}_m(\lambda')$ must be $0$. Such singular RDMs $\hat{\rho}$ imply that $\Sigma$ is not related to the relative entropy in any meaningful way as non-invertible density matrices render the expansion of Eq.~\ref{eq:def_of_susQRE} invalid. Now a classical limit of quantum systems (say $\lambda'$) often renders singular density matrices, supported on the degenerate submanifold of their ground states as $\underset{m}{\max}\{\text{Sch}(\ket{\psi^{gs}_{\lambda'}},m)\}=\#(\text{degeneracy})$. Any perturbation around this limit lifts the singularity of such $\hat{\rho}$'s, leading to $\Sigma$ being the response of relative entropy again. The behaviour of $\Sigma$ at and near the classical limit is therefore non-critical. That is why as the classical limit is approached, nearly-singular density matrices can trigger a non-universal divergence of $\Sigma$. However this is only the necessary condition for singular-density matrices to yield a non-universal divergence but not sufficient, because the interplay of $\hat{\rho}^{-1} (\partial_i\hat{\rho})^2$ can produce finite response of relative entropy at the classical limit. In this section we explore its ramifications in TFIM and three-spin Ising model.

To do this, we turn to the {\it simpler} analysis of the response when both systems are perturbed around a classical spectrum (i.e., $h \rightarrow 0$ with $J=1$ and $J \rightarrow 0$ with $h=1$). As is clear from both Fig (\ref{fig:tfim_response_one_site}) and Fig.~\ref{fig:TFIM_two_site_near_0__divergence}, $\Sigma_1^{\mathrm{TFIM}}$ remains finite while $\Sigma_2^{\mathrm{TFIM}}$ diverges as $h^{-4}$ when $h \rightarrow 0$ for the TFIM. A similar effect can be seen for the non-integrable spin chain at a fixed size of $N=15$ when ${}^{3}\Sigma_m(h)$ is calculated for different values of $h$ (see Fig.~\ref{fig:qualitative_comparison_of__different_sites_responses__N_15__3SI}) with $m=1,2$ being finite but $m=3$ showing a divergent behavior as $h \rightarrow 0$. The nature of the divergence can be seen more clearly in Fig.~\ref{fig:three_spin_three_site_near_0__divergence} and follows $h^{-4}$ as $h \rightarrow 0$. The divergence of the response as $h \rightarrow 0$ in both models appears even at small $N$ and does not require any large $N$ limit. Using the RDM of two sites in TFIM in 
Eq.~\eqref{eq:NNReducedMatrixOfTFIM} and the thermodynamic functions of the relevant expectations $G_0, ~G_{-1}$ and $G_{+1}$, one finds that around $h=0$, 
\begin{align}
 \Sigma^{\text{TFIM}}_2 (h \to 0) = \frac{8}{h^4} + \frac{5}{h^2}+ \text{regular terms}. 
\end{align}

However, for subsystems smaller than the degeneracy of the symmetry broken ground state in the thermodynamic limit, one always finds $\Sigma(h = 0) > 0$ and finite. To see that,
we consider the following perturbation as $h \to 0$,
\begin{align}
& H_2 = - \sum_{i=1}^N \sigma^x_i \sigma^x_{i+1}- h\sum_{i=1}^N \sigma^z_i = {}_2H_0 + h \ W, \\
& H_3 = - \sum_{i=1}^N \sigma^x_i \sigma^x_{i+1}\sigma^x_{i+2} - h\sum_{i=1}^N \sigma^z_i = {}_3H_0 + h \ W. 
\end{align}

At $h = 0$, the ground states of $H_2$ and $ H_3$ 
are two-fold degenerate and four-fold degenerate
respectively, with energy $E_0 = -N$. For simplicity,
we will consider the ground states given by linear
superpositions which are completely symmetric under
the symmetry $Z_2$ for $H_2$ and $Z_2 \times Z_2$
for $H_3$,
\begin{eqnarray} \ket{{}^2\psi_0} &=& \frac{1}{\sqrt 2}~ (\ket{\rightarrow\rightarrow \rightarrow \cdots} + \ket{\leftarrow\leftarrow \cdots \leftarrow}), \nonumber \\
\ket{{}^3\psi_0} &=& \frac{1}{2}~ (\ket{\rightarrow\rightarrow\rightarrow \cdots} + \ket{\leftarrow\leftarrow\rightarrow \cdots} 
\nonumber \\
&& ~~~~+ \ket{\leftarrow\rightarrow\leftarrow \cdots}+ \ket{\rightarrow\leftarrow\leftarrow \cdots}),
\end{eqnarray}
in the $\sigma^x$-basis.
When $h$ is small, the perturbation $W$ locally flips the $x$-component of the spins so that $\sigma^z_i \ket{{}^2\psi_0} = \mathcal{N} (\ket{ \cdots \rightarrow \leftarrow_i \rightarrow \cdots} + \ket{ \cdots \leftarrow \rightarrow_i \leftarrow \cdots}) := \mathcal{N} \ket{i_2}$, and, similarly, $\sigma^z_i \ket{{}^3 \psi_0} =: \mathcal{N}\ket{i_3}$. 
(The $\mathcal{N}$'s denote normalization factors).
Therefore, in first order in $h$, the excitation energy is $4$ in the TFIM and $6$ in the three-spin Ising model. Hence, the corresponding perturbed ground states become 
\begin{align}
& \ket{{}^2\psi_{0}(h)} = \mathcal{N} \Big[ \ket{\psi_0} + \frac{h}{4}\sum_{i=1}^N \ket{i_2}+\mathcal{O}(h^2) \Big], \\
& \ket{{}^3\psi_{0}(h)} = \mathcal{N} \Big[ \ket{\psi_0} + \frac{h}{6}\sum_{i=1}^N \ket{i_3}+\mathcal{O}(h^2) \Big]. 
\end{align}

Now, keeping in mind the degeneracy of ground states as $h \to 0$ in both models, the one-site RDM at $\mathcal{O}(h)$ and the corresponding response function turn out to be
\begin{align}
& {}^2\rho_{1}(h) = \begin{pmatrix} \frac{1}{2} + \frac{h}{4} & 0 \\ 0 & \frac{1}{2} - \frac{h}{4} \end{pmatrix} \implies \Sigma(h \to 0) = \frac{1}{8}, \\
& {}^3\rho_1(h) = \begin{pmatrix} \frac{1}{2} + \frac{h}{6} & 0 \\ 0 & \frac{1}{2} - \frac{h}{6} \end{pmatrix} \implies \Sigma( h \to 0) = \frac{1}{18}.
\end{align}
In a similar way, the RDM of two adjacent sites and the corresponding response function 
for three-spin Ising model is found to be
\begin{align}
 {}^3\rho_{2}(h) = \begin{pmatrix} \frac{1}{4} + \frac{h}{6} & 0 & 0 & 0 \\ 0 & \frac{1}{4} & 0 & 0 \\ 0 & 0 & \frac{1}{4} & 0 \\ 0 & 0 & 0 & \frac{1}{4} - \frac{h}{6} \end{pmatrix} \implies \Sigma( h \to 0) = \frac{1}{9}. 
\end{align}

Let us now consider the other limit where $h=1$ and $J \rightarrow 0$ for both spin models. For the TFIM, while ${}^2\Sigma_1$ is finite as $J \rightarrow 0$ (Fig.~\ref{fig:near_J_0__2_Sigma_1__}), ${}^2\Sigma_2 \sim J^{-4}$ for $n=2$ (Fig.~\ref{fig:near_J_0__2_Sigma_2__}). Similarly, for the three-spin non-integrable Ising chain, while both ${}^3\Sigma_1$ (Fig.~\ref{fig:near_J_0__3_Sigma_1__})and ${}^3\Sigma_2$ (Fig.~\ref{fig:near_J_0__3_Sigma_2__}) are finite as $J \rightarrow 0$, ${}^3\Sigma_3 \sim J^{-4}$ for $n=3$ (Fig.~\ref{fig:near_J_0__3_Sigma_3__}). To examine this feature for the dual limit of TFIM and the three-spin Ising model, consider the following Hamiltonians with periodic boundaries.
\begin{align}
& H'_2 = - \sum_{i=1}^N \sigma^z_i -J \sum_{i=1}^N \sigma^x_i \sigma^x_{i+1} = H_0 + J V_2, \\
& H'_3 = - \sum_{i=1}^N \sigma^z_i-J \sum_{i=1}^N \sigma^x_i \sigma^x_{i+1}\sigma^x_{i+2} = H_0 + J V_3. 
\end{align}

At $J=0$, the ground states of both $H'_2$ and
$H'_3$ are unique and are given by $\ket{\psi_0}= \ket{\uparrow\uparrow\uparrow \cdots}$ in the $\sigma^z$-basis with energy $E_0 = -N$ which is an unentangled product of qubits. When $J \to 0$ the perturbations $V_2$ and $V_3$ locally flip the $z$-component of the spins so that $\sigma^x_i \sigma^x_{i+1}\ket{\psi_0} = \ket{\uparrow \uparrow
\cdots \downarrow_{i}\downarrow_{i+1} \cdots \uparrow}=:\ket{\downarrow_{i}\downarrow_{i+1}}$ and $\sigma^x_i \sigma^x_{i+1}\sigma^x_{i+2}\ket{\psi_0}=\ket{\uparrow \uparrow \cdots \downarrow_{i}\downarrow_{i+1}\downarrow_{i+2} \cdots \uparrow}=:\ket{\downarrow_{i}\downarrow_{i+1}\downarrow_{i+2}}$. Therefore, to first order in $J$, the excitation energy is $4$ due to $V_2$ and $6$ due to $V_3$. Hence, the corresponding perturbed ground states become 
\begin{align}
& \ket{{}_2\psi_{0}(J)} = \mathcal{N} \Big[ \ket{\psi_0} + \frac{J}{4}\sum_{i=1}^N \ket{\downarrow_{i}\downarrow_{i+1}}+\mathcal{O}(J^2) \Big], \\
& \ket{{}_3\psi_{0}(J)} = \mathcal{N} \Big[ \ket{\psi_0} + \frac{J}{6}\sum_{i=1}^N \ket{\downarrow_{i}\downarrow_{i+1}\downarrow_{i+2}}+\mathcal{O}(J^2) \Big]. 
\end{align}
Now, due to the $\mathbb{Z}_2$ symmetry of $H'_2$ the single site reduced density matrix ${}_2\rho_1(J)$ is diagonal in the $\sigma^z$-basis whereas due to the $\mathbb{Z}_2 \times \mathbb{Z}_2$ symmetry of ${}H'_3$ the single-site and 
two-site RDM, i.e., ${}_3\rho_1(J)$ and ${}_3\rho_2(J)$, are diagonal in that basis. After fixing a particular site $(i)$, the contributions in the corresponding density matrices can be read off of the possible leading configurations of the ground state. Then the normalizer $\mathcal{N}$ is fixed by demanding $\text{tr}[\rho]=1$. Therefore, the leading behaviors at $\mathcal{O}(J^2)$ when $J \to 0$ in the $\{\ket{\uparrow_i}, \ket{\downarrow_i}\}$ basis are given by
\begin{align}
& {}_2\rho_1(J) = \begin{pmatrix} 1 - \frac{J^2}{8} & 0 \\ 0 & \frac{J^2}{8} \end{pmatrix} \implies {}^2\Sigma_1(J \to 0) = \frac{1}{4}, \\
& {}_3\rho_1(J) = \begin{pmatrix} 1 - \frac{J^2}{12} & 0 \\ 0 & \frac{J^2}{12} \end{pmatrix} \implies {}^3\Sigma_1(J \to 0) = \frac{1}{6}. 
\end{align}

Here $\rho$ is calculated up to $\mathcal{O}(J^2)$, but due to the presence of $\rho^{-1}$, only $\mathcal{O}(1)$ contribution in $\Sigma$ is correct as higher corrections in $\rho$ will change the sub-leading corrections in $\Sigma$. Similarly, for the neighboring sites $(i,i+1)$ in $H'_3$, the local Hilbert space is spanned by $\{\ket{\uparrow_i\uparrow_{i+1}},\ket{\downarrow_i\downarrow_{i+1}},\ket{\downarrow_i\uparrow_{i+1}},\ket{\uparrow_i\downarrow_{i+1}}\}$ so as before, in this basis we obtain
\begin{align}
& {}_3\rho_2(J) = \begin{pmatrix} 1 - \frac{J^2}{9} & 0 & 0 & 0 \\ 0 & \frac{J^2}{18} & 0 & 0 \\ 0 & 0 & \frac{J^2}{36} & 0 \\ 0 & 0 & 0 & \frac{J^2}{36} \end{pmatrix} \implies {}^3\Sigma_2(J \to 0) = \frac{2}{9}. 
\end{align}

At $J=0$ these reproduce the numerically observed values in Figs. [\ref{fig:near_J_0__2_Sigma_1__},\ref{fig:near_J_0__3_Sigma_1__},\ref{fig:near_J_0__3_Sigma_2__}], all of which come from a singular density matrix at $J=0$ but the leading $\mathcal{O}(J^{-2})$ contributions of $\rho^{-1}$ cancel the leading $\mathcal{O}(J^{2})$ contributions of $(\partial_J \rho)^2$ as $J \to 0$, resulting in $\mathcal{O}(1)$ responses there - even when only one eigenvalue does not vanish as $J \to 0$ in both models. However, this is not the case for $_2\Sigma_2$
and $_3\Sigma_3$ when $J \to 0$ (\textit{cf.} Fig. [\ref{fig:near_J_0__2_Sigma_2__}, \ref{fig:near_J_0__3_Sigma_3__}]) as they are singular at $J=0$ and diverge as 
\begin{equation}
_2\Sigma_2(J \to 0) \sim \frac{8}{J^4} ~~~{\rm and}~~~ _3\Sigma_3(J \to 0) \sim \frac{9}{J^4}.
\label{2233} \end{equation}

The corresponding symmetries do not make the RDMs diagonal in these cases, leading to a more intricate interplay of $\rho^{-1} ~(\partial_J \rho)^2$. We leave this for a future investigation.

\section{Discussion} 
\label{sec:discussions}

An important feature of quantum entropy that plays a key role here is the \textit{monotonicity of relative entropy} \cite{SSA_QuantumEntropy_Lieb} which directly implies the strong sub-additivity (SSA) of quantum entropy. In general, for any \textit{completely positive, trace-preserving} map $\mathcal{N}: \mathcal{B}(\mathcal{H}_{AB})\to \mathcal{B}(\mathcal{H}_{A})$ \footnote{$\mathcal{B}(\mathcal{H})$ is the subspace of all the \textit{bounded linear operators} in the Hilbert space $\mathcal{H}$. Density matrices, which are positive semi-definite, i.e., Hermitian and without any negative eigenvalue along with trace $1$, fall inside $\mathcal{B}(\mathcal{H})$} ensuring 
that~\cite{SSA_QuantumEntropy_Lieb}
\begin{align}
 & S( \ \hat{\rho} \ || \ \hat{\sigma} \ ) \geq S( \ \mathcal{N}(\hat{\rho}) \ || \ \mathcal{N}(\hat{\sigma)} \ ) \nonumber \\ 
 \implies & \Sigma_{m' \geq m}(\vec{\lambda}) \geq \Sigma_{m}(\vec{\lambda}), \ \ \forall ~m, \vec{\lambda}.
 \label{eq:monotonicity_of_relative_entropy}
\end{align}

It is also known as \textit{data processing inequality} in the quantum information literature \cite{leditzky2016thesis, M_ller_Hermes_2017, Frenkel_2023}. Since taking the partial trace preserves the positivity and the trace of density matrices as $T = \mathbbm{I}_{\mathcal{H}_A} \otimes \mathrm{Tr}_{\mathcal{H}_B}$, the QRE of a smaller subsystem is smaller due to a lesser number of distinct available modes of entanglement, at every point in the parameter space. 
This is evident from Fig. \ref{fig:qualitative_comparison_of__different_sites_responses__N_15__3SI} and Figs. [\ref{fig:response_3SI_rdm_1} $\leq$ \ref{fig:response_3SI_rdm_2} $\leq$ \ref{fig:response_3SI_rdm_3}] for all $h$.

The formalism for the metric response of relative entropy encompasses quantum systems in any spatial dimension and driven by arbitrary number of parameters as it naturally encodes the scaling of the most relevant operator driving a QPT, in the landscape of relative-entropy over the parameter space. Therefore renormalization group uniquely determines the universal scaling of this response as demonstrated in Eq.~\ref{eq:thermo_scaling_of_metric_response_gen}. For a system $H = H_0 + g \mathcal{V}$, with $\mathcal{V}$ driving the system across a QPT, consider $q$ to be size of the smallest subsystem containing the most relevant operator generated by the RG flow under coarse-graining in the thermodynamic limit. Then every density matrix of subsystems $m \geq q$ contains the most relevant scaling operator and therefore encodes the information of the critical point exhaustively. Such an information theoretic prescription requires no appeal to the nature of the order parameter associated with the phase transition.

The measure is equally applicable for all such $m$-site reduced density matrices \cite{Sachdev2011QPT, Barouch1970StatMechI, Barouch1971StatMechII} depending on the nature of the interaction driving the model under investigation, because all that is required is an assignment of invertible reduced density matrices at each point in the parameter space, which can always be obtained by either analytical solution \cite{polkovnikov}, or numerical methods \cite{Tsai2001DMRG}. 

For \textit{integrable systems} that can be mapped to non-interacting fermions, any reduced density matrix can be analytically calculated in practice using 
Eq.~\eqref{eq:derivedTwoPointCorrelations} and utilizing the available symmetries to reduce to the minimal set of distinct correlation functions, required in the operator product expansion of a reduced density matrix as in 
Eq.~\eqref{eq:op_prod_expansion}. For \textit{non-integrable systems} density matrices can be obtained through exact diagonalization or DMRG and simplicity of the formalism naturally couples with both the techniques. In fact the uniform definition for response of relative entropy allows it to naturally couple with any routine that outputs density matrices, without any additional computational cost. 


A pathology of singular/non-invertible density matrices $\hat{\rho}$ can arise in the definition of the response function at the classical limit of quantum systems, where rank-deficiency of $\hat{\rho}$ can result in a non-universal and non-extensive divergence due to the model-specific vanishing of the eigenvalues of $\hat{\rho}$. Since all the non-trivial density matrices retain full-rank during a quantum phase transition, the universality of this response remains intact and non-invertibility of density matrices never interfere with any analyses of quantum criticality.

This response encodes the geometry of the landscape of relative entropy, coordinatized by the driving parameters. With the metric response of QRE taken as a Riemannian metric, the singularity at the quantum phase transition translates to an absence of any finite-length geodesic joining 
two points of different quantum phases. The simplest of such cases with an analytical solution is the TFIM in one dimension and this
has been elaborated in Appendix~\ref{subsec:intrinsic_geom_TFIM}. Although the conclusion is general, an analytical demonstration for theories with at least two parameters remains to be addressed in future investigations.

Lastly, open quantum systems are often approximated by an evolution with an effective non-Hermitian Hamiltonian in the absence of quantum jumps where the non-equilibrium steady state displays a degenerate spectrum in the subsystems at some exceptional points in both finite and thermodynamically large systems; this is in contrast to avoided level crossings which occur in finite systems evolving unitarily \cite{heralded_magnetism}. The response of QRE is singular at the exceptional points, and the simplicity of our approach, which only requires a knowledge of some small
reduced density matrices, renders it applicable to analyze the universal characters of phase transitions in non-Hermitian quantum systems.

\section{Conclusions} \label{sec:conclusions}

In summary, an intrinsic measure, namely, the metric response of QRE $(\Sigma)$, serves as a universal identifier of a quantum phase transition in the thermodynamic limit for any subsystem. The scaling of $\Sigma$ is universal near a quantum phase transition as has been proved using both renormalization group theory and conformal field theory. It has also been demonstrated analytically in the TFIM and computationally in the three-spin Ising model. A rank-deficiency of local density matrices can lead to a non-universal divergence for subsystems beyond a fixed size, even for very small systems, but this is a consequence of the degeneracy of the symmetry-broken ground states and is not related to a gapless energy spectrum. The calculated turning points and maxima of $\Sigma$ respectively converge to the QCP and diverge in the thermodynamic limit as the derivative of the most relevant operator causes the thermodynamic singularity.

Both analytical and computational methods of calculating $\Sigma$ presented respectively for integrable and non-integrable models directly generalize to many-body quantum systems composed of arbitrary degrees of freedom and interactions. This enables further investigations on the geometric description of phase diagrams in complex quantum systems. Due to its information geometric foundation and model-agnostic definition, $\Sigma$ deserves to be investigated for $(1)$ quench or linear/periodic drive or other solvable protocols across a QCP for understanding its dynamical behavior, and $(2)$ characterizing topological phase transitions 
These not only hold the potential to reveal a novel class of protocols for optimal/desired control with the knowledge of the entire landscape of relative entropy at hand, but also open doors to examining the response of the QRE of quantum fields defined on space-times.

Hence this framework demands greater investigations both for a deeper understanding of quantum critical phenomena and for an information-theoretic pursuit of renormalization group theory. This would simultaneously deepen our understanding of universality and pave
routes for many possible implications in modern quantum technologies.

\vspace{0.6cm}
\centerline{\bf Acknowledgments}
\vspace{0.4cm}

D.S. thanks SERB, India for support through the project JBR/2020/000043. A.S. thanks Krishnendu Sengupta for discussions. 



\bibliography{references}

@article{nielsen,
  title={The Many Faces of Information Geometry},
  author={Frank Nielsen},
  journal={Notices of the American Mathematical Society},
  year={2022},
  volume={22},
  pages={69},
  url={https://api.semanticscholar.org/CorpusID:245271344},
  doi={DOI:10.1090/noti2403}
}

@Article{Kim2011InfoGeometry,
AUTHOR = {Kim, Eun-Jin},
TITLE = {Information Geometry, Fluctuations, Non-Equilibrium Thermodynamics, and Geodesics in Complex Systems},
JOURNAL = {Entropy},
volume= {23},
YEAR = {2021},
NUMBER = {11},
pages= {1393},
URL = {https://www.mdpi.com/1099-4300/23/11/1393},
PubMedID = {34828093},
ISSN = {1099-4300},
doi={ https://doi.org/10.3390/e23111393}
}

@article{EunJinKimGeodNonEqComplex,
  title = {Geometric structure and geodesic in a solvable model of nonequilibrium process},
  author = {Kim, Eun-Jin and Lee, UnJin and Heseltine, James and Hollerbach, Rainer},
  journal = {Phys. Rev. E},
  volume = {93},
  issue = {6},
  pages = {062127},
  numpages = {16},
  year = {2016},
  month = {Jun},
  publisher = {American Physical Society},
  doi = {10.1103/PhysRevE.93.062127},
  url = {https://link.aps.org/doi/10.1103/PhysRevE.93.062127}
}

@article{damski,
   title={Fidelity susceptibility of the quantum Ising model in a transverse field: The exact solution},
   volume={87},
   ISSN={1550-2376},
   url={http://dx.doi.org/10.1103/PhysRevE.87.052131},
   DOI={10.1103/physreve.87.052131},
   number={5},
   pages={052131},
   journal={Phys. Rev. E},
   publisher={American Physical Society (APS)},
   author={Damski, Bogdan},
   year={2013}
   }

@article{polkovnikov,
  title={Geometry and non-adiabatic response in quantum and classical systems},
  author={Michael H. Kolodrubetz and Dries Sels and Pankaj Mehta and Anatoli Polkovnikov},
  year={2017},
  volume={697},
  pages={1},
journal = {Physics Reports},
  url={https://api.semanticscholar.org/CorpusID:119195349},
  doi={https://doi.org/10.1016/j.physrep.2017.07.001}
}

@article{Kolodrubetz_2013,
   title={Classifying and measuring geometry of a quantum ground state manifold},
   author={Kolodrubetz, Michael and Gritsev, Vladimir and Polkovnikov, Anatoli},
   DOI={10.1103/physrevb.88.064304},
   volume={88},
   pages={064304},
   journal={Phys. Rev. B},
   year={2013}
}

@article{Zanardi2006Fidelity,
  title = {Information-Theoretic Differential Geometry of Quantum Phase Transitions},
  author = {Zanardi, Paolo and Giorda, Paolo and Cozzini, Marco},
  journal = {Phys. Rev. Lett.},
  volume={99},
  pages={100603},
  year = {2007},
  doi = {10.1103/PhysRevLett.99.100603}
}

@article{Zanardi2007Geometry,
  author    = {L. C. Venuti and Paolo Zanardi},
  title     = {Quantum Critical Scaling of the Geometric Tensors},
  journal   = {Phys. Rev. Lett.},
  volume={99},
  pages={095701},
  year      = {2007},
  doi={https://doi.org/10.1103/PhysRevLett.99.095701}
}

@article{Nielsen2005Entanglement,
  author    = {T. J. Osborne and M. A. Nielsen},
  title     = {Entanglement in a simple quantum phase transition},
  journal   = {Phys. Rev. A},
  volume={66},
  pages={032110},
  year      = {2002},
  doi       = {10.1103/PhysRevA.66.032110}
}

@article{ZhukovEvolutionOfSingleSiteEntanglement,
  author    = {Alexander V. Zhukov and Michal Kolář and Thomas F. George},
  title     = {Evolution of single-site entanglement in a non-equilibrium 1D critical Ising chain},
  journal   = {Physics Letters A},
  year      = {2007},
  volume={368},
  pages={146},
  doi       = {10.1016/j.physleta.2007.03.075}
}

@article{LarssonSingleSiteEntanglement,
  title = {Single-site entanglement of fermions at a quantum phase transition},
  author = {Larsson, Daniel and Johannesson, Henrik},
  journal = {Phys. Rev. A},
  year = {2006},
  volume={73},
  pages={042320},
  url = {https://link.aps.org/doi/10.1103/PhysRevA.73.042320}
}

@article{Tsai2001DMRG,
  author    = {Shan-Wen Tsai and J. B. Marston},
  title     = {Density-Matrix Renormalization-Group Analysis of Quantum Critical Points: I. Quantum Spin Chains},
  journal   = {Phys. Rev. B},
  volume={62},
  pages={5546},
  year      = {2001},
  doi       = {https://doi.org/10.1103/PhysRevB.62.5546}
}

@article{Heyl2017SpeedLimits,
  author    = {Markus Heyl},
  title     = {Quenching a Quantum Critical State by the Order Parameter: Dynamical Quantum Phase Transitions and Quantum Speed Limits},
  journal   = {Phys. Rev. B},
  year      = {2017},
  volume={95},
  pages={060504(R)},
  doi       = {10.1103/PhysRevB.95.060504}
}

@book{Sachdev2011QPT,
  author    = {Subir Sachdev},
  title     = {Quantum Phase Transitions},
  publisher = {Cambridge University Press},
  year      = {2011}
}

@article{Jordan1928Wigner,
  author  = {Jordan, P. and Wigner, E.},
  title   = {Über das Paulische Äquivalenzverbot},
  journal = {Zeitschrift für Physik},
  volume  = {47},
  pages   = {631--651},
  year    = {1928},
  doi     = {10.1007/BF01331938}
}

@article{Barouch1970StatMechI,
  author    = {E. Barouch and B. M. McCoy and M. Dresden},
  title     = {Statistical Mechanics of the ${XY}$ Model. {I}},
  journal   = {Phys. Rev. A},
  volume={2},
  pages={1075},
  year      = {1970},
  doi       = {10.1103/PhysRevA.2.1075}
}

@article{Barouch1971StatMechII,
  author    = {E. Barouch and B. M. McCoy},
  title     = {Statistical Mechanics of the ${XY}$ Model. {II}},
  journal   = {Phys. Rev. A},
  volume={3},
  pages={786},
  year      = {1971},
  doi       = {10.1103/PhysRevA.3.786}
}

@article{Lieb1961SolubleModels,
  author    = {Elliott H. Lieb and Theodore D. Schultz and Daniel C. Mattis},
  title     = {Two soluble models of an antiferromagnetic chain},
  journal   = {Annals of Physics},
  volume={16},
pages={407},
  year      = {1961},
  doi       = {10.1016/0003-4916(61)90115-4}
}

@article{sen3spinising,
  author    = {Adithi Udupa and Samudra Sur and Sourav Nandy and Arnab Sen and Diptiman Sen},
  title     = {Weak universality, quantum many-body scars, and anomalous infinite-temperature autocorrelations in a one-dimensional spin model with duality},
  journal = {Phys. Rev. B},
  volume = {108},
  issue = {21},
  pages = {214430},
  numpages = {21},
  year = {2023},
  month = {Dec},
  publisher = {American Physical Society},
  doi = {10.1103/PhysRevB.108.214430},
  url = {https://link.aps.org/doi/10.1103/PhysRevB.108.214430}
}

@article{quantumisingforbeginners,
	title={{The quantum Ising chain for beginners}},
	author={Glen Bigan Mbeng and Angelo Russomanno and Giuseppe E. Santoro},
	journal={SciPost Phys. Lect. Notes},
	pages={82},
	year={2024},
	publisher={SciPost},
	doi={10.21468/SciPostPhysLectNotes.82},
	url={https://scipost.org/10.21468/SciPostPhysLectNotes.82},
}

@article{Quspin_2017,
   title={QuSpin: a Python package for dynamics and exact diagonalisation of  quantum many body systems part {I}: spin chains},
   volume={2},
   pages={003},
   ISSN={2542-4653},
   url={http://dx.doi.org/10.21468/SciPostPhys.2.1.003},
   DOI={10.21468/scipostphys.2.1.003},
   number={1},
   journal={SciPost Physics},
   publisher={Stichting SciPost},
   author={Weinberg, Phillip and Bukov, Marin},
   year={2017},
   month=feb}

@ARTICLE{SSA_QuantumEntropy_Lieb,
AUTHOR = {Vershynina, A.  and Carlen, E. A and Lieb, E. H.},
TITLE   = {{S}trong {S}ubadditivity of {Q}uantum {E}ntropy},
YEAR    = {2013},
JOURNAL = {Scholarpedia},
VOLUME  = {8},
NUMBER  = {4},
PAGES   = {30920},
DOI     = {10.4249/scholarpedia.30920},
NOTE    = {revision \#169781}
}

@book{doCarmoDiffGeo,
  title = {Differential Geometry of Curves and Surfaces},
  author = {do Carmo, Manfredo P.},
  year = {1976},
  publisher = {Prentice-Hall},
  location = {Englewood Cliffs, NJ},
  edition = {Revised and Updated Second Edition},
  isbn = {0486806995},
  url = {https://books.google.com/books/about/Differential_Geometry_of_Curves_and_Surf.html?id=uXF6DQAAQBAJ},
  note = {Revised and updated republication by Dover Publications, 2016}
}

@article{reduced_fidelity_susceptibility,
author = {Ma Jian and Xu Lei and Wang Xiao-Guang},
title = {Reduced Fidelity Susceptibility in One-Dimensional Transverse Field Ising Model},
doi = {10.1088/0253-6102/53/1/36},
url = {https://dx.doi.org/10.1088/0253-6102/53/1/36},
year = {2010},
month = {jan},
publisher = {},
volume = {53},
number = {1},
pages = {175},
journal = {Communications in Theoretical Physics},
}

@article{Venuti_PRA_div_loc_ent_measures__,
   author={Campos Venuti, L. and Degli Esposti Boschi, C. and Roncaglia, M. and Scaramucci, A.},
   title={Local measures of entanglement and critical exponents at quantum phase transitions},
   volume={73},
   ISSN={1094-1622},
   url={http://dx.doi.org/10.1103/PhysRevA.73.010303},
   DOI={10.1103/physreva.73.010303},
   number={1},
   pages={010303(R)},
   journal={Phys. Rev. A},
   publisher={American Physical Society (APS)},
   year={2006}
   }

@manual{strimmer2023statistics,
  author       = {Korbinian Strimmer},
  title        = {Statistical Methods: Likelihood, Bayes and Regression},
  year         = {2023},
  month        = {6},
  day          = {6},
  organization = {Department of Mathematics, University of Manchester},
  note         = {Lecture notes for MATH20802},
  url          = {https://strimmerlab.github.io/publications/lecture-notes/MATH20802/math20802-script-a4.pdf},
  urldate      = {2025-05-25}
}

@article{Albuquerque__2010,
  author = {Albuquerque, A. Fabricio and Alet, Fabien and Sire, Cl\'ement and Capponi, Sylvain},
  title = {Quantum critical scaling of fidelity susceptibility},
  journal = {Phys. Rev. B},
  volume = {81},
  issue = {6},
  pages = {064418},
  numpages = {12},
  year = {2010},
  month = {Feb},
  publisher = {American Physical Society},
  doi = {10.1103/PhysRevB.81.064418},
  url = {https://link.aps.org/doi/10.1103/PhysRevB.81.064418}
}

@article{Seshadreesan_2018,
   title={Rényi relative entropies of quantum Gaussian states},
   volume={59},
   ISSN={1089-7658},
   url={http://dx.doi.org/10.1063/1.5007167},
   DOI={10.1063/1.5007167},
   number={7},
   pages={072204},
   journal={J. Math. Phys.},
   publisher={AIP Publishing},
   author={Seshadreesan, Kaushik P. and Lami, Ludovico and Wilde, Mark M.},
   year={2018}
   }

@article{Horodecki_2009,
   title={Quantum entanglement},
   volume={81},
   ISSN={1539-0756},
   url={http://dx.doi.org/10.1103/RevModPhys.81.865},
   DOI={10.1103/revmodphys.81.865},
   number={2},
   journal={Rev. Mod. Phys.},
   publisher={American Physical Society (APS)},
   author={Horodecki, Ryszard and Horodecki, Paweł and Horodecki, Michał and Horodecki, Karol},
   year={2009},
   month=jun, pages={865} }

@article{Frenkel_2023,
   title={Integral formula for quantum relative entropy implies data processing inequality},
   volume={7},
   ISSN={2521-327X},
   url={http://dx.doi.org/10.22331/q-2023-09-07-1102},
   DOI={10.22331/q-2023-09-07-1102},
   journal={Quantum},
   publisher={Verein zur Forderung des Open Access Publizierens in den Quantenwissenschaften},
   author={Frenkel, Péter E.},
   year={2023},
   month=sep, pages={1102} }

@misc{leditzky2016thesis,
      title={Relative entropies and their use in quantum information theory}, 
      author={Felix Leditzky},
      year={2016},
      eprint={1611.08802},
      archivePrefix={arXiv},
      primaryClass={quant-ph},
      url={https://arxiv.org/abs/1611.08802}, 
}

@article{M_ller_Hermes_2017,
   title={Monotonicity of the Quantum Relative Entropy Under Positive Maps},
   volume={18},
   ISSN={1424-0661},
   url={http://dx.doi.org/10.1007/s00023-017-0550-9},
   DOI={10.1007/s00023-017-0550-9},
   number={5},
   journal={Annales Henri Poincaré},
   publisher={Springer Science and Business Media LLC},
   author={Müller-Hermes, Alexander and Reeb, David},
   year={2017},
   month=jan, pages={1777} }

@misc{lu2025estimatingquantumrelativeentropies,
      title={Estimating quantum relative entropies on quantum computers}, 
      author={Yuchen Lu and Kun Fang},
      year={2025},
      eprint={2501.07292},
      archivePrefix={arXiv},
      primaryClass={quant-ph},
      url={https://arxiv.org/abs/2501.07292}, 
}

@article{Islam_2015,
   title={Measuring entanglement entropy in a quantum many-body system},
   volume={528},
   ISSN={1476-4687},
   url={http://dx.doi.org/10.1038/nature15750},
   DOI={10.1038/nature15750},
   number={7580},
   journal={Nature},
   publisher={Springer Science and Business Media LLC},
   author={Islam, Rajibul and Ma, Ruichao and Preiss, Philipp M. and Eric Tai, M. and Lukin, Alexander and Rispoli, Matthew and Greiner, Markus},
   year={2015},
   month=dec, pages={77} }

@article{AbaninDemlerPRL2012,
  title = {Measuring Entanglement Entropy of a Generic Many-Body System with a Quantum Switch},
  author = {Abanin, Dmitry A. and Demler, Eugene},
  journal = {Phys. Rev. Lett.},
  volume = {109},
  issue = {2},
  pages = {020504},
  numpages = {5},
  year = {2012},
  month = {Jul},
  publisher = {American Physical Society},
  doi = {10.1103/PhysRevLett.109.020504},
  url = {https://link.aps.org/doi/10.1103/PhysRevLett.109.020504}
}

@book{Cardy_1996, place={Cambridge}, series={Cambridge Lecture Notes in Physics}, title={Scaling and Renormalization in Statistical Physics}, publisher={Cambridge University Press, Cambridge}, author={Cardy, John}, year={1996}, collection={Cambridge Lecture Notes in Physics}}

@article{L_Turban_1982,
doi = {10.1088/0022-3719/15/4/006},
url = {https://doi.org/10.1088/0022-3719/15/4/006},
year = {1982},
month = {feb},
publisher = {},
volume = {15},
number = {4},
pages = {L65},
author = {L Turban},
title = {Self-dual {I}sing chain in a transverse field with multispin interactions},
journal = {J. Phys. C}
}

@article{Pfeuty_1982__PhaseTrans_w_MultipleInteractions,
  title = {Phase transitions in systems with multispin interactions},
  author = {Penson, K. A. and Jullien, R. and Pfeuty, P.},
  journal = {Phys. Rev. B},
  volume = {26},
  issue = {11},
  pages = {6334(R)},
  numpages = {0},
  year = {1982},
  month = {Dec},
  publisher = {American Physical Society},
  doi = {10.1103/PhysRevB.26.6334},
  url = {https://link.aps.org/doi/10.1103/PhysRevB.26.6334}
}

@article{Maritan_1st_2nd_PhaseTrans_w_MultipleInteractions,
  title = {First- and second-order phase transitions in a system with multispin interactions},
  author = {Maritan, A. and Stella, A. and Vanderzande, C.},
  journal = {Phys. Rev. B},
  volume = {29},
  issue = {1},
  pages = {519--521},
  numpages = {0},
  year = {1984},
  month = {Jan},
  publisher = {American Physical Society},
  doi = {10.1103/PhysRevB.29.519},
  url = {https://link.aps.org/doi/10.1103/PhysRevB.29.519}
}

@article{BELAVIN1984_CFT,
title = {Infinite conformal symmetry in two-dimensional quantum field theory},
journal = {Nuclear Physics B},
volume = {241},
number = {2},
pages = {333-380},
year = {1984},
issn = {0550-3213},
doi = {https://doi.org/10.1016/0550-3213(84)90052-X},
url = {https://www.sciencedirect.com/science/article/pii/055032138490052X},
author = {A.A. Belavin and A.M. Polyakov and A.B. Zamolodchikov},
}

@article{Cardy_Nightingale_1986_CFT,
  title = {Conformal invariance, the central charge, and universal finite-size amplitudes at criticality},
  author = {Bl\"ote, H. W. J. and Cardy, John L. and Nightingale, M. P.},
  journal = {Phys. Rev. Lett.},
  volume = {56},
  issue = {7},
  pages = {742--745},
  numpages = {0},
  year = {1986},
  month = {Feb},
  publisher = {American Physical Society},
  doi = {10.1103/PhysRevLett.56.742},
  url = {https://link.aps.org/doi/10.1103/PhysRevLett.56.742}
}

@article{Vidal_extraction_CFT_data_2017,
  title = {Extraction of conformal data in critical quantum spin chains using the Koo-Saleur formula},
  author = {Milsted, Ashley and Vidal, Guifre},
  journal = {Phys. Rev. B},
  volume = {96},
  issue = {24},
  pages = {245105},
  numpages = {13},
  year = {2017},
  month = {Dec},
  publisher = {American Physical Society},
  doi = {10.1103/PhysRevB.96.245105},
  url = {https://link.aps.org/doi/10.1103/PhysRevB.96.245105}
}

@article{Vidal_CFT_&_OPEs_in_spin_systems_2020,
  title = {Conformal Fields and Operator Product Expansion in Critical Quantum Spin Chains},
  author = {Zou, Yijian and Milsted, Ashley and Vidal, Guifre},
  journal = {Phys. Rev. Lett.},
  volume = {124},
  issue = {4},
  pages = {040604},
  numpages = {5},
  year = {2020},
  month = {Jan},
  publisher = {American Physical Society},
  doi = {10.1103/PhysRevLett.124.040604},
  url = {https://link.aps.org/doi/10.1103/PhysRevLett.124.040604}
}

@book{Fradkin_2013_field_thery_in_condensed_matter, place={Cambridge}, edition={2}, title={Field Theories of Condensed Matter Physics}, publisher={Cambridge University Press, Cambridge}, author={Fradkin, Eduardo}, year={2013}}

@article{heralded_magnetism,
  title = {Heralded Magnetism in Non-Hermitian Atomic Systems},
  author = {Lee, Tony E. and Chan, Ching-Kit},
  journal = {Phys. Rev. X},
  volume = {4},
  issue = {4},
  pages = {041001},
  numpages = {13},
  year = {2014},
  month = {Oct},
  publisher = {American Physical Society},
  doi = {10.1103/PhysRevX.4.041001},
  url = {https://link.aps.org/doi/10.1103/PhysRevX.4.041001}
}

@book{cardy1988FiniteSizeScaling,
  title={Finite-size Scaling},
  author={Cardy, J.L.},
  isbn={9780444871107},
  lccn={88025216},
  series={Current physics},
  url={https://books.google.co.in/books?id=lqPvAAAAMAAJ},
  year={1988},
  publisher={North-Holland}
}


\appendix
\onecolumngrid 

\section{From spins to fermions and back} \label{subsec:jordanwignerfromLSM}

The first two of the subsections will follow a similar route taken in \cite{Lieb1961SolubleModels} making use of the derivation of Eq.~C-5 therein; but for our context of the quantum Ising chain in transverse magnetic field\footnote{Ref.~\cite{Lieb1961SolubleModels} presents a general method of calculating the two-site correlations of a free fermions in 
one dimension, but the specific Eq.~C-5 is derived for the $XY$ chain, defined in Eq.~2.1 therein.}. Spin chains defined by the Hamiltonian like Eq.~\eqref{eq:TFIM_ham} and 
Eq.~\eqref{XYChainWithTransverseField} can be mapped to theories of non-interacting fermions where in general only quadratic interactions in the fermionic degrees of freedom, contribute \cite{Jordan1928Wigner, polkovnikov, Lieb1961SolubleModels, Sachdev2011QPT}. Such a Hamiltonian can be analytically diagonalized to obtain the entire spectrum of the theory so that calculations of correlations are often simpler in fermionic language \cite{Lieb1961SolubleModels}.

A class of many-body spin-$\frac{1}{2}$ systems is solved by a transformation from spin-$\frac{1}{2}$ degrees of freedom to spinless fermionic ones, known as Jordan-Wigner transformation \cite{Jordan1928Wigner, Lieb1961SolubleModels, polkovnikov}. Here the fermionic version of Eq.~\eqref{eq:TFIM_ham} is derived as it elucidates the transformation between different quantum degrees of freedom, and also sheds crucial light on the calculation of the two-point correlation function required for our analysis. In this section we exhaustively establish the momentum quantizations for all the fermionic parity sectors for both even and odd sizes of periodic TFIM.
The calculations will be done for a periodic spin chain to maximally harness the translation symmetry throughout the analysis. We start with the spin Hamiltonian
\begin{align}
& \quad \quad \quad \quad \quad \quad \quad \quad \quad \quad \quad H = - \sum_{j=1}^{L} \Big[ J \ \sigma_j^x \sigma_{j+1}^x + h \ \sigma_j^z \Big], \ \ \ \text{with} \\
&\sigma^0_i \equiv \begin{pmatrix} 1 & 0 \\ 0 & 1 \end{pmatrix} \equiv \mathbf{1}_{2 \times 2} , \ \ \ \sigma^1_i \equiv \begin{pmatrix} 0 & 1 \\ 1 & 0 \end{pmatrix} \equiv \sigma^x_i , \ \ \ \sigma^2_i \equiv \begin{pmatrix} 0 & -i \\ i & 0 \end{pmatrix} \equiv \sigma^y_i, \ \ \ \sigma^3_i \equiv\begin{pmatrix} 1 & 0 \\ 0 & -1 \end{pmatrix} \equiv \sigma^z_i.
 \label{PauliSigmaMatrices}
\end{align}
To find the spectrum of this system exactly we first consider the raising and lowering operators
\begin{align*}
& \sigma^x_i \equiv a_i+a^{\dagger}_i, \ \ \sigma^y_i \equiv i(a_i-a^{\dagger}_i), \ \ \sigma^z_i \equiv 1- 2a^{\dagger}a_i, \\
& \{a_i, a_i^{\dagger}\} = 1, \ \ (a_i)^2 = 0 = (a_i^{\dagger})^2 ~~{\rm for} ~~i = 1, 2, \cdots, N, \ \ \text{and} \ [a_{i}^{\dagger}, a_j] = [a_{i}, a_j] = [a_{i}^{\dagger}, a_j^{\dagger}] = 0 
~~{\rm for}~~ i \neq j.
\end{align*}
This implies
\begin{equation}
H = - \sum_{i=1}^{N} J \left[ \left( a_i^\dagger a_{i+1} + \ a_i^\dagger a_{i+1}^\dagger \right) + \text{H.c.} \right] +h \sum_{i=1}^{N} \ (1-2a_i^\dagger a_{i}).
\end{equation}
They partly resemble bosonic operators for different sites but act like fermions on the same site. Therefore it is not possible to diagonalize $\mathcal{H}_{}$ directly with a canonical transformation. They are also known as \textit{hard-core bosons} \cite{quantumisingforbeginners} due to the constraint $(a^\dagger_i)^2 \ket{\psi}=0$ implying no more than 
one boson per site. To resolve this Jordan and Wigner came up with an elegant way of defining fermionic degrees of freedom with the appropriate properties as 
\begin{align}
&\{ c_i \to a_i \} \ : \ \ \ c_i = \exp[i\pi \ \sum_{j=1}^{i-1}a_j^{\dagger}a_j] ~ a_i, \ \ \ {\rm and} \ \ \ c_i^{\dagger} = a_i^{\dagger} ~\exp[-i\pi \ \sum_{j=1}^{i-1}a_j^{\dagger}a_j], \\
& \{ a_i \to c_i \} \ : \ \ \ a_i = \exp[-i\pi \ \sum_{j=1}^{i-1}c_j^{\dagger}c_j] ~ c_i, \ \ \ {\rm and} \ \ \ a_i^{\dagger} = c_i^{\dagger} ~ \exp[i\pi \ \sum_{j=1}^{i-1}c_j^{\dagger}c_j], \\
&\{c_i, c_j^{\dagger}\} = \delta_{ij}, \ \ (c_i)^2 = 0 = (c_i^{\dagger})^2, \ \ {\rm and} \ \ \{c_{i}, c_j\} = 0 = \{c_{i}^{\dagger}, c_j^{\dagger}\} ~~{\rm for}~~ i, j = 1, 2, \cdots , N. \nonumber
\end{align}
So that now these $\{c_i\}$'s and $\{c_i^{\dagger}\}$'s are fermionic operators. Now for the case of open boundaries we have 
\begin{equation} H^{\text{OBC}} = - \sum_{i=1}^{L-1} \left[ J \ (c_i^\dagger c_{i+1} + c_i^\dagger c_{i+1}^\dagger) + \text{H.c.} \ +h \ (1-2c^\dagger_i c_i )\right].
\end{equation}
But the Hamiltonian for the periodic spin chain does not simply translate to the periodicity in the fermionic degrees, rather displays a 
non-trivial dependence on the fermionic parity of the dual state of a many-body spin state. This does not manifest in any the thermodynamic consequences, but results in different quantization rules for the Brillouin zone. Periodic boundary condition in the original spin system requires $\sigma^\alpha_{N+1}=\sigma^\alpha_{1}$ for $\alpha=x,y$. Correspondingly the boundary terms in the hard-core boson Hamiltonian transform in fermionic language in the following way,
\begin{align}
&a_N^{\dagger}a_i = - c_N^{\dagger}c_i \cdot \exp(i\pi \mathcal{\hat{N}}) \neq c_N^{\dagger}c_i, \ \ \ a_N^{\dagger}a_i^{\dagger} = - c_N^{\dagger}c_i^{\dagger} \cdot \exp(i\pi \mathcal{\hat{N}}) \neq c_N^{\dagger}c_i^{\dagger}, \ \ \text{where}~ \ \mathcal{\hat{N}} \equiv \sum_{j=1}^{N}c_j^{\dagger}c_j = \frac{1}{2}\sum_{j=1}^{N}(1-\sigma^z_i). \nonumber \\
& \implies H^{\text{PBC}} = H^{\text{OBC}}_{} + H^{\text{B}}, \ \ \ \text{where} \ \ H^{\text{B}} = \left[ (c_N^\dagger c_1 +  c_N^\dagger c_1^\dagger) + \text{H.c.} \right] \left( e^{i\pi \mathcal{\hat{N}}} + 1 \right). \label{eq:Hgammahinfermions}
\end{align}
With $N =$ the size of the periodic spin chain, and $\mathcal{N} = \braket{gs|\mathcal{\hat{N}}|gs}=$ number of fermionic excitation in the ground state. $H^{\text{PBC}} $ is the Hamiltonian of our interest. Note that even though the number of fermions $\mathcal{N}$ is not conserved by $H_{\text{PBC}}$, the parity of a state $\hat{P} = e^{i\pi \hat{\mathcal{N}}} = (-1)^{\hat{\mathcal{N}}}$ is a \textit{constant of motion} with eigenvalue $\pm 1$ \cite{quantumisingforbeginners}; for even parity $p=0$ and or odd parity $p=1$ of any state under investigation. For a periodic spin chain, $\hat{\sigma}^\alpha_{N+1}=\hat{\sigma}^\alpha_{1}$ so that it either maps to anti-periodic boundary condition (ABC) of the fermionic spectrum where $\mathcal{N} = \text{even} \implies c_{N+1}=-c_1$ or to periodic boundary condition (PBC) of the fermionic spectrum with $\mathcal{N} = \text{odd} \implies c_{N+1}=c_1$. The projectors on the subspaces with even and odd number of particles are 
\begin{equation} \hat{P}_{even}=\frac{1}{2}(e^{i\pi \hat{\mathcal{N}}}+1) = \hat{P}_{0} \ \ {\rm and} \ \ \hat{P}_{odd}=\frac{1}{2}(1-e^{i\pi \hat{\mathcal{N}}}) = \hat{P}_{1}. \end{equation}
With these projectors one can define two fermionic Hamiltonians acting on the $2^{N-1}$ dimensional even/odd parity subspaces of the full Hilbert space.
\begin{equation} H_0 = \hat{P}_{0} \ H_{\text{PBC}} \ \hat{P}_{0} \ \ 
{\rm and} \ \ H_1 = \hat{P}_{1} \ H_{ \text{PBC}} \ \hat{P}_{1} \ \ \implies \ \ \hat{H}_{\text{PBC}} = 
\begin{pmatrix}
 \ \ \ \hat{H}_0 & \big| & 0 \\
 \hline
 0 & \big| & \hat{H}_1 \ \ \ 
\end{pmatrix}. \end{equation}
Now to incorporate the information of parity in a Hamiltonian that acts on the entire Hilbert space, consider
\begin{align}
& \mathbb{H}_{p} = H_{\text{OBC}} + (-1)^p J \left[ (c_N^\dagger c_1 + c_N^\dagger c_1^\dagger) + \text{H.c.} \right], \nonumber \\
& \implies \hat{\mathbb{H}}_{\text{p=\{0,1\}}} = 
- J\sum_{j=1}^{N} \left( c^{\dagger}_j c_{j+1} 
+ \ c^{\dagger}_j c^{\dagger}_{j+1} 
+ \text{H.c.} \right) 
+ h \sum_{j=1}^{N} \left( 1- 2 c^\dagger_j c_j \right).
\end{align}
In this way the boundary terms are compactly treated as $c^{N+1} \equiv (-1)^{p+1} c_1$ for both the fermionic parity sectors. For a finite-sized translationally invariant system, the momentum is quantized differently for different parity sectors and where the allowed momentum values become a set $\mathcal{K}$ of equally distant $N$ angles between $[-\pi, \pi]$ . Due to the symmetry, translation operator and the Hamiltonian is simultaneously diagonalizable. This defines a Fourier transformation in the momentum space with the standard definition of Dirac-delta function. Periodic boundary allow us to translationally invariant modes to define the Fourier transform as

\begin{equation}
c_k = \frac{1}{\sqrt{N}} \sum_{j=1}^{N} c_j \ e^{i j k} \ ~{\rm and} \ \ c_j = \frac{1}{\sqrt{N}} \sum_{j=1}^{N} c_k \ e^{-i j k} \ ~\text{with}~ \ k \in \mathcal{K} \ \ \to \ \ \epsilon_{\mathbf{k}}(h) = \sqrt{(h- \cos k)^2 + \sin^2 k}. \end{equation}

Now depending on the parity sector and the size of the system, the index set of allowed momentum quanta changes for all finite systems, nevertheless converges in the thermodynamic limit. 
Hence, in order for the Fourier transformation to be well-defined for both finite systems with correct convergence for the thermodynamically large system, the necessary requirements for all the combinations of fermionic ground-state parity and sizes of spin chains are shown in Table~\ref{tab:allowed_momentum_quanta_ALL}.

\begin{center}
\begin{table}[htbp]
\centering
\renewcommand{\arraystretch}{3} 
\begin{tabular}{ | c | c | c| }
 \hline
 \textbf{JW: TFIM}$\to H^{(c^\dagger, c)}_{\text{free}}$ &
 $N=\text{even size of spin chain}$ &
 $N=\text{odd size of spin chain}$ \\
 \hline
 $\mathcal{N}=\text{even gs parity}$ &
 \makecell[c]{$c_{N+1} = -c_1 \implies e^{ikN} = -1$ \\ $\implies kN = (2m-1)\pi \in \mathcal{K}^0_{\text{even}},$ for some $m \in \mathbb{Z}$ \\ $\mathcal{K}^0_{\text{even}} := \{ \frac{(2m-1) \pi }{N} : m \in [-\frac{N}{2}+1, \frac{N}{2}] \subset \mathbb{Z} \}$} &
 \makecell[c]{$c_{N+1} = -c_1 \implies e^{ikN} = -1$ \\ $\implies k = (2m-1) \pi \in \mathcal{K}^0_{\text{odd}},$ for some $m \in \mathbb{Z}$ \\ $\mathcal{K}^0_{\text{odd}} := \{ \frac{(2m-1) \pi }{N} : m \in [-\frac{N-1}{2}, \frac{N-1}{2}] \subset \mathbb{Z} \}$} \\
 \hline
 $\mathcal{N}=\text{odd gs parity}$ &
 \makecell[c]{$c_{N+1} = c_1 \implies e^{ikN} = 1$ \\ $\implies k = 2m \pi \in \mathcal{K}^1_{\text{even}},$ for some $m \in \mathbb{Z}$ \\ $\mathcal{K}^1_{\text{even}} := \{ \frac{2m \pi }{N} : m \in [-\frac{N}{2}+1, \frac{N}{2}] \subset \mathbb{Z} \}$} &
 \makecell[c]{$c_{N+1} = c_1 \implies e^{ikN} = 1$ \\ $\implies k = 2m \pi \in \mathcal{K}^1_{\text{odd}},$ for some $m \in \mathbb{Z}$ \\ $\mathcal{K}^1_{\text{odd}} := \{ \frac{2m \pi }{N} : m \in [-\frac{N-1}{2}, \frac{N-1}{2}] \subset \mathbb{Z} \}$} \\
 \hline
\end{tabular}
\caption{Boundary conditions and momentum quantization for TFIM of all sizes and ground-state fermionic parity.}
\label{tab:allowed_momentum_quanta_ALL}
\end{table}
\end{center}

The recurring feature to be noticed here is the presence of $k=0$ and $k= \pm \pi$. Except for these three, every $k$-mode is paired with its counterpart for $-k$ with the same energy, thanks to the symmetry $\epsilon_{-k}(h)=\epsilon_{\mathbf{k}}(h)$. To complete the notations compactly, let us define $\mathcal{\bar{K}}=\{k \in \mathcal{K} : k \neq 0, \pm \pi \}$ and $\mathcal{\bar{K}}^+=\{k \in \mathcal{\bar{K}} : k > 0 \}$ for all the four $\mathcal{K}$'s. This way, for all the $\{k, -k\}$ pairs, that are not at the center or the boundary of the Brillouin zone, the terms in the Hamiltonian can be grouped together where the evolution of each mode is generated by the Hamiltonian
\begin{equation}
\tilde{H}_k = (h- J \cos k)(c^\dagger_k c_k - c_{-k}c^\dagger_{-k}) + iJ \sin k \ (c_{-k}c_k - c^\dagger_k c^\dagger_{-k} ). \end{equation}

Now the boundary of the Brillouin zone contributes by either $\tilde{H}^{\text{B}}_{0}$, $\tilde{H}^{\text{B}}_{\pi}$ or $\tilde{H}^{\text{B}}_{-\pi}$, depending on the fermionic parity and the size of the spin system. For all the different combinations, the Hamiltonians take the forms shown in Table~\ref{tab:hamiltonians_ALL}. 

\begin{center}
\begin{table}[htbp]
\centering
\renewcommand{\arraystretch}{2} 
\begin{tabular}{ | c | c | c| } 
 \hline
 \textbf{JW: TFIM}$\to H^{(c^\dagger, c)}_{\text{free}}$& $N= \text{even size of spin chain}$ & $N= \text{odd size of spin chain}$ \\ 
 \hline
 $\mathcal{N}= \text{even gs parity}$ & $\mathbb{H}_0 = H^{\text{P}}_{0, \text{even}} = 2\sum_{ \mathcal{\bar{K}}^{0, +}_{\text{even}}} \tilde{H_k}$ & $\mathbb{H}_0 = H^{\text{P}}_{0, \text{odd}} + H^{\text{UP}}_{0, \text{odd}} = 2\sum_{\mathcal{\bar{K}}^{0, +}_{\text{odd}}} \tilde{H_k} + \tilde{H}^{\text{B}}_{-\pi}$ \\ 
 \hline
 $\mathcal{N}= \text{odd gs parity}$ & $\mathbb{H}_1 = H^{\text{P}}_{1, \text{even}} + H^{\text{UP}}_{1, \text{even}} = 2\sum_{\mathcal{\bar{K}}^{1,+}_{\text{even}}} \tilde{H_k} + (H^{\text{B}}_{0} + \tilde{H}^{\text{B}}_{\pi})$ & $\mathbb{H}_1 = H^{\text{P}}_{1, \text{odd}} + H^{\text{UP}}_{1, \text{odd}} = 2\sum_{\mathcal{\bar{K}}^{1, +}_{\text{odd}}} \tilde{H_k} + \tilde{H}^{\text{B}}_{0}$ \\ 
 \hline
\end{tabular}
\caption{Hamiltonians for TFIM of all sizes and ground-state fermionic parity.}
\label{tab:hamiltonians_ALL}
\end{table}
\end{center}

The notation emphasizes the paired terms in the fermionic Hamiltonian and the unpaired terms, which arise from the center or the boundary of the Brillouin zone. The unpaired terms are as follows,
\begin{eqnarray}
\tilde{H}^{\text{B}}_{\phi} &=& -h \ (1-2\hat{n}_\phi) + 2 J \cos \phi \ \hat{n}_\phi = 2(h + J \cos \phi)\hat{n}_\phi - h, \ \ \ \text{where}~~ \ \phi = -\pi, 0, \pi. 
\nonumber \\
\implies \tilde{H}^{\text{B}}_{0} &=& 2(h+J)\hat{n}_0 - h, \ \ {\rm and} \ \ \tilde{H}^{\text{B}}_{\pm \pi} = 2(h-J)\hat{n}_{\pm \pi} - h\textbf{}.
\end{eqnarray}

We now note two important consequences.

\begin{enumerate}
 \item The presence of only $\{c^\dagger_k c^\dagger_{-k}, \ c_{-k}c_k, \ c^{\dagger}_k c_k, \ c^{\dagger}_{-k} c_{-k} \}$ when $k \in \mathcal{\bar{K}}^+$, ensures that excitations can arise only in pairs at the symmetric momentum values around $k=0$ therefore considering only the paired terms minima of the ground state energy is achieved by even number of fermionic excitations in the ground state; and 
 \item The boundary contributions only care about the number of fermions at that mode which can only be $0$ or $1$, which can only increase the number of electron by at most $1$, flipping the overall fermionic parity.
\end{enumerate}

Together these simplify the ground state energy from the previous table, which are as follows.

\begin{enumerate}
 \item For $N$ \textbf{even} and $\mathcal{N}$ \textbf{even} : Boundaries or the center of the Brillouin zone do not manifest anyway,
 \item For $N$ \textbf{odd} and $\mathcal{N}$ \textbf{even} : we must have $\hat{n}_0=0$ due to the presence of $\tilde{H}^{\text{B}}_{0}$ in order to preserve even parity,
 \item For $N$ \textbf{odd} and $\mathcal{N}$ \textbf{odd} : we must have $\hat{n}_\pi=1$ due to the presence of $\tilde{H}^{\text{B}}_{\pi}$ in order to preserve odd parity, and lastly,
 \item For $N$ \textbf{even} and $\mathcal{N}$ \textbf{odd} : both the boundary terms at $k= 0$ and $k = \pi$ contribute so that they can be written as
\begin{equation} \tilde{H}^{\text{B}}_{0}+\tilde{H}^{\text{B}}_{\pi} =( 2(h+J)\hat{n}_0 - h) + ( 2(h+J)\hat{n}_\pi - h)= 2h (\hat{n}_0+\hat{n}_\pi-1) + 2J(\hat{n}_0-\hat{n}_\pi). \end{equation}

For $h>0$ this is minimized by $\hat{n}_\pi=1$
and $\hat{n}_0=0$. Hence the additional unpaired fermionic mode contributes by the presence of a unpaired fermionic excitation at the boundary of the Brillouin zone so that the ground state energy receives the corresponding contribution 
\begin{equation} \delta \tilde{E}^{\text{B}, h^+}_{0 \ \& \ \pi} = \text{min}^{h>0}_{\{n_0, n_\pi\}}(\tilde{H}^{\text{B}}_{0}+\tilde{H}^{\text{B}}_{\pi}) = -2J, \end{equation}
whereas for $h<0$ this is also minimized by $\hat{n}_\pi=1$ and $\hat{n}_0=0$ which makes a different contribution to the ground state energy given by
\begin{equation} \delta \tilde{E}^{\text{B}, h^-}_{0 \ \& \ \pi} = \text{min}^{h<0}_{\{n_0, n_\pi\}}(\tilde{H}^{\text{B}}_{0}+\tilde{H}^{\text{B}}_{\pi}) = 2(h-J). \end{equation}
\end{enumerate}

The above equations essentially reproduce Eqs.~(43) and (71) from Ref.~\cite{quantumisingforbeginners}, also extending it further for momentum-space quantization of odd chains. Therefore, a TFIM with even $N$ has the fermionic representation
\begin{equation}
 H = 2\sum_{ \mathcal{\bar{K}}^{0, +}_{\text{even}}} \tilde{H_k}.
\end{equation}

\section{Diagonalizing free-fermionic Hamiltonians} \label{subsec:DiagQuadHam}

A general Hamiltonian which is quadratic in fermionic 
operators has the form 
\begin{equation}
 H = \sum_{i,j} [c_i^{\dagger}A_{ij}c_j + \frac{1}{2}(c_i^{\dagger}B_{ij}c_j^{\dagger} + \text{H.c.})], \label{GenQuadFermHam}
\end{equation} 
where the $c_i^{\dagger}$ and $c_i$ are fermionic creation and annihilation operators. The Hermiticity of the Hamiltonian requires that $\bar{A}_{ij} = A_{ji}$ while the anti-commutation of the $\{c_i\}$'s require that $\bar{B}_{ij} = -B_{ji}$. In our case of Eq.~\eqref{eq:TFIM_ham} and 
Eq.~\eqref{XYChainWithTransverseField}, $A$ and $B$ are real. Now diagonalizing this in terms of fermionic degrees amounts to finding the linear transformation 
\begin{align}
&\gamma_k = \sum_i \left( g_{ki} c_i + h_{ki} c_i^\dagger \right), \ \ {\rm and} \ \ \gamma_k^\dagger = \sum_i \left( g_{ki} c_i^\dagger + h_{ki} c_i \right). \ \ 
\label{finalcanonicalfermions}
\end{align}
Hence the Hamiltonian becomes
\begin{align}
& H= \sum_k \epsilon_{\mathbf{k}} \ \gamma^\dagger_{k} \gamma_{k} + \text{constant}, ~\ \text{requiring} \ \ [\eta_k, \ H] = \Lambda_k \ \eta_k,
\end{align}
just as in Eq.~\eqref{XYChainWithTransverseField}. This restricts the possibilities of the $\{g_{ki}$
and $h_{ki}\}$'s because 
\begin{equation}
 \Lambda_{k}g_{k i} = \sum_{j} (g_{k j}A_{j i} - h_{k j}B_{j i}) ~~~{\rm and}~~~ \Lambda_{i}h_{k i} = \sum_{j} (g_{k j}B_{j i} - h_{k j}A_{j i}).
\end{equation} 
They are simplified by introducing the linear combinations $\mathbf{\phi}_{k i} = g_{k i} + h_{k i}$ and $\mathbf{\psi}_{k i} = g_{k i} - h_{k i}$.
In terms of these the coupled equations become $\phi_k(\mathbf{A} - \mathbf{B}) = \Lambda_k\psi_k, \ 
{\rm and} \ \ \psi_k(\mathbf{A} + \mathbf{B}) = \Lambda_k\phi_k$. They are decoupled by using 
\begin{align}
 \phi_k(\mathbf{A} - \mathbf{B})(\mathbf{A} + \mathbf{B}) = \Lambda_k^2\phi_k \ \ \ \ {\rm and} \ \ \ \ \psi_k(\mathbf{A} + \mathbf{B})(\mathbf{A} - \mathbf{B}) = \Lambda_k^2\psi_k.
 \label{transformationequation} 
\end{align}
Since $\mathbf{A}$ is symmetric and $\mathbf{B}$ is antisymmetric, we have $(\mathbf{A} + \mathbf{B})^T = \mathbf{A} - \mathbf{B}$. 
Hence both $(\mathbf{A} - \mathbf{B})(\mathbf{A} + \mathbf{B})$ and $(\mathbf{A} + \mathbf{B})(\mathbf{A} - \mathbf{B})$ are symmetric and at least positive semi-definite. Thus all the $\Lambda_k$'s are real and it is possible to choose all the $\phi_k$'s and $\psi_k$'s to be real as well as orthogonal. If the $\phi_k$'s are normalized vectors $(\sum_i \phi_{k i}^2 = 1)$, then the $\psi_k$'s are also automatically normalized when $\Lambda_k$ $\neq 0$ or can be so chosen when $\Lambda_k = 0$. This ensures that
\begin{equation} \sum_i (g_{k i}g_{k' i} + h_{k i}h_{k' i}) = \delta_{k k'} \ \ \ {\rm and} \ \ \ \sum_i (g_{k i}h_{k' i} - g_{k' i}h_{k i}) = 0. \end{equation}
which are
the necessary and sufficient conditions that the $\eta_k$'s and $\eta_k^\dagger$'s be canonical Fermi operators. Therefore, from the specifics of the theory, as in case $H$, $A$ and $B$ matrices will encode the spacial interactions across the system, which will provide the solutions of $\{\mathbf{\phi}_{k i}, \mathbf{\psi}_{k i}\}$ guaranteeing the exact diagonalization.

For TFIM, we have the following for $\mathbf{A}$, $\mathbf{B}$ and $Z = ( \mathbf{A} - \mathbf{B})( \mathbf{A} + \mathbf{B})$

\begin{align}
& \mathbf{A} = - \begin{pmatrix} 
2h & 1 & & & 1 \\
1 & 2h & 1 & & 0 \\
& \ddots & \ddots & \ddots & \\
0 & & 1 & 2h & 1 \\
1 & & & 1 & 2h
\end{pmatrix}, \ \mathbf{B} = \begin{pmatrix} 
0 & -1 & & & 1 \\
1 & 0 & -1 & & 0 \\
& \ddots & \ddots & \ddots & \\
0 & & 1 & 0 & -1 \\
-1 & & & 1 & 0
\end{pmatrix} , \ Z = \begin{pmatrix} 
h^2 + 1 & -h & 0 & \dots & 0 &-h \\
-h & h^2+ 1 & -h & \dots & 0 &0 \\
0 & -h & h^2+ 1 & \dots &\vdots & 0\\
\vdots & \ddots & \ddots & \vdots & \vdots&\vdots\\
0 & 0 & \dots & 0 & h^2+ 1 & -h \\
-h & 0 & \dots & \dots &-h & h^2+ 1
\end{pmatrix}.
\label{eq:A_and_B_TFIM}
\end{align}
This transforms $H$ into Eq.~\eqref{eq:Hgammahinfermions}, which is equivalent to having the above $\mathbf{A}$ and $ \mathbf{B}$ matrices. These matrices are sufficient to determine the complete solution of $\{\mathbf{\phi_{k}}, \mathbf{\psi_{k}}\}$ due to $\mathbf{\phi_{k}} (\mathbf{A}-\mathbf{B})(\mathbf{A}+\mathbf{B}) = \epsilon_{\mathbf{k}}^2 \ \mathbf{\phi_{k}}$ and $\psi_k(\mathbf{A} + \mathbf{B}) = \epsilon_{\mathbf{k}}\phi_k$ with
\begin{eqnarray}
\mathbf{u}_{\mathbf{k} j} = \begin{pmatrix} \sin jk \\ \cos jk \end{pmatrix}, 
\mathbf{v}_{\mathbf{k} j } = \begin{pmatrix} (\cos k -h)\sin kj + \sin k \cos kj \\ ( \cos k -h)\cos kj - \sin k \sin kj \end{pmatrix}, \nonumber \\ 
\implies \mathbf{\phi_{k}}^{T} = \frac{1}{\sqrt{N} }\begin{pmatrix} \vdots \\ \mathbf{u}_{\mathbf{k} j} \\ . \\ \end{pmatrix}, \ 
\mathbf{\psi_k}^{T} = \frac{1}{\epsilon_{\mathbf{k}}\sqrt{N} }\begin{pmatrix} \vdots \\ \mathbf{v}_{\mathbf{k} j} \\ . \\ \end{pmatrix}. \label{eq:PhiPsiSS}
\end{eqnarray}

With the upper and lower solutions of $(\mathbf{u}_{\mathbf{k} j}, \mathbf{v}_{\mathbf{k} j})$ respectively for $k>0$ and $k \leq 0$. As evident in Eq.~\eqref{finalcanonicalfermions} and the related discussions in Appendix \ref{subsec:DiagQuadHam}, these $\{\mathbf{\phi_{k}}, \mathbf{\psi_{k}}\}$'s define the correct canonical fermionic operators. 

\section{Two-point correlations in TFIM}\label{subsec:twopointcorrelationsinTFIM}

The two-point correlation functions are required to construct the two-site reduced density matrix of any quantum system defined discretely, i.e., on lattices or graphs. In the language of operator product expansion any two-point correlation function of a many spin-$\frac{1}{2}$ system can be expressed as (see Eqs.~18-19 in \cite{Nielsen2005Entanglement} for more detail)
\begin{align}
\hat{\rho}_{i j} := \text{tr}_{\hat{ij}}(\hat{\rho}) = \frac{1}{4} \sum_{\alpha, \beta = 0} p_{\alpha \beta} \ \hat{\sigma}_{i}^{\alpha} \otimes \hat{\sigma}_{j}^{\beta} ~~{\rm for}~~ i, j = 1, 2, \cdots, N~~ \text{with} \ p_{\alpha \beta} = \text{tr}(\hat{\rho}_{i j} \hat{\sigma}_{i}^{\alpha} \hat{\sigma}_{j}^{\beta}) = \langle \hat{\sigma}_{i}^{\alpha} \ \hat{\sigma}_{j}^{\alpha} \rangle,
\label{eq:op_prod_expansion}
\end{align}
and $\{ \hat{\sigma}_{i}^{\alpha} \ | \ \alpha \in [0, 1, 2, 3]\}$ is the basis of the space spanning the Hilbert space of a single site. Therefore to obtain the basis of the Hilbert space spanning $K$ spin-$\frac{1}{2}$ degrees of freedom, in general one needs to take the linear combination of $4^K$-many tensor products of the $\{ \hat{\sigma}_{i}^{\alpha} \} $ bases, weighted by the expectation value of the corresponding operator since we are interested in correlations of 
nearest-neighbor spins, we call $\langle \hat{\sigma}_{i}^{\alpha} \ \hat{\sigma}_{i+1}^{\alpha} \rangle$ (for $\alpha \in \{ x, y, z \}$) the \textit{NN}-correlations for brevity. The \textit{NN-reduced density matrix} $ \hat{\rho}_{\text{NN}}^{(2)}$ is a special case of $\hat{\rho}_{i j} $ when $j=i + 1$, therefore
\begin{equation}
 \hat{\rho}_{\text{NN}}^{(2)}:= \hat{\rho}_{i, i+1} = \frac{1}{4} \sum_{\alpha, \beta = 0} \langle \hat{\sigma}_{i}^{\alpha} \ \hat{\sigma}_{i+1}^{\alpha} \rangle \ \hat{\sigma}_{i}^{\alpha} \otimes \hat{\sigma}_{i+1}^{\beta}.
 \label{rhoNN} 
\end{equation} 
Now, to completely obtain the NN-reduced density matrix, many of the correlations vanish due to symmetry for our case of $H$. We first note that the only non-vanishing single-site correlation is $\langle \sigma^z \rangle$ and for the two-site correlations, non-vanishing ones come in the form of $\langle \hat{\sigma}_{i}^{\alpha} \ \hat{\sigma}_{i+1}^{\alpha} \rangle$ for $\alpha \in \{x, y, z\}$. This is solely due to the presence of the global invariance of $H$ under $\{ \sigma^x \to -\sigma^x \}$ and/or $\{ \sigma^y \to -\sigma^y \}$. A widely accepted measure of long-range order in 1-dimensional quantum spin chains as been proposed in \cite{Lieb1961SolubleModels} in which the two-point correlations are derived for an 
antiferromagnetic $XY$ chain using 
\begin{align}
&\rho_{lm} = \langle\Psi_0 | \mathbf{S}_l \cdot \mathbf{S}_m | \Psi_0\rangle, \ \ \\
\implies ~~& \rho_{lm}^x = \langle\Psi_0 | \sigma_l^x \sigma_m^x | \Psi_0\rangle = \langle\Psi_0 | (a_l^\dagger + a_l)(a_m^\dagger + a_m) | \Psi_0\rangle, \nonumber \\
&\rho_{lm}^y = \langle\Psi_0 | \sigma_l^y \sigma_m^y | \Psi_0\rangle = \langle\Psi_0 | (a_l^\dagger - a_l)(a_m^\dagger - a_m) | \Psi_0\rangle, \nonumber \\
&\rho_{lm}^z = \langle\Psi_0 | \sigma_l^z \sigma_m^z | \Psi_0\rangle = \langle\Psi_0 | (2a_l^\dagger a_l - 1)(2a_m^\dagger a_m - 1) | \Psi_0\rangle. \nonumber 
\end{align}
We now consider $\rho_{lm}^x$ written in terms of fermionic operators $c_i$'s as 
\begin{equation}
 \rho_{lm}^x = \langle\Psi_0 | (c_l^\dagger + c_l) ~ \exp(i \pi \sum_{j=l}^{m-1} c_j^\dagger c_j) ~ (c_m^\dagger + c_m) | \Psi_0\rangle.
\end{equation} 
We now observe that $\exp(i \pi c_j^\dagger c_j) = (c_j^\dagger + c_j)(c_j^\dagger - c_j)$. This readily allows us to break the correlation functions into chains
of alternative terms of $(c_j^\dagger + c_j)$ and $(c_j^\dagger - c_j)$. Hence defining \begin{equation}
 \mathcal{A}_j=c_j^\dagger + c_j \ \ {\rm and} \ \ \ \mathcal{B}_j=c_j^\dagger - c_j.
\end{equation} 
allows us to express the correlations in the following way,
\begin{equation}
 \rho_{lm}^x = \langle\Psi_0 | \mathcal{B}_l \ \mathcal{A}_{l+1} \ \mathcal{B}_{l+2} \ \mathcal{A}_{l+3} \dots \mathcal{A}_{m-1} \ \mathcal{B}_{m-1} \ \mathcal{A}_{m}| \Psi_0\rangle.
\end{equation} 
Further observing that $\exp(i \pi c_j^\dagger c_j) = -(c_j^\dagger - c_j)(c_j^\dagger + c_j)$ we find that
\begin{equation}
 \rho_{lm}^y = (-1)^{m-l}\langle\Psi_0 | \mathcal{A}_l \ \mathcal{B}_{l+1} \ \mathcal{A}_{l+2} \ \mathcal{B}_{l+3} \dots \mathcal{B}_{m-1} \ \mathcal{A}_{m-1} \ \mathcal{B}_{m}| \Psi_0\rangle.
\end{equation} 
Finally using $2a^{\dagger}_ia_i - 1 = -(a^{\dagger}_i+a_i)(a^{\dagger}_i-a_i) = -(c^{\dagger}_i+c_i)(c^{\dagger}_i-c_i)$ we find
\begin{equation}
 \rho_{lm}^z = \langle\Psi_0 | \mathcal{A}_l \ \mathcal{B}_{l} \ \mathcal{A}_{m} \ \mathcal{B}_{m}| \Psi_0\rangle.
\end{equation} 
To evaluate these expectation values we make use of the well-known Wick theorem in quantum field theory to reduce the computation of vacuum expectation values because only the pair-contractions contribute because the operators anti-commute with themselves. In general if $\mathcal{O}_1, \mathcal{O}_2, \dots \mathcal{O}_{2n}$ are such operators then \cite{Lieb1961SolubleModels} 
\begin{equation}
 \langle\Psi_0 | \mathcal{O}_1 \ \mathcal{O}_2 \ \dots \ \mathcal{O}_{2n} | \Psi_0\rangle = \sum_{pairs} (-1)^p \ \prod_{pairs} \text{contractions of the pairs},
 \label{WickTheorem}
\end{equation} 
where the contractions are defined as $ \langle \mathcal{O}_i \ \mathcal{O}_j \rangle = \langle\Psi_0 | \mathcal{O}_i \ \mathcal{O}_j | \Psi_0\rangle$ with $p$ being the signature of the permutation, for a given pairing, necessary to bring operators of the same pair next to one another from the original order. This simplifies the computation of the correlation functions substantially as the only relevant correlations required to calculate the $\rho^{\alpha}_{lm}$'s are 
\begin{equation}
 \langle \mathcal{A}_i \mathcal{A}_j \rangle = \sum_{k} \phi_{ki} \ \phi_{kj} = \delta_{ij} = - \langle \mathcal{B}_i \mathcal{B}_j \rangle = - \sum_{k} \psi_{ki} \ \psi_{kj}, \ \ {\rm and} \ \ \ \langle \mathcal{B}_i \mathcal{A}_j \rangle = - \langle \mathcal{A}_j \mathcal{B}_i \rangle = - \sum_{k} \psi_{ki} \ \phi_{kj} =: G_{ij}.
 \label{basicCorrelations}
\end{equation}
An important property of $G_{ij}$ for a cyclic problem is that $G_{ij} = G_{i-j}$; this will be used later. The most straightforward pairing contributing to $\rho^x_{lm}$ is of the kind $\langle B_l A_{l+1}\rangle\langle B_{l+1}A_{l+2}\rangle \cdots \langle B_{m-1}A_m\rangle$. All other pairings can be obtained from this one by permuting the $A$'s among themselves with the $B$'s fixed. Because the number of crossings of $B$'s by $A$'s is then always even, the sign associated with a given permutation is $(-1)^{p'}$, where $p'$ is the signature of the permutation of the $A$'s. Thus $\rho^x_{lm} = \sum_p (-1)^{p'} G_{l,P(l+1)} \cdots G_{m-1,P(m)}$. Similarly $\rho^y_{lm}$ and $\rho^z_{lm}$ can be calculated using the Wick theorem to obtain 
\begin{equation} 
\rho^x_{lm}= \begin{vmatrix} 
G_{l,l+1} & G_{l,l+2} & \cdots & G_{lm} \\
\vdots & & & \vdots \\
G_{m-1,l+1} & \cdots & & G_{m-1,m}
\end{vmatrix}, \ \ \ \ \ \rho^y_{lm} = \begin{vmatrix}
G_{l+1,l} & G_{l+1,l+1} & \cdots & G_{l+1,m-1} \\
\vdots & & & \vdots \\
G_{ml} & \cdots & & G_{m,m-1}
\end{vmatrix},
\label{eq:derivedTwoPointCorrelations}
\end{equation}
and
\begin{equation*}
 \rho^z_{lm} = (\langle \mathcal{A}_l \mathcal{B}_l\rangle\langle \mathcal{B}_m \mathcal{B}_m\rangle - \langle \mathcal{A}_l \mathcal{B}_m\rangle\langle \mathcal{A}_m \mathcal{B}_l\rangle) = G_{0}^2 - G_{m-l}G_{l-m}, 
 ~~{\rm for} ~~l < m \leq N.
\label{TwoPointSpinCorrelationsInFermionicSolutions}
\end{equation*}
Thus, both $\rho^x_{lm}$ and $\rho^y_{lm}$ are particular sub-determinants of $\det \mathbf{G}$. Now it is immediate that the \textit{NN}-correlations are only the simplest of these general two-point correlations. In case of that, we simply need to consider what happens when $m = l+1$ for all $l$.
As discussed in Eq.~\eqref{rhoNN} we are interested in $\rho^{\alpha}_{l, \ l+1} = \langle \hat{\sigma}_{i}^{\alpha} \ \hat{\sigma}_{i+1}^{\alpha} \rangle, \ \ \alpha \in \{x, y, z\}$ for our analysis. But the technology can be used to calculate any correlation function of any system that can be mapped to a theory of quadratic fermions. We then have
\begin{equation}
 \rho^{x}_{l, \ l+1} = G_{l+1,l} =: G_{-1}, \ \ \rho^{y}_{l, \ l+1} = G_{l,l+1} =: G_{+1} \ \ {\rm and} \ \ \rho^{z}_{l, \ l+1} = G_{0}^2 - G_{-1}G_{+1}, \ \ \ {\rm for} ~~l = 1,2,\cdots,N.
 \label{NNcorrelations}
\end{equation}

This is the main object of our interest. From the detail of a given model, say in our case of $H$, the matrices $A$ and $B$ are determined for Eq.~\eqref{GenQuadFermHam} which gives the solution of $\{\phi_{ik}, \psi_{ik}\}$'s as in 
Eq.~\eqref{transformationequation}. These then determine all the basic correlations of the theory using 
Eq.~\eqref{basicCorrelations}.

\section{Thermodynamic singularities of correlation functions and the response of relative entropy in TFIM}
\label{subsec:singular_contributions_section}

The exact solution of TFIM allows us to determine the singular contributions of the correlation functions and their derivatives. Since we focus on correlations contained in a single site and a pair of adjacent sites in a translationally invariant TFIM, results at $h=1$ for finite but large $N$ would allow us to determine the square logarithmic divergence of this response function. Now since the correlation functions remain finite, the derivatives of the correlations are solely responsible for the divergence of the response of QRE. All the summands being symmetric around $k=0$, the singular contributions are due to $\int_{\pi/N}^\pi dk/k$ in the derivatives of the correlations because they identically diverge at $k=0$ making the analysis ill-defined. So making use of $\sum_k \to (N/\pi) \int_{0}^{\pi} dk$ for the correlations, and $\sum_k \to (N/\pi) \int_{\pi/N}^{\pi} dk$ for their derivatives, we find in the limit
$N \to \infty$,
\begin{align}
& G_0 = \braket{\sigma^z_i}(h,N) = \frac{1}{N} \sum_{k \in \mathcal{K}^0_{\text{even}}} \frac{h- \cos k}{\sqrt{1+h^2- 2h \cos k}} \to G_0\Big\vert^{h=1}_{N = \infty} = \frac{2}{\pi}, \\
& G_{xx} = \braket{\sigma^x_{i}\sigma^x_{i+1}}(h,N)= - \frac{1}{N} \sum_{k \in \mathcal{K}^0_{\text{even}}} \frac{(h- \cos k) \cos k - \sin^2k}{\sqrt{1+h^2- 2h \cos k}} \to G_{xx}\Big\vert^{h=1}_{N = \infty} = \frac{2}{\pi}, \\
& G_{yy} = \braket{\sigma^y_{i}\sigma^y_{i+1}}(h,N)= - \frac{1}{N} \sum_{k \in \mathcal{K}^0_{\text{even}}} \frac{(h- \cos k) \cos k + \sin^2 k}{\sqrt{1+h^2- 2h \cos k}} \to G_{yy}\Big\vert^{h=1}_{N = \infty} = -\frac{2}{3\pi}, \\
& G_{zz} = \braket{\sigma^z_{i}\sigma^z_{i+1}}(h,N) = G_0^2 - G_{xx}G_{yy} = \braket{\sigma^z_i}^2- \braket{\sigma^x_{i}\sigma^x_{i+1}}\braket{\sigma^y_{i}\sigma^y_{i+1}} \to G_{zz}\Big\vert^{h=1}_{N = \infty} = \frac{16}{3\pi^2}.
\end{align}

Also for the derivatives of these correlations, we can expand around $k=0$ to isolate the singular and regular contributions to exactly derive the logarithmic divergence of those derivatives. We note that all the integrands are symmetric around $k=0$, therefore the series expansion contain $|k|$ instead of $k$, but since we are integrating from $\pi/N \to 0$ to $\pi$, we omit the absolute value symbol for brevity. We then obtain

\begin{align}
& \frac{d G_0}{dh}\Big\vert^{h=1}_{N \to \infty} = \lim_{N \to \infty} \frac{1}{N} \sum_{k \in \mathcal{K}^0_{\text{even}}} \frac{\cos^2 \frac{k}{2}}{2|\sin \frac{k}{2}|} \sim \frac{1}{\pi} \int_{\pi/N}^\pi dk \ \Big[ \underbrace{\frac{1}{k}}_\text{singular part} \underbrace{ - \frac{5k}{24} + \frac{67k^3}{5760} - \frac{19 k^5}{193536}+ \mathcal{O}(k^7)}_\text{regular part} \Big], \nonumber 
\end{align}
\begin{align}
& \implies \frac{d G_0}{dh}\Big\vert^{h=1}_{N \to \infty, \ \text{sing}} = \frac{1}{\pi} \int_{\pi/N}^\pi \frac{dk}{k} = \frac{1}{\pi} \ln[\frac{\pi}{\pi/N}] = \frac{\ln N}{\pi}, \ \ {\rm and} \ \ \frac{d G_0}{dh}\Big\vert^{h=1}_{N \to \infty, \ \text{regular}} = \frac{5 \pi}{48 }(-1+\frac{1}{ N^2}) + \mathcal{O}(\frac{1}{N^3}). \\
& \frac{d G_{xx}}{dh}\Big\vert^{h=1}_{N \to \infty} = \lim_{N \to \infty} \frac{1}{N} \sum_{k \in \mathcal{K}^0_{\text{even}}} \frac{\cos^2 \frac{k}{2}}{2|\sin \frac{k}{2}|} \sim \frac{1}{\pi} \int_{\pi/N}^\pi dk \ \Big[\underbrace{\frac{1}{k}}_\text{singular part} \underbrace{ - \frac{5k}{24} + \frac{67k^3}{5760} - \frac{19 k^5}{193536}+ \mathcal{O}(k^7)}_\text{regular part} \Big], \nonumber \\
& \implies \frac{d G_{xx}}{dh}\Big\vert^{h=1}_{N \to \infty, \ \text{sing}} = \frac{1}{\pi} \int_{\pi/N}^\pi \frac{dk}{k} = \frac{1}{\pi} \ln[\frac{\pi}{\pi/N}] = \frac{\ln N}{\pi}, \ \ {\rm and} \ \ \frac{d G_{xx}}{dh}\Big\vert^{h=1}_{N \to \infty, \ \text{regular}} = \frac{5 \pi}{48 }(-1+\frac{1}{ N^2}). \\
& \frac{d G_{yy}}{dh}\Big\vert^{h=1}_{N \to \infty} = \lim_{N \to \infty} \frac{1}{N} \sum_{k \in \mathcal{K}^0_{\text{even}}} \frac{2\cos^2 k+\cos k-1}{4|\sin \frac{k}{2}|} \sim \frac{1}{\pi} \int_{\pi/N}^\pi dk \ \Big[\underbrace{\frac{1}{k}}_\text{singular part} \underbrace{ - \frac{29 \ k}{24} + \frac{1747 \ k^3}{5760} - \frac{30839 \ k^5}{967860}+ \mathcal{O}(k^7)}_\text{regular part} \Big], \nonumber \\
& \implies \frac{d G_{yy}}{dh}\Big\vert^{h=1}_{N \to \infty, \text{sing}} = \frac{1}{\pi} \int_{\pi/N}^\pi \frac{dk}{k} = \frac{1}{\pi} \ln[\frac{\pi}{\pi/N}] = \frac{\ln N}{\pi}, \ {\rm and} \ \frac{d G_{yy}}{dh}\Big\vert^{h=1}_{N \to \infty, \text{regular}} = \frac{29 \pi}{48 }(-1+\frac{1}{ N^2})+ \mathcal{O}(\frac{1}{N^3}), \\
& \text{finally \ } \ \frac{d G_{zz}}{dh}\Big\vert^{h=1}_{N \to \infty} = 2 G_0\Big\vert^{h=1}_{N \to \infty} \cdot \frac{d G_{0}}{dh}\Big\vert^{h=1}_{N \to \infty}- G_{xx}\Big\vert^{h=1}_{N \to \infty} \cdot \frac{d G_{yy}}{dh}\Big\vert^{h=1}_{N \to \infty}- G_{yy}\Big\vert^{h=1}_{N \to \infty} \cdot \frac{d G_{xx}}{dh}\Big\vert^{h=1}_{N \to \infty} \\
& \implies \frac{d G_{zz}}{dh}\Big\vert^{h=1}_{N \to \infty, \ \text{sing}} = \frac{16 \ln N }{3 \pi^2}, \ 
{\rm and} \ \frac{d \ G_{zz}}{dh}\Big\vert^{h=1}_{N \to \infty, \ \text{reg}} = - \frac{14}{9} - \frac{14}{9N^2} + \mathcal{O}(\frac{1}{N^3}).
\end{align}

To obtain the thermodynamic singularity of the response of QRE in one and (neighboring) two sites of TFIM with finite but large size, the singularity caused by the $k=0$ pole of the derivative of correlation functions are sufficient. Such poles of the correlation functions come with a coefficient $\propto |h-1|$, therefore all the correlation functions are regular at the QPT. Therefore to determine the singular contributions of the response at $h=1$, the correlation functions can be replaced by their thermodynamic values while allowing finite size to manifest only in the divergence of the derivatives of correlation functions.

For a single site in the TFIM, 
Eq.~\eqref{eq:TFImOneSiteSuscept} implies that
\begin{align}
& \Sigma_{1}^\text{TFIM}(h=1,N \to \infty) = \frac{1}{2} \frac{\Big( \frac{d G_0}{dh}\Big\vert^{h=1}_{N \to \infty} \Big)^2}{1- \Big( G_0\Big\vert^{h=1}_{N = \infty} \Big)^2} = \frac{(5\pi^2 - 48\ln N)^2}{4608\,(-4 + \pi^2)} - \frac{5\pi^2(5\pi^2 - 48\ln N)}{2304\,(-4 + \pi^2)\,N^2} + \mathcal{O}(\frac{1}{N^3}), \nonumber \\
& \implies \Sigma_{1}^\text{TFIM}(h=1,N \to \infty)_{\text{sing}} = A_1 \ \ln^2 N, \ \ \ \ \text{where} \ \ \ \ A_1 = \frac{48^2}{4608(\pi^2 - 4)} \approx 0.0851846.
\label{eq:singularity_of_one_site_response}
\end{align}

For the case of two neighboring sites in the TFIM, the density matrices defined in Eq.~\eqref{eq:NNReducedMatrixOfTFIM} introduce all the non-vanishing correlations in that range, which along with their derivatives, make the formal expression of the response lengthy. Hence we present the singular and leading regular contributions at $h=1$,
\begin{align}
& \Sigma_{2}^\text{TFIM}(h=1,N \to \infty) \nonumber \\
& = \frac{(-1 + N^2)^2 \pi^4 (-15424 + 2061 \pi^2) - 96 N^2 (-1 + N^2) \pi^2 (-1184 + 99 \pi^2) \ln N + 2304 N^4 (-64 + 9 \pi^2) \ln^2 N}{1152 N^4 (256 - 112 \pi^2 + 9 \pi^4)} \nonumber
\\
& \approx 1.81971 \left(8.37385 + 3.42731 \ln N + \ln^2 N \right) - \frac{6.23672 \left(4.88654 + \ln N \right)}{N^2} + \mathcal{O}(\frac{1}{N^3}) \nonumber \\
& \implies \Sigma_{2}^\text{TFIM}(h=1,N \to \infty)_{\text{sing}} = 1.81971 \ln^2 N.
\label{eq:singularity_of_two_site_response}
\end{align}

To understand the nature of the divergence near the vicinity of critical point of large systems, $\mathcal{O}(|h-1|)$ contributions play the leading role since the momentum modes $k=\pm \pi/N$ dominate the finite-size behaviors.
\begin{align}
\frac{d G_0}{dh} \Big \vert_{N \to \infty}^{h \to 1} = \frac{-12(-1 + h)^2(-1 + N^2) + (73 - 90h + 33h^2)\pi^2 \ \ln N}{16\pi^3}.
\label{eq:dhmz_squared_near_h_1}
\end{align}

Therefore to determine the turning points $h_m(N)$ of $(\partial_h G_0)^2$, this limit is sufficient.
\begin{align}
h_m(N) = \frac{4 - 4N^2 + 15\pi^2 \ln N}{4 - 4N^2 + 11\pi^2 \ln N} = \ 1 - \frac{\pi^2 \ln N}{N^2} + \mathcal{O}(\frac{1}{N^4}).
\label{eq:turning_point_of__dh_mz_squared}
\end{align}

In the same way the floating exponents of the response function for a single site in TFIM can be explained. For that notice the summand within $\partial_h G_0$ and $\partial_h^2 G_0$ takes the following form near $h=1$,
\begin{align}
& \partial_h G_0 = \sum_{k \in \mathcal{K}^{0}_{\text{even}}} \mathcal{S}_k, \ \ \text{where}~ \ \mathcal{S}_k= \frac{\sin^2 k}{(2- 2\cos k)^{3/2}}+\frac{2\sin^2 k (\cos k -1)(h-1)}{(2- 2\cos k)^{5/2}} + \mathcal{O}(|h-1|^2), \nonumber \\
& \partial_h^2 G_0 = \sum_{k \in \mathcal{K}^{0}_{\text{even}}} \mathcal{S}'_k, \ \ \text{where}~ \ \mathcal{S}'_k= \frac{3\sin^2 k (\cos k -1) }{(2- 2\cos k)^{5/2}} - \frac{12 \sin^2 k (-3+5 \cos k)(h-1)}{(2- 2\cos k)^{7/2}} + \mathcal{O}(|h-1|^2). \nonumber
\end{align}

Due to the singularity at $k=0$, contributions nearest to it dominate its behavior when $N \to \infty$. Therefore taking into account only the singular contribution we find
\begin{align}
& \mathcal{S}_{k}^{\text{sing}} = \mathcal{S}_k \Big \vert_{k \to 0} = \frac{5-3h}{2k} \implies \partial_h G_0^{\text{sing}} = \frac{( 5 - 3h )\ln N}{2 \pi},\label{eq:singularity_of_dh_G0} \\
& \mathcal{S}_{k}' \ ^{\text{sing}} = \mathcal{S}_k'\Big \vert_{k \to 0} = 3\frac{1-h}{k^3} + \frac{33h - 45}{8k} \implies \partial_h^2 G_0^{\text{sing}} = -\frac{3(4(h-1)(N^2-1) + (15-11h)\pi^2 \ln N)}{8 \pi^3}. \label{eq:singularity_of_ddh_G0}
\end{align}

Since the response function for a single site in TFIM is $\Sigma^{\text{TFIM}}_1 = (1/2)(\partial_h G_0)^2/(1-G_0^2)$, its peak location is the solution of the equation
\begin{align}
\frac{G_0 (\partial_h G_0)^3}{(1-G_0^2)^2} + \frac{\partial_h G_0 \ \partial_h^2 G_0}{1-G_0^2} = 0.
\label{eq:single_site_response_turning_point_equation}
\end{align}

When $N \to \infty$, only the derivatives of correlation functions are singular and therefore the values of the correlation function can be replaced by their thermodynamic values, i.e., $G_0 \to 2/\pi$ and all the derivatives can be replaced by their corresponding singular contributions as in Eq.~\eqref{eq:singularity_of_dh_G0} and \eqref{eq:singularity_of_ddh_G0}. The solution near $h=1$ represents the asymptotic behavior of that peak location of for $N \gg 1$ with the following expression
\begin{align}
 h^c_1(N) = 1- \frac{\pi^2 ( 3 \pi^2 -12 - 4 \ln N) \ln N}{3(\pi^2-4)N^2} + \mathcal{O}(\frac{1}{N^3}) = 1+ \frac{4\pi^2 \ln^2 N}{3(\pi^2-4)N^2} - \frac{\pi^2 \ln N}{N^2} + \mathcal{O}(\frac{1}{N^3}).
\label{eq:floating_exponent_of_single_site_response} 
\end{align}

\section{Intrinsic geometry of relative entropy in TFIM} \label{subsec:intrinsic_geom_TFIM}

The local description of the quadratic form $S(\hat{\rho}_{ \vec{\lambda}} \ || \ \hat{\rho}_{\vec{\lambda}'})$ in the parameter space endows it with a Riemannian metric with the standard parameters as local coordinates that encapsulates the uncertainty of how the entanglement Hamiltonian varies - revealing the intrinsic geometry of entanglement spectra. The relation between the more well-known fidelity susceptibility and Fubini-Study metric comes from the local description of $1-\braket{\psi_{\vec{\lambda}} |\psi_{\vec{\lambda}'}}$ which satisfies the conditions $(1,2,3)$ (\textit{cf.} discussion after Eq.~\eqref{eq:def_of_metric_response_QRE}). Such an information theoretic scheme of constructing a metric response may serve as a non-perturbative measure of fluctuations of entanglement spectra in complex quantum systems without any appeal to the underlying order parameter. Therefore when sufficient smoothness of the eigenspectra are guaranteed, the parameter space can be considered being a Riemannian manifold locally described by the intrinsic line element $ds = \sqrt{\Sigma_{ij} \ d \lambda_i \ d \lambda_j}$. The corresponding first fundamental form $ds^2$ is made up of all the fluctuations with respect to individual parameters as $\Sigma_{ii}d\lambda_i^2$ and cross-fluctuations between the parameters as $\Sigma_{i\neq j}d\lambda_i\lambda_j$. Near a QCP the entangled modes in complementary subsystems are expected to fluctuate significantly under negligible change of the parameters; thus making any subsystem maximally distinguishable between infinitesimally separated points near the QCP. So one can ask how deep-rooted is this geometric view of the fluctuations of entanglement spectra. 

 We focus on the case of thermodynamically large TFIM to facilitate an analytical description with the simplest fundamental consequence of the intrinsic geometry of QRE and to examine its behavior during the QPT. Now Symmetry ensures $\braket{\sigma^x}=0=\braket{\sigma^y}\neq \braket{\sigma^z}$ so the single-site reduced density matrix takes the form of Eq.~\eqref{eq:singleSiteReduceddensityMatrix} where the transverse magnetization is given by 
\begin{align}
& m_z\equiv \braket{\sigma^z}_{\text{thermo}} = \frac{1}{2 \pi}\int^{\pi}_{-\pi} \frac{dk \ (h- \cos k)}{\sqrt{h^2+1-2 h \cos k}} \label{eq:thermo_mz_def_EliptInt} \\
&= \text{sign}(h+1)\frac{ (h + 1) \mathrm{E}(\frac{4 h }{(h + 1)^2}) + (h - 1) \mathrm{K}(\frac{4 h }{(h + 1)^2})}{\pi h}. \nonumber 
\end{align}
Here $\mathrm{K}(x)$ and $\mathrm{E}(x)$ are the complete elliptic integrals of respectively first and second kind, and $\text{sign}(h+1)$ does not manifest non-trivially since the response is a function of $m_z^2$ and $(\partial_h m_z)^2$. For the paradigmatic TFIM with one driving parameter parameter $h$, the one-dimensional line element can be thought to rescale an abstract distance with
respect to any change of that parameter for a thermodynamically large system, therefore defining a distance between corresponding points in the parameter space.

\begin{align}
& \Sigma_1^{\text{TFIM}} = \frac{1}{2}\frac{(\partial_h m_z)^2}{1-m_z^2} \implies ds = \sqrt{\Sigma_1^{\text{TFIM}}} \ dh \implies d(h_1,h_2)=\int^{h_+}_{h_-}\sqrt{\Sigma_1^{\text{TFIM}}} \ dh, \nonumber \\
& \text{where} \ \ h_+ = \max(h_1, h_2) \ ~{\rm and}~ \ h_- = \min(h_1, h_2). \nonumber
\end{align}
 
Due to the fundamental theorem of calculus the anti-derivative exists only when the integrand $\sqrt{\Sigma_1^{\text{TFIM}}}$ is a continuous function on the closed interval $[h_+, h_-]$. But in the thermodynamic limit, $\Sigma_1^{\text{TFIM}}$ is discontinuous at $|h|=1$ which is indicative of the second order QPT between paramagnet $\leftrightarrow$ ferromagnet
(\textit{cf.} Fig. \ref{fig:tfim_response_one_site} whose extrapolation for $N \to \infty$ diverges as $h\to h_c = \pm 1$). Therefore $d(h_1,h_2)$ is defined if and only if the interval $[h_1, h_2]$ does not contain $h = \pm 1$. For any such interval excluding the QCP,
one has the following expression for the distance function which satisfies the three necessary conditions of a distance, i.e., $(1)$ non-negative, $(2)$ symmetric, and $(3)$ satisfying triangular inequality, i.e., $d(a,b) \leq d(a,c)+d(c,b)$ (for one
dimension, $(3)$ is trivially satisfied, by becoming an equality). 
Hence we obtain
\begin{align}
 d(h_1,h_2) = \frac{1}{\sqrt{2}} \Big( \sin^{-1}[ m_z(h_+)] - \sin^{-1}[ m_z(h_-)] \Big). 
\end{align}

Note that this expression only works when the
interval $[h_1,h_2]$ does not contain either
of the critical points $h_c = \pm 1$. Since $- \frac{\pi}{2} \leq \sin^{-1}(x) \leq \frac{\pi}{2}$ for any $x \in [-1,1]$, $d(h_1,h_2)$ is bounded for parameters residing within either of the quantum phases; this is a consequence of the fact that
$-1 \le m_z(h) \le 1$ for all
$h$. Numerically summing up the integrand over a domain including either of the critical points $h_c = \pm 1$ diverges since $\Sigma_1^{\text{TFIM}}(h \to h_c) \to \infty$ due to the gapless spectra at any second order QPT such as this. Therefore due to the 
non-analyticity of the response function $\Sigma^{\text{TFIM}}_1$ at $h = \pm 1$, two different sides of it denote points in the parameter space corresponding to two different quantum phases; points between which do not have a well-defined analytical distance - uniquely identifies the distinct quantum phases as disjoint subspaces of the parameter space. 

For a TFIM with a finite number of sites, the distance between points on the entire parameter space of $h$ is well-defined without any restriction, further re-affirming the absence of any QPT for finite systems. Also the formalism allows one to explore the richer intrinsic geometry QRE of models with arbitrary number of parameters, potentially characterizing QPTs as geometric singularities in the entropy landscape; we leave this for future investigations.

\section{An efficient routine for exact diagonalization}
\label{subsec:bitmask_algo}
\begin{algorithm}[H]
\caption{Maximal basis reduction via Bitmasking for efficient exact diagonalization of local Ising chains}
\begin{algorithmic}[1]
\Require System size \(N\), number of sublattices \(q\), desired symmetry quantum numbers \(\{d_i\}_{i=1}^q\) for the \(D_i\) operators, momentum \(k\) (if applicable), and full integer basis \(\mathcal{B}\).
\For{\(i=1,\ldots,q\)}
 \State \textbf{Define pattern} \(P_i \in \{0,1\}^q\) such that only one pair of adjacent entries multiply to \(+1\) (e.g., for \(q=3\): \(P_1=(1,1,0)\), \(P_2=(0,1,1)\), \(P_3=(1,0,1)\)).
 \State \textbf{Extend} \(P_i\) to length \(N\) and \textbf{Form bitmask}:
 \[P_i^{(N)} \gets \underbrace{P_i\,P_i\,\cdots\,P_i}_{\lfloor N/q \rfloor \text{ times}} \, P_i[1:(N \bmod q)] \ \ \implies \ \ M_i = \sum_{j=0}^{N-1} P_i^{(N)}(j) \, 2^j \] \ \ 
\EndFor
\State \textbf{Define translation operator} \(T\): For any \(x\in\mathcal{B}\), let \(T(x)\) be the circular (periodic) bit-rotation by one site.
\For{each state \(x\in \mathcal{B}\)}
 \For{\(i=1,\ldots,q\)}
 \State Compute \(x_i \gets x \oplus M_i\).
 \EndFor
 \If{\(\forall\, i:\, x_i\) exhibits eigenvalue \(d_i\) \textbf{and} \(T(x)\) has momentum \(k\)}
 \State Include \(x\) in the reduced basis \(\mathcal{B}_{\mathrm{red}}\).
 \EndIf
\EndFor

\Return \(\mathcal{B}_{\mathrm{red}}\) and the map \(\{x \mapsto \{T(x),\, x\oplus M_i\}_{i=1}^q\}\).
\end{algorithmic}
\label{alg:maximal_basis_reduction_with_bitmasks}
\end{algorithm}

Here $x \oplus M_i$ performs bit-wise XOR representing the action of the symmetries $D_i$ in the $\sigma^x$-basis.

\section{Derivatives of local expectations in the three-spin Ising model}
\label{subsec:derivativeops}

In a many-body quantum system near a QCP, operators that have non-zero overlaps with the primary operators of the underlying CFT display singularities in the thermodynamic limit in their derivatives with respect to the driving parameter. The generic local operators that contribute in the construction of corresponding reduced density matrices overlap with the primary operators \cite{Vidal_extraction_CFT_data_2017, Vidal_CFT_&_OPEs_in_spin_systems_2020} as has been observed in Fig. \ref{fig:three_spin_Ising_three_RDMs} because $\sigma^z$ is the most relevant operator driving the QPT in both the TFIM and the
three-spin Ising model \cite{Venuti_PRA_div_loc_ent_measures__}. However 
Eq.~\eqref{eq:derivative_of_local_operator_expectation} results in either a strictly slower divergence or an $\mathcal{O}(1)$ number for the operators that only overlap with derivative descendants of the underlying CFT. 

\begin{figure*}[htbp]
 \centering
 \makebox[\textwidth]{ 
 \begin{minipage}[t]{0.33\textwidth}
 \centering
 \includegraphics[width=\textwidth, keepaspectratio]{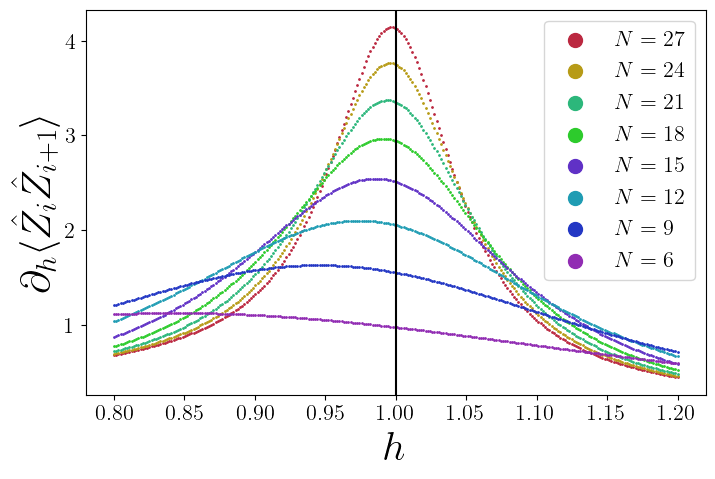}
 \subcaption{}
 \label{fig:dh_ZZ_3SI}
 \end{minipage}
 \hfill
 \begin{minipage}[t]{0.33\textwidth}
 \centering
 \includegraphics[width=\textwidth, keepaspectratio]{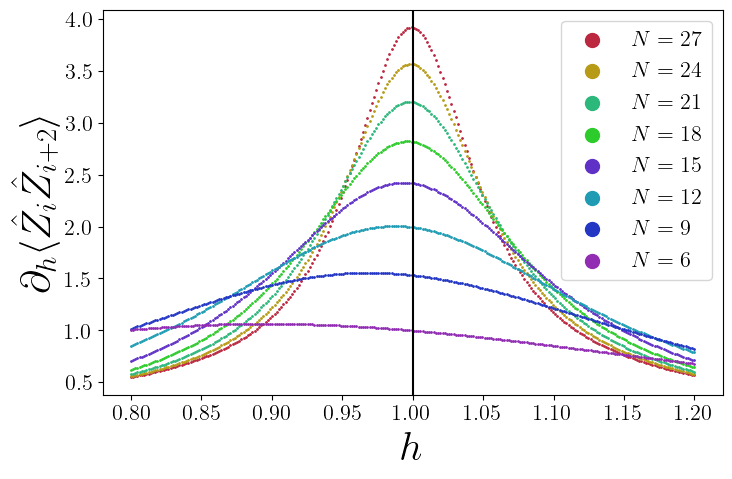}
 \subcaption{}
 \label{fig:dh_Z0Z_3SI}
 \end{minipage}
 \hfill
 \begin{minipage}[t]{0.33\textwidth}
 \centering
 \includegraphics[width=\textwidth, keepaspectratio]{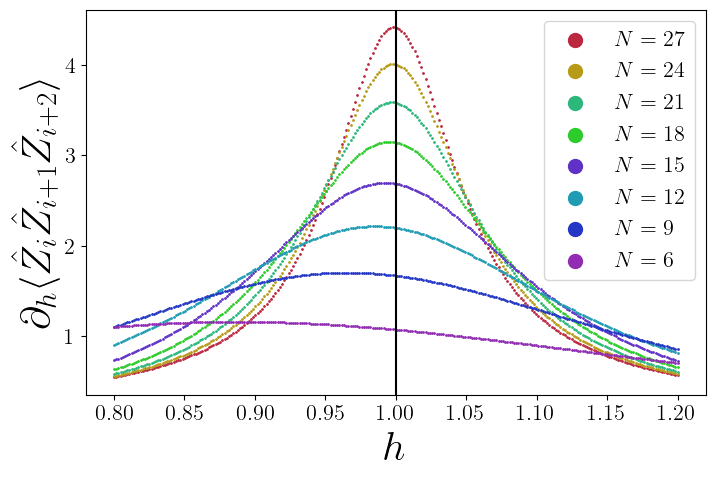}
 \subcaption{}
 \label{fig:dh_ZZZ_3SI}
 \end{minipage}
 \hfill
 }
 \makebox[\textwidth]{ 
 \begin{minipage}[t]{0.33\textwidth}
 \centering
 \includegraphics[width=\textwidth, keepaspectratio]{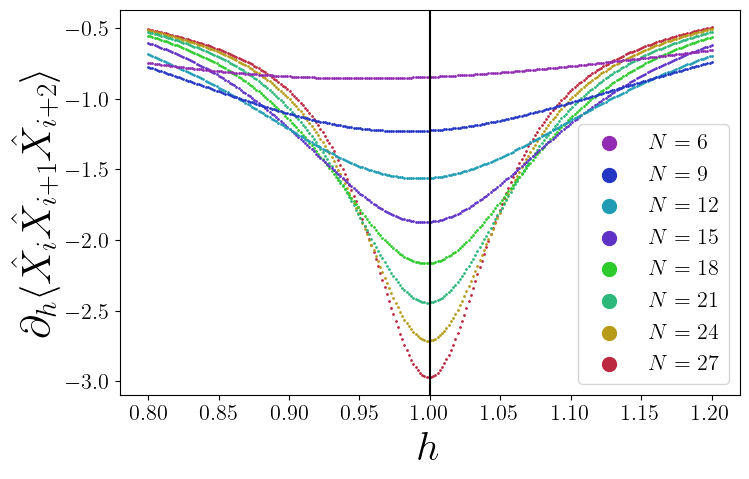}
 \subcaption{}
 \label{fig:dh_XXX_3SI}
 \end{minipage} 
 \hfill
 \begin{minipage}[t]{0.33\textwidth}
 \centering
 \includegraphics[width=\textwidth, keepaspectratio]{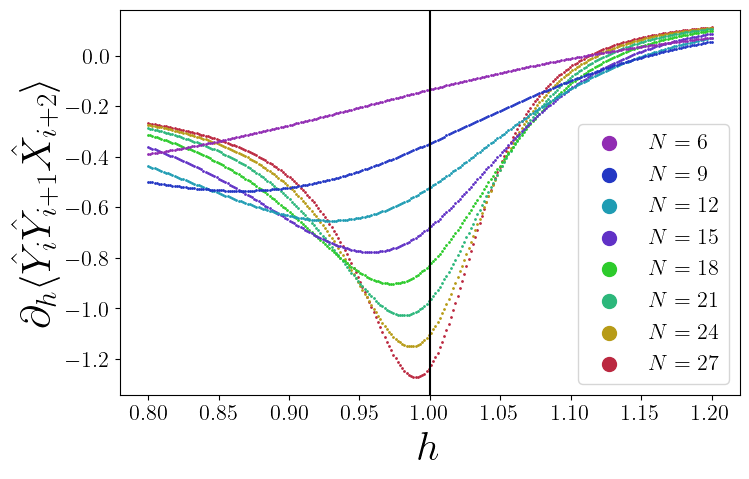}
 \subcaption{}
 \label{fig:dh_YYX_3SI}
 \end{minipage} 
 \hfill
 \begin{minipage}[t]{0.33\textwidth}
 \centering
 \includegraphics[width=\textwidth, keepaspectratio]{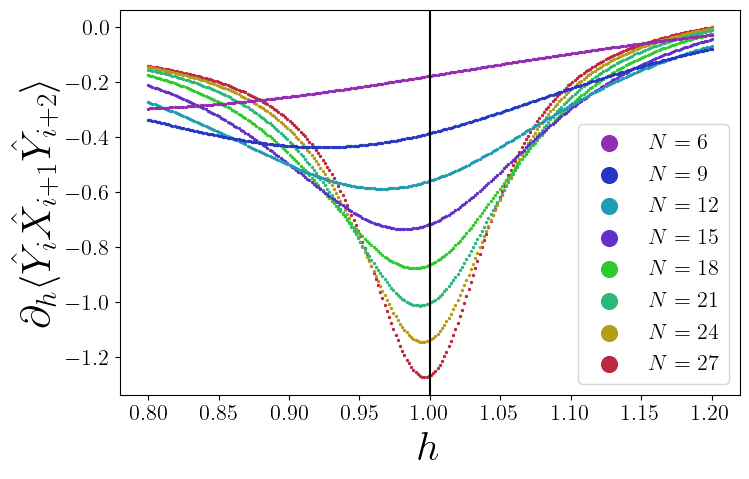}
 \subcaption{}
 \label{fig:dh_YXY_3SI}
 \end{minipage} 
 }
 \makebox[\textwidth]{ 
 \begin{minipage}[t]{0.33\textwidth}
 \centering
 \includegraphics[width=\textwidth, keepaspectratio]{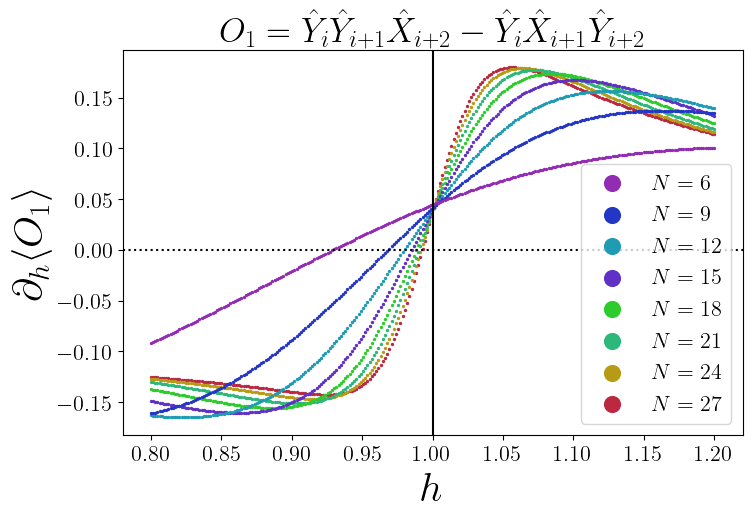}
 \subcaption{}
 \label{fig:dh_O1_3SI}
 \end{minipage}
 
 \begin{minipage}[t]{0.33\textwidth}
 \centering
 \includegraphics[width=\textwidth, keepaspectratio]{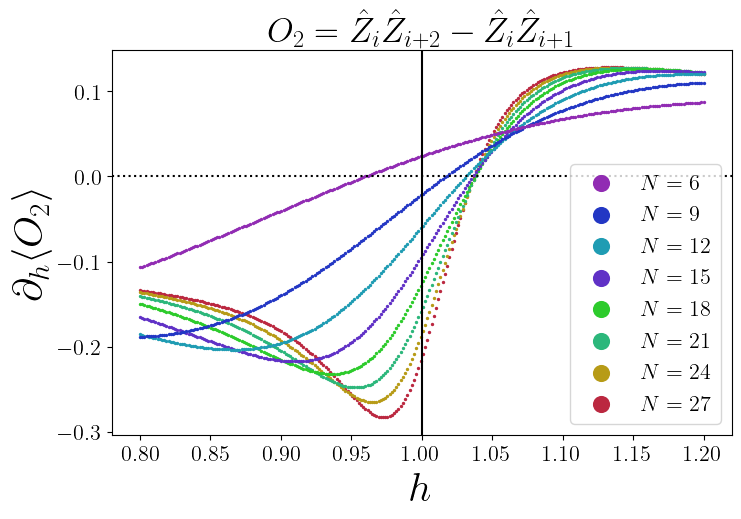}
 \subcaption{}
 \label{fig:dh_O2_3SI}
 \end{minipage} 
 \hfill
 \begin{minipage}[t]{0.33\textwidth}
 \centering
 \includegraphics[width=\textwidth, keepaspectratio]{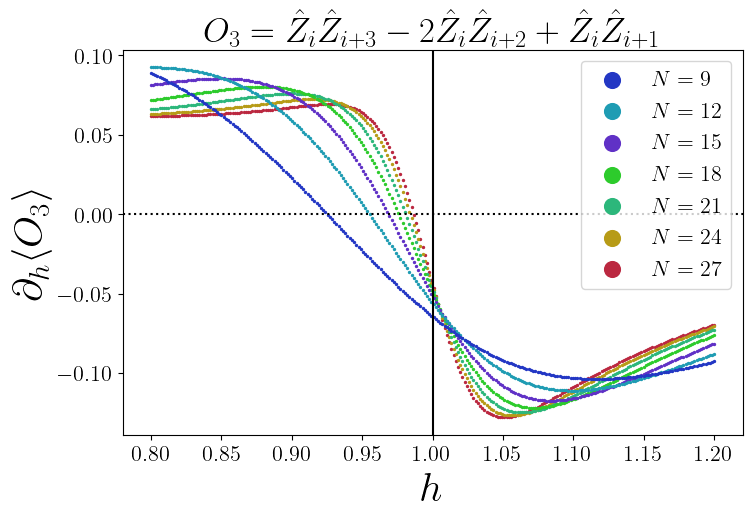}
 \subcaption{}
 \label{fig:dh_O3_3SI}
 \end{minipage} 
 }
\caption{(a-f) Derivatives of some local correlations which contribute to each local
reduced density matrix of size $m=1,2,3$
for the three-spin Ising model. (g-i) Derivatives of the expectation values of operators with
respect to $h$ which only overlap with the derivative descendants in the four-state Potts model. Here $X_i, ~Y_i$ and $Z_i$ denote the Pauli operators $\sigma^x_i, ~\sigma^y_i$ and $\sigma^z_i$ respectively.}
\label{fig:dh_local_correlations_apndx}
\end{figure*}

In the following, Fig. \ref{fig:dh_local_correlations_apndx} presents this discrepancy, for the general local operators in the three-spin Ising model that constitute the local reduced density matrices of $1, ~2$ and $3$ consecutive sites, and operators that correspond to lattice derivatives in the thermodynamic limit which are strictly less relevant operators than the most relevant one.

\end{document}